\documentclass[12pt]{iopart}
\usepackage{graphicx}
\usepackage{amssymb}

\begin{document}

\title[Recent developments in no-core shell-model calculations]{Recent developments in no-core shell-model calculations}

\author{Petr Navr{\'a}til$^1$, Sofia Quaglioni$^1$, Ionel Stetcu$^2$ and Bruce R. Barrett$^3$}
\address{$^1$Lawrence Livermore National Laboratory, P.O. Box 808,
L-414, Livermore, CA 94551, USA}
\address{$^2$Department of Physics, University of Washington, Box 351560, Seattle, Washington, 98195-1560, USA}
\address{$^3$Department of Physics, PO Box 210081, University of Arizona, Tucson, AZ 85721, USA}
\ead{navratil1@llnl.gov}
\date{\today}

\begin{abstract}
We present an overview of recent results and developments of the no-core shell model (NCSM), 
an {\it ab initio} approach to the nuclear many-body problem for light nuclei. In this approach, we start from realistic two-nucleon or two- plus three-nucleon interactions. Many-body calculations 
are performed using a finite harmonic-oscillator (HO) basis. To facilitate convergence for realistic inter-nucleon interactions that generate strong short-range correlations, we derive effective interactions by unitary transformations that are tailored to the HO basis truncation. For soft realistic interactions this might not be necessary. If that is the case, the NCSM calculations are variational. In either case, the {\it ab initio} NCSM preserves translational invariance of the nuclear many-body problem. In this review, we, in particular, highlight results obtained with the chiral two- plus three-nucleon interactions. We discuss efforts to extend the applicability of the NCSM to heavier nuclei and larger model spaces using importance-truncation schemes and/or use of effective interactions with a core. We outline an extension of the {\it ab initio} NCSM to the description of nuclear reactions by the resonating group method technique. A future direction of the approach, the {\it ab initio} NCSM with continuum, which will provide a complete description of nuclei as open systems with coupling of bound and continuum states, is given in the concluding part of the review.
\end{abstract}

\maketitle

\section{Introduction}

A major outstanding problem in nuclear physics is to calculate properties of finite nuclei
starting from the basic interactions among nucleons. This problem has two parts. First, the
basic interactions among nucleons are complicated. They are not uniquely defined and there
is clear evidence that more than just two-nucleon forces are important. Second, the nuclear many-body
problem is very difficult to solve. This is a direct consequence of the complex nature
of the inter-nucleon interactions. Both short-range and medium-range correlations among 
nucleons are important and for some observables long-range correlations also play a 
significant role. 

In this review, we focus mainly on the second part of the problem, namely on the solution 
of the many-nucleon problem. The two-nucleon interactions we take as an input provided 
to us by other theorists. We do, however, utilize three- and many-nucleon calculations
to determine parameters of three-nucleon interactions. We also note that, in general, few-nucleon and 
many-nucleon calculations provide feedback to those constructing potentials.

Various methods have been used to solve the few-nucleon problem 
in the past. The Faddeev method \cite{Fad60}
has been successfully applied to solve the three-nucleon bound-state as well as the scattering
problem for different nucleon-nucleon (NN) potentials \cite{CPFG85,FPSS93,NHKG97}.
For the solution of the four-nucleon problem one can employ Yakubovsky's generalization
of the Faddeev formalism \cite{Ya67}, as done, e.g., in Refs. \cite{GH93} or \cite{CC98}.
Alternatively, other methods have also been succesfully used, such as, 
the correlated hyperspherical harmonics expansion method \cite{VKR95,BLO99}
or the Green's function Monte Carlo method (GFMC) \cite{GFMC}. 
Recently, a benchmark calculation by seven different methods
was performed for a four-nucleon bound state problem \cite{benchmark} giving the same result within error.
However, there are few approaches that can be successfully applied to solve the bound-state
problem in systems of more than four nucleons, when realistic inter-nucleon interactions are used.
These include the Green's function Monte Carlo method, which is capable of solving the nuclear many-body problem with realistic interactions for systems of up to $A=12$ and the coupled cluster method \cite{CCM,HM99,KoDe04,WlDe05,Ha08}, which is applicable typically to closed-shell and nearby nuclei.

The solution of the nuclear many-body problem is still more complex 
when scattering or nuclear reactions are considered. 
For $A=3$ and 4 nucleon systems, the Faddeev and Faddeev-Yakubovsky as well
as the hyperspherical harmonics (HH) \cite{Pisa} or the Alt, Grassberger and Sandhas (AGS) 
\cite{Deltuva} methods are applicable and successful. 
However, {\em ab initio} calculations for scattering processes involving more than four 
nucleons overall are challenging and still a rare exception~\cite{GFMC_nHe4}.

Nuclei are open systems with bound states, weakly bound halo states, unbound resonances
as well as scattering states.
A realistic {\it ab initio} description of light nuclei with predictive power must
have a capability to describe all the above classes of states within a unified framework.
Coupling to the continuum cannot be neglected.

In this review, we describe the {\it ab initio} no-core shell model 
(NCSM) \cite{NaVa00}, another method for solving the nuclear many-body problem developed 
recently and applicable to light nuclei up to $A=16$ and beyond. 
The first no-core shell model calculations~\cite{ZBV} were performed with G-matrix-based two-body
interactions~\cite{BHM71}. Later, the Okubo-Lee-Suzuki procedure~\cite{Okubo,LS80}
was implemented to derive two-body effective interactions for the  NCSM~\cite{NB96}. 
This resulted in the elimination of the purely phenomenological parameter used to
define the G-matrix starting energy. A truly {\it ab initio} formulation of
the approach was presented in Ref.~\cite{three_NCSM}, where convergence to the exact
bound-state solutions was demonstrated for the $A=3$ system. 
Here, in the first part of the review, we discuss the NCSM in its standard formulation, applicable to the nuclear bound-state problem.  In Sect.~\ref{sec:NCSM}, we briefly present the NCSM formalism. In Sect.~\ref{sec:chiral_NN_NNN}, we show recent results obtained with the chiral
NN plus three-nucleon (NNN) interactions. Calculations of radii, moments and transitions of He, Li and Be isotopes using different realistic NN potentials are discussed in Sect.~\ref{sec:Mom}.
In Sect.~\ref{sec:Ext}, we describe efforts to extend the applicability
of the NCSM to heavier nuclei. In the second part of the review in Sections~\ref{sec:overlap}-\ref{sec:NCSMC}, we discuss in detail new developments and outline future efforts to extend the NCSM by including continuum states to describe unbound states, scattering and nuclear reactions in a unified framework. Conclusions are given in Sect.~\ref{sec:Concl}.

\section{{\it Ab initio} no-core shell model}\label{sec:NCSM}

In the {\it ab initio} no-core shell model, we consider a system 
of $A$ point-like non-relativistic nucleons that interact by realistic two- 
or two- plus three-nucleon interactions. By the term ``realistic two-nucleon
interactions'', we mean NN potentials that fit nucleon-nucleon phase shifts 
with high precision up to a certain energy, typically up to 350 MeV. A realistic
NNN interaction includes terms related to two-pion exchanges with an intermediate
delta excitation. In the NCSM, all the nucleons are considered active, 
there is no inert core like in
standard shell model calculations. Hence, the ``no-core'' in the name of the approach.

There are two other major features in addition to the employment of realistic 
NN or NN+NNN interactions. The first one is
the use of the harmonic oscillator (HO) basis, truncated by a chosen maximal total HO energy
of the $A$-nucleon system. The reason behind the choice of the HO basis is the fact that
this is the only basis that allows for the use of single-nucleon coordinates and, consequently, 
the second-quantization representation, without violating the translational invariance
of the system. The powerful techniques based on the second quantization and developed 
for standard shell model calculations can then be utilized. Therefore, the ``shell model''
in the name of the approach. As a downside, one has to face the consequences of the incorrect
asymptotic behavior of the HO basis. 

The second feature comes as a result of the basis truncation.
Standard, accurate NN potentials, such as the Argonne V18 (AV18) \cite{AV18}, CD-Bonn 2000 \cite{cdb2k}, INOY (inside non-local outside Yukawa) \cite{INOY} and, to some extent, also the chiral N$^3$LO \cite{N3LO}, generate strong short-range correlations that cannot be accomodated even in a reasonably large HO basis.  
In order to account for these short-range correlations and to speed up convergence with the basis enlargement, we construct an effective interaction
from the original, realistic NN or NN+NNN potentials by means of a unitary transformation.
The effective interaction depends on the basis truncation and by construction becomes
the original, realistic NN or NN+NNN interaction as the size of the basis approaches infinity.

Recently, a new class of soft potentials has been developed, mostly by means of unitary transformations
of the standard, accurate NN potentials mentioned above. These include the $V_{low {\it k}}$ \cite{Vlowk}, the Similarity Renormalization Group (SRG) \cite{SRG} and the UCOM \cite{UCOM} NN potentials. A different class of soft phenomenological NN potential used in some NCSM calculations are the JISP potentials based on inverse scattering~\cite{JISP}. These soft potentials are to some extent already renormalized for the purpose of simplifying many-body calculations. Therefore, we can perform convergent NCSM calculations with these potentials unmodified, or ``bare.'' In fact, the chiral N$^3$LO NN potential \cite{N3LO} can also be used bare with some success. NCSM calculations with bare potentials are variational with the HO frequency and the basis truncation parameter as variational parameters.

\subsection{Hamiltonian}

The starting Hamiltonian of the {\it ab initio} NCSM is
\begin{equation}\label{ham}
H_A= 
\frac{1}{A}\sum_{i<j}\frac{(\vec{p}_i-\vec{p}_j)^2}{2m}
+ \sum_{i<j}^A V_{{\rm NN}, ij} + \sum_{i<j<k}^A V_{{\rm NNN}, ijk} \; ,
\end{equation}
where $m$ is the nucleon mass, $V_{{\rm NN}, ij}$ is the NN interaction, and
$V_{{\rm NNN}, ijk}$ is the three-nucleon interaction. In the NCSM, we employ a large
but finite HO basis. When soft NN potentials are used, it is often feasible to employ 
a sufficiently large basis to reach convergence with the Hamiltonian (\ref{ham}).

On the other hand, if realistic nuclear interactions 
that generate strong short-range correlations are used in Eq. (\ref{ham}),
we must derive an effective interaction appropriate for the basis truncation.
To facilitate the derivation of the effective interaction, we modify the
Hamiltonian (\ref{ham}) by adding to it the center-of-mass (CM) HO Hamiltonian
$H_{\rm CM}=T_{\rm CM}+ U_{\rm CM}$, where
$U_{\rm CM}=\frac{1}{2}Am\Omega^2 \vec{R}^2$,
$\vec{R}=\frac{1}{A}\sum_{i=1}^{A}\vec{r}_i$.
The effect of the HO CM Hamiltonian will later be subtracted
out in the final many-body calculation. Due to the translational invariance of the
Hamiltonian (\ref{ham}), the HO CM Hamiltonian has in fact no effect on the intrinsic
properties of the system. 
The modified Hamiltonian can be cast into the form
\begin{eqnarray}\label{hamomega}
H_A^\Omega &=& H_A + H_{\rm CM}=\sum_{i=1}^A h_i + \sum_{i<j}^A V_{ij}^{\Omega,A}
+\sum_{i<j<k}^A V_{{\rm NNN}, ijk} 
\nonumber \\ 
&=& \sum_{i=1}^A \left[ \frac{\vec{p}_i^2}{2m}
+\frac{1}{2}m\Omega^2 \vec{r}^2_i
\right] 
+ \sum_{i<j}^A \left[ V_{{\rm NN}, ij}
-\frac{m\Omega^2}{2A}
(\vec{r}_i-\vec{r}_j)^2
\right] 
\nonumber \\ 
&&
+ \sum_{i<j<k}^A V_{{\rm NNN}, ijk} \; .
\end{eqnarray}

\subsection{Basis}

In the {\it ab initio} NCSM, we use a HO basis that allows preservation of translational symmetry of the nuclear self-bound system, even if single-nucleon coordinates are utilized. This is possible as long as the basis is truncated  by a maximal total HO energy of the $A$-nucleon system. A further advantage is that the HO wave functions have important transformation properties~\cite{Moshinsky} that facilitate and simplify calculations. A single-nucleon HO wave function can be written as
\begin{equation}\label{HOwave}
\varphi_{nlm}(\vec{r};b)=R_{nl}(r;b)Y_{lm}(\hat{r})  \; ,
\end{equation}
with $R_{nl}(r,b)$, the radial HO wave function, and $b$, the HO length parameter 
related to the HO frequency $\Omega$ as $b=\sqrt{\frac{\hbar}{m\Omega}}$, with $m$ the nucleon mass.

Because the NN and NNN interactions depend on relative coordinates and/or momenta, the
natural coordinates in the nuclear problem are the relative, or Jacobi, coordinates.
For the present purposes we consider just a single set
of Jacobi coordinates (a more general discussion can be found in Ref.~\cite{Jacobi_NCSM}):
%
\begin{eqnarray}\label{jacobiam11}
\vec{\xi}_0 &=& \sqrt{\frac{1}{A}}\left[\vec{r}_1+\vec{r}_2
                                   +\ldots +\vec{r}_A\right]
\; , \\
\vec{\xi}_1 &=& \sqrt{\frac{1}{2}}\left[\vec{r}_1-\vec{r}_2
                                                     \right]
\; , \\
\vec{\xi}_2 &=& \sqrt{\frac{2}{3}}\left[\frac{1}{2}
                 \left(\vec{r}_1+\vec{r}_2\right)
                                   -\vec{r}_3\right]
\; , \\
&\ldots & \nonumber
\\
\vec{\xi}_{A-1} &=& \sqrt{\frac{A-1}{A}}\left[\frac{1}{A-1}
      \left(\vec{r}_1+\vec{r}_2 + \ldots+ \vec{r}_{A-1}\right)
                                   -\vec{r}_{A}\right]
\; ,
\end{eqnarray}
%
Here, $\vec{\xi}_0$ is proportional to the center of mass of the
$A$-nucleon system. On the other hand, $\vec{\xi}_\rho$ is proportional
to the relative position of the $\rho+1$-st nucleon and the
center of mass of the $\rho$ nucleons. 

\subsubsection{Antisymmetrization of Jacobi-coordinate HO basis}
  
As nucleons are fermions, we need to construct an antisymmetrized basis.
The way to do this, when the Jacobi-coordinate HO basis is used, is extensively discussed 
in Refs.~\cite{three_NCSM,Jacobi_NCSM,four_NCSM}.
Here we briefly illustrate how to do this for the simplest case of three nucleons.

One starts by introducing a HO basis that depends on the Jacobi coordinates
$\vec{\xi}_1$ and $\vec{\xi}_2$, defined in Eqs. (5) and (6), 
e.g.,
\begin{equation}\label{hobas}
|(n l s j t; {\cal N} {\cal L} {\cal J}) J T \rangle \; .
\end{equation}
Here $n, l$ and ${\cal N}, {\cal L}$ are the HO quantum numbers
corresponding to the harmonic oscillators associated with the coordinates 
(and the corresponding momenta) $\vec{\xi}_1$ and $\vec{\xi}_2$, respectively. 
The quantum numbers $s,t,j$ describe the spin, isospin and angular momentum
of the relative-coordinate two-nucleon channel of nucleons 1 and 2, while 
${\cal J}$ is the angular momentum of the third nucleon relative to the
center of mass of nucleons 1 and 2. The $J$ and $T$ are the total angular 
momentum and the total isospin, respectively.
Note that the basis (\ref{hobas}) is antisymmetrized with respect
to the exchanges of nucleons 1 and 2, as the two-nucleon channel
quantum numbers are restricted by the condition $(-1)^{l+s+t}=-1$.
It is not, however, antisymmetrized with respect to the exchanges of nucleons
$1\leftrightarrow 3$ and $2\leftrightarrow 3$.
In order to construct a completely antisymmetrized basis, one needs to 
obtain eigenvectors of the antisymmetrizer 
\begin{equation}\label{antisymm3}
{\cal X}=\frac{1}{3}(1+{\cal T}^{(-)}+{\cal T}^{(+)}) \; ,
\end{equation}
where ${\cal T}^{(+)}$ and ${\cal T}^{(-)}$ are the cyclic and the anti-cyclic 
permutation operators, respectively. The antisymmetrizer ${\cal X}$
is a projector satisfying ${\cal X} {\cal X}={\cal X}$. 
When diagonalized in the basis (\ref{hobas}), its eigenvectors
span two eigenspaces. One, corresponding to the eigenvalue 1, is formed
by physical, completely antisymmetrized states and the other, corresponding
to the eigenvalue 0, is formed by spurious states. There are about 
twice as many spurious states as the physical ones \cite{NBG99}.

Due to the antisymmetry with respect to the exchanges $1\leftrightarrow 2$,
the matrix elements in the basis (\ref{hobas}) of the antisymmetrizer 
${\cal X}$ can be evaluated simply as 
$\langle {\cal X} \rangle = \frac{1}{3}\langle 1-2 P_{2,3}\rangle $,
where $P_{2,3}$ is the transposition operator corresponding to the exchange
of nucleons 2 and 3.
Its matrix element can be evaluated in a straightforward way (see e.g., Ref.~\cite{three_NCSM})
\begin{eqnarray}\label{t13t23}
&&\langle (n_1 l_1 s_1 j_1 t_1; {\cal N}_1 {\cal L}_1  
{\cal J}_1) J T | P_{2,3} |  
(n_2 l_2 s_2 j_2 t_2; {\cal N}_2 {\cal L}_2  
{\cal J}_2) J T\rangle 
\nonumber \\
&=& \delta_{N_1,N_2} \hat{t}_1 \hat{t}_2
\left\{ \begin{array}{ccc} \textstyle{\frac{1}{2}} & \textstyle{\frac{1}{2}} 
               & t_1 \\
             \textstyle{\frac{1}{2}}  &  T  & t_2
\end{array}\right\}
\nonumber \\
&& \times \sum_{LS} \hat{L}^2 \hat{S}^2
\hat{j}_1 \hat{j}_2 \hat{\cal J}_1 \hat{\cal J}_2 \hat{s}_1 \hat{s}_2
 (-1)^L
          \left\{ \begin{array}{ccc} l_1   & s_1   & j_1   \\
          {\cal L}_1  & \textstyle{\frac{1}{2}}  & {\cal J}_1 \\
                                     L   & S  & J
\end{array}\right\}
          \left\{ \begin{array}{ccc} l_2   & s_2   & j_2   \\
       {\cal L}_2  & \textstyle{\frac{1}{2}}   & {\cal J}_2 \\
                                     L   & S  & J
\end{array}\right\}
\nonumber \\
&& \times
\left\{ \begin{array}{ccc} \textstyle{\frac{1}{2}} & \textstyle{\frac{1}{2}} 
               & s_1 \\
             \textstyle{\frac{1}{2}}  &  S  & s_2
\end{array}\right\}
\langle n_1 l_1 {\cal N}_1 {\cal L}_1 L 
| {\cal N}_2 {\cal L}_2 n_2 l_2 L \rangle_{\rm 3} \; ,
\end{eqnarray}
where $N_i=2n_i+l_i+2{\cal N}_i+{\cal L}_i, i= 1,2$; 
$\hat{j}=\sqrt{2j+1}$; 
and $\langle n_1 l_1 {\cal N}_1 {\cal L}_1  L 
| {\cal N}_2 {\cal L}_2 n_2 l_2 L \rangle_{\rm 3}$
is the general HO bracket for two particles with mass 
ratio 3, as defined, e.g., in Ref. \cite{Tr72}.
The expression (\ref{t13t23}) can be derived by examining 
the action of $P_{2,3}$ on the basis states (\ref{hobas}).
That operator changes the state 
$|nl(\vec{\xi}_1),{\cal NL}(\vec{\xi}_2), L\rangle$ 
to $|nl(\vec{\xi'}_1),{\cal NL}(\vec{\xi'}_2), L\rangle$,
where $\vec{\xi'}_i, i=1,2$ are defined as $\vec{\xi}_i, i=1,2$
but with the single-nucleon indexes 2 and 3 exchanged. The primed Jacobi 
coordinates can be expressed as an orthogonal transformation
of the unprimed ones, see e.g., Ref.~\cite{three_NCSM}. 
Consequently, the HO wave functions
depending on the primed Jacobi coordinates can be expressed
as an orthogonal transformation of the original HO wave functions.
Elements of the transformation are the generalized HO brackets
for two particles with the mass ratio $d$, with $d$ determined
from the orthogonal transformation of the coordinates, see e.g. Ref.~\cite{Tr72}.

The resulting antisymmetrized states can be classified 
and expanded in terms of the original basis (\ref{hobas}) as follows
\begin{equation}\label{abas3}
|N i J T\rangle = \sum \langle nlsjt; {\cal NLJ}||N i J T\rangle 
|(nlsjt;{\cal NLJ}) JT\rangle \; ,
\end{equation}
where $N=2n+l+2{\cal N}+{\cal L}$ and 
where we have introduced an
additional quantum number $i$ that distinguishes states with
the same set of quantum numbers $N, J, T$, e.g.,
$i=1,2, \ldots r$ with 
$r$ the total number of antisymmetrized states for a given $N, J, T$.
The symbol $\langle nlsjt; {\cal NLJ}||N i J T\rangle$ is a coefficient
of fractional parentage.

\subsubsection{Slater determinant basis}

A generalization to systems of more than three nucleons can be done as shown,
e.g. in Ref.~\cite{Jacobi_NCSM}. It is obvious, however, that as we increase the number 
of nucleons, the antisymmetrization becomes more and more involved.
Consequently, in standard shell model calculations one utilizes antisymmetrized
wave functions constructed in a straightforward way as Slater determinants of 
single-nucleon wave functions depending on single-nucleon coordinates $\varphi_i(\vec{r}_i)$. 
It follows from the transformations of HO wave functions that the use of a Slater determinant basis 
constructed from single nucleon HO wave functions, such as,
\begin{equation}\label{HOwavesp}
\varphi_{nljm m_t}(\vec{r},\sigma,\tau;b)=R_{nl}(r;b)
(Y_{l}(\hat{r})\chi(\sigma))^{(j)}_m\chi(\tau)_{m_t}  \; ,
\end{equation}
results in eigenstates of a translationally invariant Hamiltonian that factorize
as products of a wave function depending on relative coordinates and a wave function 
depending on the CM
coordinates. This is true as long as the basis truncation is done by a chosen maximum
of the sum of all HO excitations, i.e., $\sum_{i=1}^A(2n_i+l_i)\leq N_{totmax}$.
In Eq.~(\ref{HOwavesp}), $\sigma$ and $\tau$ are spin and isospin coordinates of the nucleon, respectively.
The physical eigenstates of a translationally invariant Hamiltonian can then be selected
as eigenstates with the CM in the $0\hbar\Omega$ state:
\begin{eqnarray}\label{state_relation}
&&\langle \vec{r}_1 \ldots \vec{r}_A \sigma_1 \ldots \sigma_A \tau_1 \ldots \tau_A 
| A \lambda J M T M_T\rangle_{\rm SD} 
\nonumber \\
&=& 
\langle \vec{\xi}_1 \ldots \vec{\xi}_{A-1}
\sigma_1 \ldots \sigma_A \tau_1 \ldots \tau_A 
| A \lambda J M T M_T\rangle \varphi_{000}(\vec{\xi}_0;b) \; .
\end{eqnarray}
For a general single-nucleon wave function this factorization is not possible. 
The use of any other single-nucleon wave function than the HO wave function will result in the mixing 
of CM and internal motion.

In the {\it ab initio} NCSM calculations, we use both the Jacobi-coordinate HO basis
and the single-nucleon Slater determinant HO basis. One can choose whichever
is more convenient for the problem to be solved. One can also mix the two types of bases.
In general, for systems of $A\leq4$, the Jacobi coordinate basis is more efficient, as one can
perform the antisymmetrization easily. The CM degrees of freedom can be explicitly removed
and a coupled $J^\pi T$ basis can be utilized with matrix dimensions of the order of thousands.
For systems with $A>4$, it is, in general, more efficient to use the Slater determinant
HO basis. In fact, we use the so-called m-scheme basis with conserved 
quantum numbers $M=\sum_{i=1}^Am_i$,
parity $\pi$ and $M_T=\sum_{i=1}^A m_{ti}$. The antisymmetrization is trivial, but the dimensions
can be huge, as the CM degrees of freedom are present, and no $J T$ coupling is considered.
The advantage is the possibility to utilize the powerful second-quantization technique,
shell model codes, transition density codes and so on.

As mentioned above, the model space truncation is always done using the condition
$\sum_{i=1}^A(2n_i+l_i)\leq N_{totmax}$. Often, instead of $N_{totmax}$, we introduce
the parameter $N_{\rm max}$ that measures the maximal allowed HO excitation energy above
the unperturbed ground state. For $A=3,4$ systems $N_{\rm max}=N_{totmax}$. For the $p$-shell nuclei
they differ, e.g. for $^6$Li,  $N_{\rm max}=N_{totmax}-2$, for $^{12}$C, $N_{\rm max}=N_{totmax}-8$, {\it etc}.

\subsection{Effective interaction}
\label{sec:effHam}

In the {\it ab initio} NCSM calculations we use a truncated HO basis,
as discussed in previous sections. The inter-nucleon interactions
act, however, in the full space. As long as one uses soft potentials, such as 
the $V_{low {\it k}}$, SRG, UCOM or JISP, convergent NCSM results can be obtained. Such NCSM calculations are variational with the HO frequency and the basis truncation parameter $N_{\rm max}$ acting as variational parameters. 

However, the situation is different when standard NN potentials that generate strong short-range correlations, such as AV18, CD-Bonn 2000, and INOY, are used, or when a not-large-enough $N_{\rm max}$ 
truncation can be reached with the chiral N$^3$LO NN potential (in particular, when 
it is used in combination with the chiral NNN interaction). 
In order to obtain meaningful results in the truncated (or model) space, 
the inter-nucleon interactions need to be renormalized. 
We need to construct an effective Hamiltonian 
with the inter-nucleon interactions replaced by effective interactions.
By meaningful results we understand results as close as possible to the 
full space exact results for a subset of eigenstates. Mathematically 
we can construct an effective Hamiltonian that exactly reproduces
the full space results for a subset of eigenstates. In practice, we cannot in general
construct this exact effective Hamiltonian for the $A$-nucleon problem
we want to solve. However, we can construct an effective Hamiltonian that is exact
for a two-nucleon system or for a three-nucleon system or even for a four-nucleon system.
The corresponding effective interactions can then be used in the $A$-nucleon calculations.
Their use, in general, improves the convergence of the problem to the exact full space result
with the increase of the basis size. By construction, these effective interactions
converge to the full-space inter-nucleon interactions, therefore, guaranteeing
convergence to the exact solution, when the basis size approaches the infinite full space.

In our approach we employ the so-called Okubo or Lee-Suzuki similarity transformation
method \cite{Okubo,LS80,S82SO83}, which yields 
a starting-energy independent hermitian effective interaction.
We first recapitulate general formulation and basic results
of this method. Applications of this method for computation of two- or
three-body effective interactions are described afterwards. 

\subsubsection{Lee-Suzuki similarity transformation method}

Let us consider an ${\it arbitrary}$ Hamiltonian $H$ with the eigensystem
$E_k, |k\rangle$, i.e.,
\begin{equation}\label{schreq}
H|k\rangle = E_k |k\rangle \; .
\end{equation}
Let us further divide the full space into the model space defined by 
a projector $P$ and the complementary space defined by a projector
$Q$, $P+Q=1$.
A similarity transformation of the Hamiltonian $e^{-\omega} H e^\omega$ 
can be introduced with a transformation operator $\omega$ satisfying 
the condition $\omega= Q \omega P$. The transformation operator
is then determined from the requirement of decoupling of the Q-space and
the model space as follows
\begin{equation}\label{decoupl}
Q e^{-\omega} H e^\omega P = 0 \; .
\end{equation}
Using a Feshbach construction, one can show that the particular
choice of the decoupling condition (\ref{decoupl}) ensures that the
effective Hamiltonian is energy independent \cite{Navratil:1993fesh}.
If we denote the model space basis states as $|\alpha_P\rangle$,
and those which belong to the Q-space, as $|\alpha_Q\rangle$,
then the relation $Q e^{-\omega} H e^\omega P |k\rangle = 0$,
following from Eq. (\ref{decoupl}), will be satisfied for a particular
eigenvector $|k\rangle$ of the Hamiltonian (\ref{schreq}),
if its Q-space components can be expressed as a combination
of its P-space components with the help of the transformation operator 
$\omega$, i.e.,
\begin{equation}\label{eigomega}  
\langle\alpha_Q|k\rangle=\sum_{\alpha_P}
\langle\alpha_Q|\omega|\alpha_P\rangle \langle\alpha_P|k\rangle \; .
\end{equation}
If the dimension of the model space is $d_P$, we may choose a set
${\cal K}$ of $d_P$ eigenevectors, 
for which the relation (\ref{eigomega}) 
will be satisfied. Under the condition that the $d_P\times d_P$ 
matrix defined by the matrix elements $\langle\alpha_P|k\rangle$ for $|k\rangle\in{\cal K}$
is invertible, the operator $\omega$ can be determined from 
(\ref{eigomega}) as
\begin{equation}\label{omegasol}
\langle\alpha_Q|\omega|\alpha_P\rangle = \sum_{k \in{\cal K}}
\langle\alpha_Q|k\rangle\langle\tilde{k}|\alpha_P\rangle \; ,
\end{equation}  
where we denote by tilde the inverted matrix of $\langle\alpha_P|k\rangle$, e.g.,
$\sum_{\alpha_P}\langle\tilde{k}|\alpha_P\rangle\langle\alpha_P
|k'\rangle = \delta_{k,k'}$, for $k,k'\in{\cal K}$.

The hermitian effective Hamiltonian defined on the model space $P$
is then given by \cite{S82SO83}
\begin{equation}\label{hermeffomega}
\bar{H}_{\rm eff}
=\left[P(1+\omega^\dagger\omega)P\right]^{1/2}
PH(P+Q\omega P)\left[P(1+\omega^\dagger\omega)
P\right]^{-1/2} \; .
\end{equation}
By making use of the properties of the operator $\omega$,
the effective Hamiltonian $\bar{H}_{\rm eff}$ can be rewritten
in an explicitly hermitian form as
\begin{eqnarray}\label{exhermeff}
\bar{H}_{\rm eff}
&=&\left[P(1+\omega^\dagger\omega)P\right]^{-1/2}
(P+P\omega^\dagger Q)H(Q\omega P+P)
\nonumber \\
&&\times\left[P(1+\omega^\dagger\omega)P\right]^{-1/2} \; .
\end{eqnarray}
With the help of the solution for $\omega$ (\ref{omegasol})
we obtain a simple expression for the matrix elements of 
the effective Hamiltonian
\begin{eqnarray}\label{effham}
\langle \alpha_P | \bar{H}_{\rm eff} |\alpha_{P'}\rangle
&=& \sum_{\alpha_{P''}}\sum_{\alpha_{P'''}} 
\sum_{kk'k'' \in{\cal K}}\langle \alpha_P |\tilde{k}'' \rangle \langle \tilde{k}'' 
|\alpha_{P''}\rangle
\nonumber \\
&&\times
\langle \alpha_{P''} |\tilde{k}\rangle E_k 
\langle\tilde{k}|\alpha_{P'''}\rangle
\langle \alpha_{P'''} | \tilde{k}' \rangle \langle \tilde{k}'
|\alpha_{P'}\rangle \; .
\end{eqnarray}
with all the summations over the Q-space basis states removed.
The effective Hamiltonian (\ref{effham}) reproduces the
eigenenergies $E_k, k\in{\cal K}$ in the model space.

It has been shown \cite{UMOA} that the hermitian effective 
Hamiltonian (\ref{exhermeff}) can be obtained
directly by a unitary transformation of the original Hamiltonian:
\begin{equation}\label{UMOAtrans}
\bar{H}_{\rm eff} = Pe^{-S} H e^{S}P \; ,
\end{equation}
with an anti-hermitian operator $S={\rm arctanh}(\omega-\omega^\dagger)$.
The transformed Hamiltonian then satisfies decoupling conditions
$Qe^{-S} H e^{S}P=Pe^{-S} H e^{S}Q=0$.

We can see from Eqs.~(\ref{effham}) that in order
to construct the effective Hamiltonian we need to know a subset of exact
eigenvalues and model space projections of a subset of exact eigenvectors.
This may suggest that the method is rather impractical. Also, it follows
from Eq.~(\ref{effham}) that the effective Hamiltonian contains many-body terms,
in fact for an $A$-nucleon system, all terms up to $A$-body will in general 
appear in the effective Hamiltonian, even if the original Hamiltonian
consisted of just two-body or two- plus three-body terms.

\subsubsection{Two-body effective interaction in the NCSM}\label{sec:tbeff}

In the {\it ab initio} NCSM we use the above effective interaction theory
as follows. Since the two-body part dominates the $A$-nucleon 
Hamiltonian (\ref{hamomega}), it is reasonable to expect that a two-body
effective interaction that takes into account full space two-nucleon correlations
would be the most important part of the exact effective interaction. If the NNN
interaction is taken into account, a three-body effective interaction that
takes into account full space three-nucleon correlations would be a good
approximation to the exact $A$-body effective interaction. We construct
the two-body or three-body effective interaction by application of 
the above described Lee-Suzuki procedure to a two-nucleon or three-nucleon system.
The resulting effective interaction is then exact for the two- or three-nucleon
system. It is an approximation of the {\it exact} $A$-nucleon effective interaction.

Using the notation of Eq.(\ref{hamomega}), the two-nucleon effective interaction is obtained as
\begin{equation}\label{V2eff}
V_{\rm 2eff,12} = P_2[e^{-S_{12}}(h_1+h_2+V^{\Omega,A}_{12})e^{S_{12}}-(h_1+h_2)]P_2   \; ,
\end{equation}
with $S_{12}={\rm arctanh}(\omega_{12}-\omega_{12}^\dagger)$ and $P_2$ is a two-nucleon
model space projector. The two-nucleon model space is defined by a truncation $N_{\rm 12max}$
corresponding to the $A$-nucleon $N_{\rm max}$. For example, for $A=3,4$, $N_{\rm 12max}=N_{\rm max}$,
for $p$-shell nuclei with $A>5$ $N_{\rm 12max}=N_{\rm max}+2$. The operator $\omega_{12}$
is obtained with the help of Eq. (\ref{omegasol}) from exact solutions of the Hamiltonian
$h_1+h_2+V^{\Omega,A}_{12}$, which are straightforward to find. 
In practice, we actually do not need to calculate
$\omega_{12}$, rather we apply Eqs.~(\ref{effham}) with the two-nucleon
solutions to directly calculate $P_2e^{-S_{12}}(h_1+h_2+V^{\Omega,A}_{12})e^{S_{12}}P_2$. 
To be explicit, the two-nucleon calculation is done with 
\begin{equation}\label{hamomega2}
H_2^\Omega = H_{02}+V^{\Omega,A}_{12}=
\frac{\vec{p}^2}{2m}
+\frac{1}{2}m\Omega^2 \vec{r}^2
+ V_{NN}(\sqrt{2}\vec{r})-\frac{m\Omega^2}{A}\vec{r}^2 \; ,
\end{equation}
where $\vec{r}=\sqrt{\frac{1}{2}}(\vec{r}_1-\vec{r}_2)$ and
$\vec{p}=\sqrt{\frac{1}{2}}(\vec{p}_1-\vec{p}_2)$ and
where $H_{02}$ differs from $h_1+h_2$ by the omission of 
the center-of-mass HO term of nucleons 1 and 2. Since $V^{\Omega,A}_{12}$ acts 
on relative coordinate, the $S_{12}$ is independent of the two-nucleon 
center of mass and the two-nucleon center-of-mass Hamiltonian cancels
out in Eq.~(\ref{V2eff}). We can see that for $A>2$ the solutions of
(\ref{hamomega2}) are bound. The relative-coordinate two-nucleon 
HO states used in the calculation are characterized by
quantum numbers $|nlsjt\rangle$ with the radial and orbital
HO quantum numbers corresponding to coordinate $\vec{r}$ 
and momentum $\vec{p}$. Typically, we solve the two-nucleon
Hamiltonian (\ref{hamomega2}) for all two-nucleon channels up
to $j=8$. For the channels with higher $j$ only the kinetic-energy 
term is used in the many-nucleon calculation.
The model space $P_2$ is defined by the maximal number of allowed HO 
excitations $N_{\rm 12max}$ from the condition $2n+l\leq N_{\rm 12max}$.
In order to construct the operator $\omega$ (\ref{omegasol})
we need to select the set of eigenvectors ${\cal K}$.
We select the lowest states
obtained in each channel. It turns out that these states also have
the largest overlap with the $P_2$ model space. Their number is given 
by the number of basis states satisfying $2n+l\leq N_{\rm 12max}$. 

The two-body effective Hamiltonian used in the $A$-nucleon calculation is then
\begin{equation}\label{Ham_A_Omega_2eff}
H^\Omega_{A, {\rm eff}}=\sum_{i=1}^A h_i 
+\sum_{i<j}^A V_{{\rm 2eff}, ij} \; .
\end{equation}
At this point
we also subtract the $H_{\rm CM}$ and, if the Slater determinant basis is to be used, 
we add the Lawson projection term 
$\beta(H_{\rm CM}-\frac{3}{2}\hbar\Omega)$ to shift the spurious
CM excitations. Eigenenergies of physical states are independent of the parameter $\beta$.

\subsubsection{Three-body effective interaction in the NCSM}\label{sec:thrbeff}

An improvement over the two-body effective interaction approximation
is the use of the three-body effective interaction that takes into account
the full space three-nucleon correlations. If the NNN interaction is included,
the three-body effective interaction approximation is rather essential
for $A>3$ systems. First, let us consider the case with no NNN interaction.
The three-body effective interaction can be calculated as
\begin{eqnarray}\label{v3eff_2b}
V_{{\rm 3eff},123}^{\rm NN}&=&
P_3 \left[e^{-S^{\rm NN}_{123}}(h_1+h_2+h_3+V_{12}^{\Omega,A}
+V_{13}^{\Omega,A}+V_{23}^{\Omega,A})e^{S^{\rm NN}_{123}}\right.
\nonumber \\
&&\left.-(h_1+h_2+h_3)\right] P_3 \; . 
\end{eqnarray}
Here, $S^{\rm NN}_{123}={\rm arctanh}(\omega_{123}-\omega_{123}^\dagger)$ and $P_3$ is a three-nucleon
model space projector. The $P_3$ space contains all three-nucleon states up to the highest 
possible three-nucleon excitation, which can be found in the $P$ space
of the $A$-nucleon system. For example,
for $A=6$ and $N_{\rm max}=6$ ($6\hbar\Omega$) space we have $P_3$ defined 
by $N_{\rm 123max}=8$. Similarly, for the $p$-shell nuclei with $A\geq 7$
and $N_{\rm max}=6$ ($6\hbar\Omega$) space we have $N_{\rm 123max}=9$. 
The operator $\omega_{123}$
is obtained with the help of Eq. (\ref{omegasol}) from exact solutions of the Hamiltonian
$h_1+h_2+h_3+V_{12}^{\Omega,A}+V_{13}^{\Omega,A}+V_{23}^{\Omega,A}$,
which are found using the antisymmetrized three-nucleon Jacobi coordinate HO basis. 
In practice, we again do not need to calculate
$\omega_{123}$, rather we apply Eqs.~(\ref{effham}) with the three-nucleon
solutions. The three-body effective interaction is then used in $A$-nucleon calculations
using the effective Hamiltonian
\begin{equation}\label{Ham_A_Omega_eff_NN}
H^\Omega_{A, {\rm eff}}=\sum_{i=1}^A h_i 
+ \frac{1}{A-2}\sum_{i<j<k}^A V_{{\rm 3eff}, ijk}^{\rm NN} \; ,
\end{equation}
where the $\frac{1}{A-2}$ factor takes
care of over-counting the contribution from the two-nucleon interaction.

If the NNN interaction is included, we need to calculate in addition to (\ref{v3eff_2b})
the following effective interaction
\begin{eqnarray}\label{v3eff}
V_{{\rm 3eff},123}^{\rm NN+NNN}&=&P_3 \left[ e^{-S^{\rm NN+NNN}_{123}}(h_1+h_2+h_3+V_{12}^{\Omega,A}
+V_{13}^{\Omega,A}+V_{23}^{\Omega,A} \right.
\nonumber \\
&&\left.
+V_{{\rm NNN}, 123})e^{S^{\rm NN+NNN}_{123}}
-(h_1+h_2+h_3)\right] P_3 \;  .
\end{eqnarray}
This three-body effective interaction is obtained using full space solutions
of the Hamiltonian 
$h_1+h_2+h_3+V_{12}^{\Omega,A}+V_{13}^{\Omega,A}+V_{23}^{\Omega,A}+V_{{\rm NNN}, 123}$.
The three-body effective interaction contribution from the NNN interaction 
we then define as 
\begin{equation}\label{v3eff_3b}
V_{{\rm 3eff},123}^{\rm NNN}\equiv V_{{\rm 3eff},123}^{\rm NN+NNN}
-V_{{\rm 3eff},123}^{\rm NN} \; .
\end{equation}
The three-body effective Hamiltonian used in the $A$-nucleon calculation is then
\begin{equation}\label{Ham_A_Omega_eff}
H^\Omega_{A, {\rm eff}}=\sum_{i=1}^A h_i 
+ \frac{1}{A-2}\sum_{i<j<k}^A V_{{\rm 3eff}, ijk}^{\rm NN}
+\sum_{i<j<k}^A V_{{\rm 3eff}, ijk}^{\rm NNN} \; .
\end{equation}
As in the case of the two-body effective Hamiltonian~(\ref{Ham_A_Omega_2eff}),
we subtract the $H_{\rm CM}$ and, if the Slater determinant basis is to be used, 
we add the Lawson projection term $\beta(H_{\rm CM}-\frac{3}{2}\hbar\Omega)$.

It should be noted that all the effective interaction calculations are performed
in the Jacobi coordinate HO basis. As discussed above, the two-body effective interaction 
is performed in the $|nlsjt\rangle$ basis and the three-body effective interaction 
in the $|N i J T\rangle$ basis (\ref{abas3}). In order to perform the $A$-nucleon
calculation in the Slater determinant HO basis, as is typically done for $A>4$,
the effective interaction needs to be transformed to the single-nucleon HO basis. 
This is done with help of the HO wave function transformations.
The details for the three-body case, in particular, are given in Refs.~\cite{NO03}
and \cite{Nogga06}. 

It should also be noted that one may attempt to separate the two-body and the three-body parts of the $V_{{\rm 3eff}}^{\rm NN}$ (\ref{v3eff_2b}). This has not been done yet in the NCSM calculations as the current implementation (\ref{Ham_A_Omega_eff_NN}) proved robust (as also demonstrated in the next section).  In recent one-dimensional model calculations with SRG evolved interactions such a separation has been achieved and shown to be useful~\cite{Ju09}. It should be also explored within the NCSM, although care must be taken to avoid introducing spurious model-space effects.

\subsection{Effective operators}
\label{sec:effOp}

Besides spectra, other properties of the nuclear states are of interest, as they impose a strong test on the theoretical wave functions. For consistency, the same unitary transformation used to compute the effective interaction should be employed in order to obtain effective operators in the model spaces used to diagonalize the effective Hamiltonian. 

In addition to consistency, another motivation for implementing the renormalization of general operators is the long standing effective charge problem in the phenomenological shell model. Arising from the inevitable truncation of the Hilbert space, the relatively large effective charges were found to be essential in the overall description of the transition strength. However, previous perturbation theory attempts to describe phenomenological charges needed to obtain correct transition strengths have been unsuccessful \cite{osnes}. On the other hand, investigations within the framework of the NCSM have reported  some progress in explaining the large values of the effective charges \cite{navratil1997}. 

The renormalization of effective operators is much more involved than the renormalization of the Hamiltonian. In order to ensure energy independence of the effective operator, the decoupling condition (\ref{decoupl}) has to be supplemented with the Hermitian conjugate \cite{Navratil:1993fesh}. This transformation, however, has the advantage that the effective Hamiltonian produced is Hermitian, as discussed in Sec. \ref {sec:effHam} and written out explicitly in Eq. (\ref{exhermeff}).

A general tensor operator can change the spin and isospin. Hence, the renormalization of a rank $\Delta J$, $\Delta T$ tensor operator writes as

\begin{equation}
O_{eff}^{(\Delta J;\Delta T)}=\frac{P_{J'T'} +P_{J'T'}\omega_{J'T'}^\dagger Q_{J'T'}}{\sqrt{P_{J'T'}+\omega_{J'T'}^\dagger\omega_{J'T'}}}
O^{(\Delta J;\Delta T)}\frac{P_{JT}+Q_{JT}\omega_{JT} P_{JT}}{\sqrt{P_{JT}+\omega_{JT}^\dagger\omega_{JT}}}\;,
\label{effOp}
\end{equation}
where we have shown explicitly the possible change in spin and isospin of the initial ($J$, $T$) and final ($J'$, $T'$) states, respectively. Equation (\ref{effOp}) is the generalization of (\ref{exhermeff}), and shows the complexity of the renormalization of a a tensor operator compared with the renormalization of a scalar operator, as the tensor operator allows for the  the possible mixture of different spin and isospin quantum numbers. Finally, because the transformation is a scalar, the effective operator preserves the tensor character of the starting operator.

The simplest approximation for the unitary transformation is at the two-body cluster level. Because of the complexity of the renormalization, the two-body cluster is the only one developed so far for general operators \cite{brucefest,Ionel,coulomb_ren}. Under this approximation, the transformation becomes
\begin{equation}
S_2\approx \sum_{i>j=1}^A S_{ij},
\end{equation}
with $S_{ij}=\mathrm{arctanh}(\omega_{ij}-\omega_{ij}^\dagger)$. Applying the operator identity
\begin{equation}
e^{-S_2}Oe^S_2=O+[O,S_2]+\frac{1}{2!}[[O,S_2],S_2]+...
\end{equation}
to transform a general one-body operator $O^{(1)}=\sum_{i=1}^A O_i$, one obtains
\begin{equation}
{\cal O}^{(1)}=O^{(1)}+\sum_{i>j=1}^A[O_i+O_j,S_{ij}]+\sum_{i>j}^A[[O_i+O_j,S_{ij}],S_{ij}]+...,
\end{equation}
where we have retained only the one- and two-body terms, neglecting higher body contributions, such as $[O_i,S_{jk}]$, with $i\neq j$ and $i\neq k$. Resummation of the commutators yields

\begin{eqnarray}
\lefteqn{P_2 O_{\mathrm{2eff}} P_2=P_2\sum_i O_i P_2\nonumber}\\
& & +P_2\sum_{i>j=1}^A\left[e^{-S_{ij}}\left(O_i+O_j\right)e^{S_{ij}}-
\left(O_i+O_j\right)\right]P_2.
\label{oneb}
\end{eqnarray}

Analogously, for a general two-body operator
\begin{equation}
P_2 O_{\mathrm{2eff}} P_2=P_2\sum_{i>j=1}^Ae^{-S_{ij}}O_{ij}e^{S_{ij}}P_2,
\label{twob}
\end{equation}
and, in particular, the effective Hamiltonian is given by
\begin{eqnarray}
\lefteqn{P_2 H_{\mathrm{2eff}}P_2=P_2\sum_{i=1}^Ah_i P_2\nonumber }\\
& &+P_2\sum_{i>j=1}^A \left[e^{-S_{ij}}\left( h_i+h_j+v_{ij}\right)e^{S_{ij}}-h_i-h_j\right]P_2,
\end{eqnarray}
which recovers expression (\ref{V2eff}) for the effective interaction.

\subsection{Convergence tests}
\label{sec:Conv}

In this subsection, we give examples of convergence of {\it ab initio} NCSM calculations.

\begin{figure}[t]
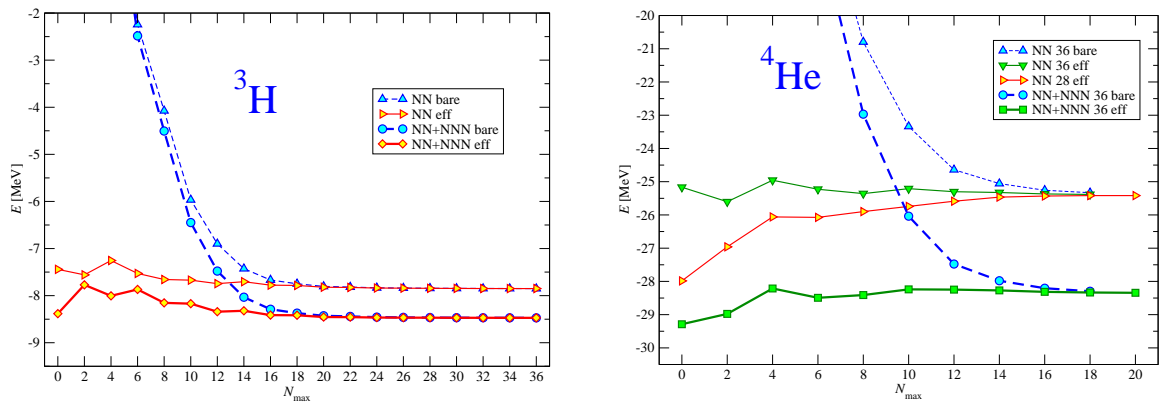

\begin{minipage}{8cm}
  \includegraphics*[width=0.9\columnwidth]
   {gs_H3N3LO3NFMT_1.eps}
\end{minipage}
\hfill
\begin{minipage}{8cm}
  \includegraphics*[width=0.9\columnwidth]
   {gs_He4N3LO3NFMT_1.eps}
\end{minipage}
\caption{$^3$H (left) and $^4$He (right) ground-state energy dependence on the size of the basis.
The HO frequencies of $\hbar\Omega=28$ MeV ($^3$H) and 28 or 36 MeV ($^4$He) were employed. 
Results with (thick lines) and without (thin lines) the NNN interaction are shown. 
The solid lines correspond to calculations with two-body ($^3$H) or three-body ($^4$He) 
effective interactions, the dashed lines to calculations with the bare interactions. 
  \label{gs_H3_He4}}
\end{figure}

First, we discuss calculations for $s$-shell nuclei.
In  Fig.~\ref{gs_H3_He4}, we show the
convergence of the $^3$H and $^4$He ground-state energies
with the size of the basis. Thin lines correspond to results obtained with 
the NN interaction only. Thick lines correspond to calculations that also include the 
NNN interaction. 
Here, we use the chiral effective field theory (EFT) NN interaction of Ref.~\cite{N3LO} and the local chiral NNN interaction that will be discussed in detail in the next section.
The solid lines correspond to $^3$H ($^4$He) calculations with two-body (three-body) effective
interaction derived from the chiral EFT potentials.
The dashed lines correspond 
to calculations with the bare, that is the original, unrenormalized chiral EFT interactions. 
In $^3$H calculations, the bare NNN interaction is added to either the bare NN (dashed thick line) or to the effective NN interaction (solid thick line). 
We observe that the convergence is faster when
the effective interactions are used. However, starting at about $N_{\rm max}=24(18)$
the convergence is reached in $^3$H($^4$He) calculations also with the bare NN interaction. 
It should be noted, however, that $p$-shell calculations with the NNN interactions
are presently feasible in model spaces up to $N_{\rm max}=6$ or $N_{\max}=8$. 
The use of the three-body effective interaction is then essential in the $p$-shell 
calculations with NN+NNN interactions. It should be noted that in calculations with the effective interaction, 
the effective Hamiltonian is different at each point, as the effective interaction depends
on the size of the model space given by $N_{\rm max}$. The calculation with the bare interaction
is a variational calculation converging from above with $N_{\rm max}$ 
and HO frequency $\Omega$ as variational parameters. The calculation with the effective
interaction is not variational. The convergence can be from above, from below or oscillatory.
This is because a part of the exact effective Hamiltonian is omitted. The calculation
without NNN interaction converges to the $^3$H ground-state energy $-7.852(5)$~MeV, well above
the experimental $-8.482$~MeV. Once the NNN interaction is added, we obtain $-8.473(5)$~MeV,
close to experiment. As discussed in the next section, the NNN parameters were tuned to
reproduce the average of the $^3$H and $^3$He binding energies. 

The rate of convergence also depends on the choice of the HO frequency. 
The $^4$He calculations without NNN interaction were done for two different HO
frequencies. It is apparent that convergence to the same result occurs in both cases.
We note that in the case of no NNN interaction,
we may use just the two-body effective interaction (two-body cluster approximation), which
is much simpler. The convergence is slower, however, see discussion in Ref.~\cite{NO02}.
We also note that $^4$He properties with the chiral EFT NN interaction that we employ
here were calculated using the two-body cluster approximation in Ref.~\cite{NC04} and
present results are in agreement with results found there.
Our $^4$He ground-state energy results are $-25.39(1)$~MeV in the NN case 
and $-28.34(2)$~MeV in the NN+NNN case. The experimental value is $-28.296$~MeV. We note that
the present {\it ab initio} NCSM $^3$H and $^4$He results obtained with the chiral EFT
NN interaction are in a perfect agreement with results obtained using the variational
calculations in the hyperspherical harmonics basis as well as with the Faddeev-Yakubovsky
calculations published in Ref.~\cite{HHnonloc}. A satisfying feature of the present
NCSM calculation is the fact that the rate of convergence is not affected
in any significant way by inclusion of the NNN interaction.

\begin{figure}[!ht]
\begin{minipage}{8cm}
 \includegraphics[scale = 0.18]{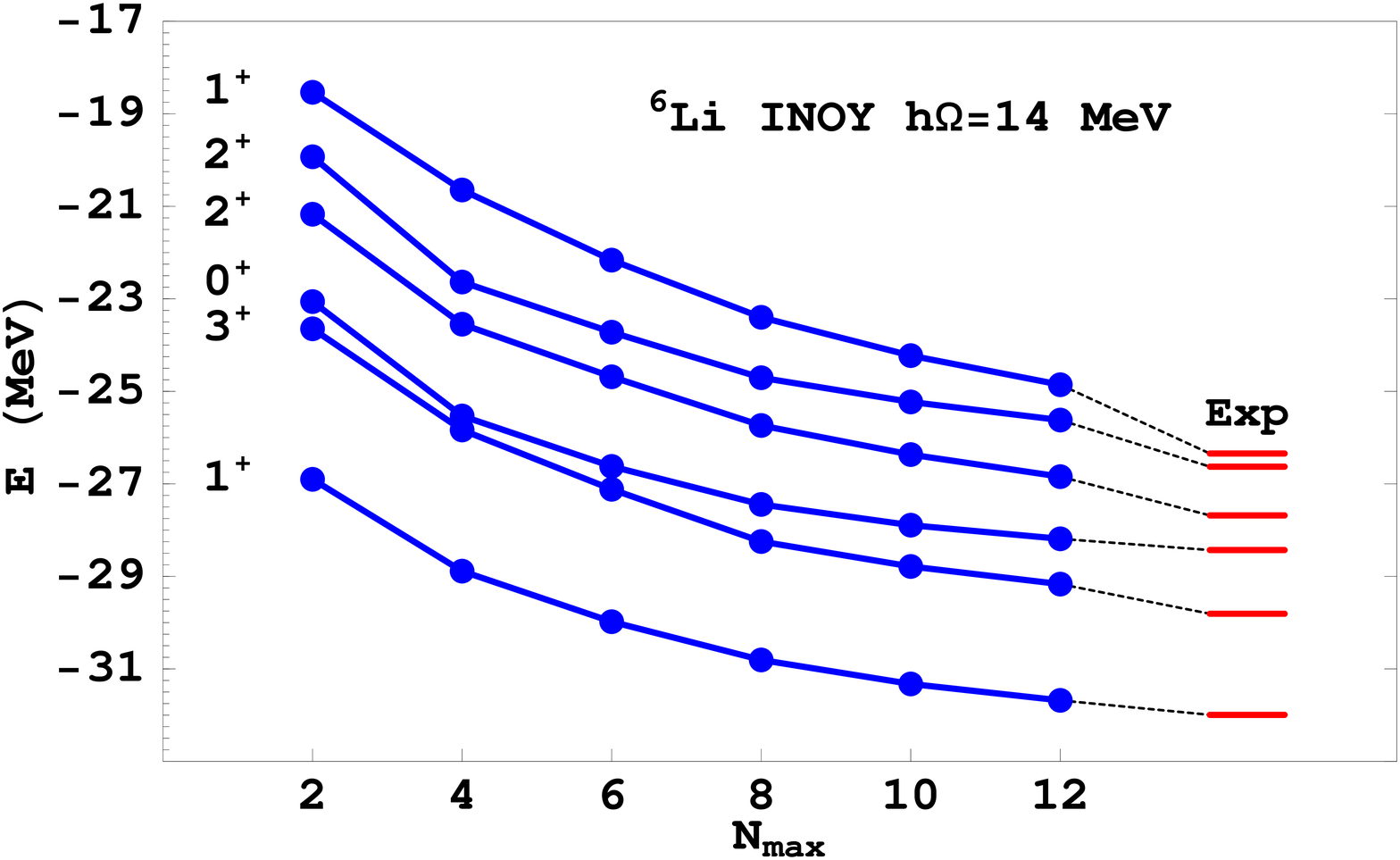}%
\end{minipage}
\begin{minipage}{8cm}
\includegraphics[width=1.0\columnwidth]{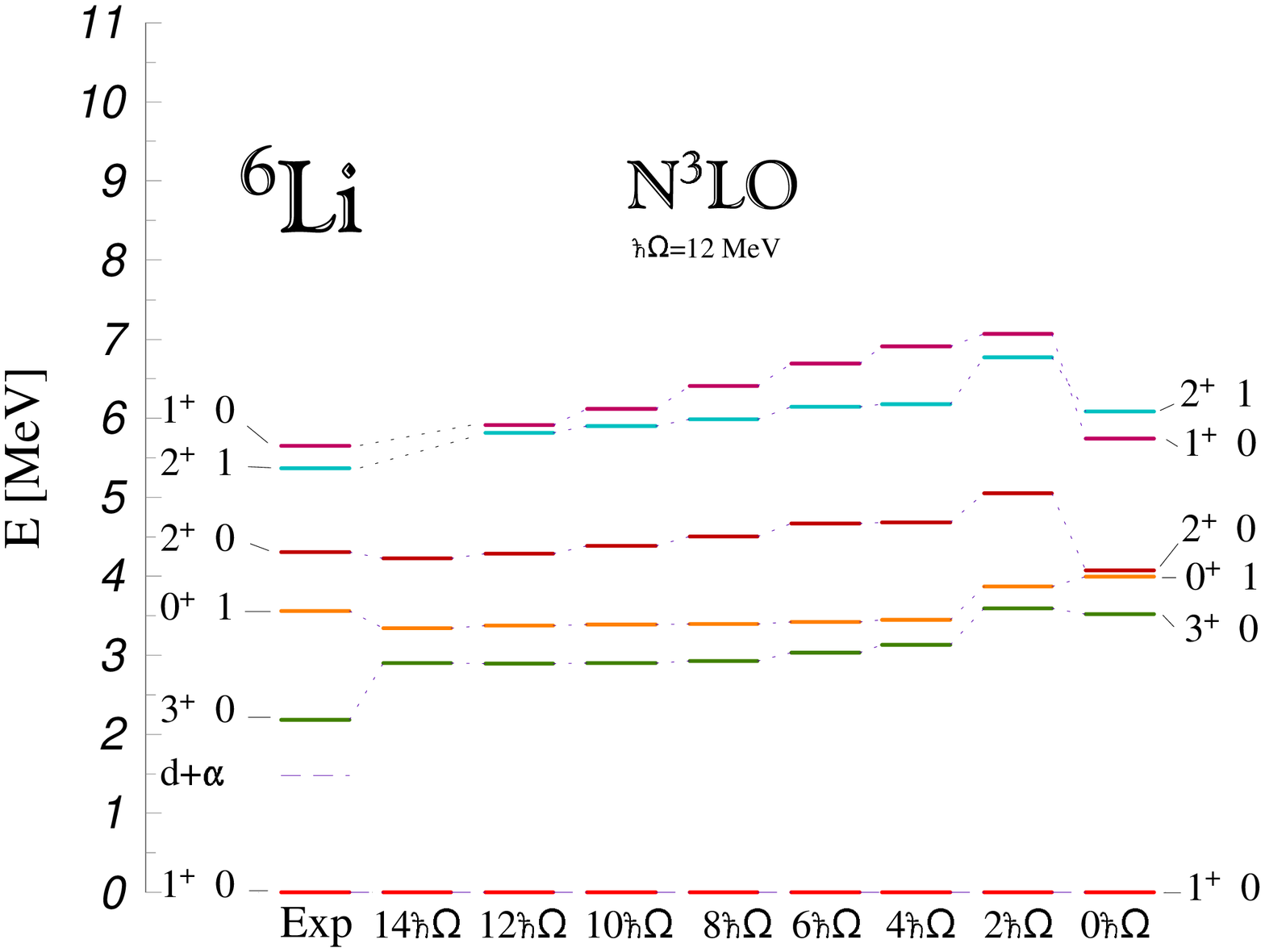}
\end{minipage}
\caption{\label{li6_exc_12}
Calculated absolute (left) and excitation (right) energies of
$^{6}$Li obtained in $0\hbar\Omega$-$12\hbar\Omega$ ($0\hbar\Omega$-$14\hbar\Omega$) basis spaces using two-body effective
interactions derived from the INOY (chiral EFT) NN potential
compared to experiment. The HO frequencies of $\hbar\Omega=14$ MeV (left) and 12 MeV (right) were used.
}
\end{figure}

As an example of convergence of {\it ab initio} NCSM calculations for $p$-shell nuclei,
we present $^6$Li results obtained using the INOY and the chiral EFT NN potential. 
The dependence of the NCSM absolute and excitation energies on the basis size
is presented in Fig.~\ref{li6_exc_12}.
The calculations were performed using the two-body effective interaction in the Slater 
determinant HO basis with the shell-model code Antoine~\cite{Antoine}. Results for other
HO frequencies were published in Refs.~\cite{NC04,FCN09}. 
As discussed in Ref.~\cite{NC04}, the convergence rate with $N_{\rm max}$ is different for different states.
In particular, the $3^+ 0$ state and the $0^+ 1$ state converge faster in the higher frequency 
calculations ($\hbar\Omega=12,13$ MeV with the chiral EFT NN potential), 
while the higher lying states converge faster 
in the lower frequency calculations ($\hbar\Omega=8,10$ MeV with the chiral EFT NN potential). 
The results on the right of Fig.~\ref{li6_exc_12}
demonstrate a good convergence of the excitation energies, in particular, for the $3^+ 0$ and $0^+ 1$
states. An interesting result is the overestimation of the $3^+ 0$ excitation energy
compared to experiment, in particular with the chiral EFT NN potential. 
It turns out that this problem is resolved once the NNN interaction is included in the Hamiltonian.
More discussion on eigenenergy convergence in p-shell nuclei NCSM calculations can be found, e.g.,  in Refs.~\cite{Na01,NCSM_He_rad,Fo08}.

In the final part of this section, we discuss the renormalization of different observables, such as electro-magnetic moments, radii, etc. The first exclusive investigation of the effective operators in the NCSM framework concentrated on implementing a procedure in which the relative states in the excluded space were restricted, and observed convergence by releasing the restriction \cite{brucefest}. The goal of the investigation in Ref. \cite{brucefest} was a test of the numerical implementation in a simple model, in which the ``full'' space was restricted to a numerically tractable size, so that all missing correlations could be exactly calculated. Even in this simple model it was found that the two-body cluster renormalization was very weak for long range operators, such as $E2$ transitions. The same results were later reported when the same method was implemented in realistic cases \cite{Ionel}.

The main goal of Ref. \cite{Ionel} was a qualitative understanding of the influence of effective operators and not a highly accurate description of the experimental data; therefore, the NNN interactions were left out. The same goal also motivated the use of rather small model spaces and of the two-body cluster approximation, given that, as expected from the convergence properties of effective operators, larger renormalization effects are expected in smaller model spaces. Furthermore, a more efficient implementation of the renormalization than the one in Ref. \cite{brucefest} was introduced, similar to the one applied to the Hamiltonian. Thus, the renormalization was implemented in relative coordinates, allowing the same treatment for general operators as for the Hamiltonian, as long as the former can be written in relative coordinates \cite{Ionel}. This implementation was tested on the deuteron, where the two-body cluster provides the exact solution. In that case, the bare quadrupole operator in $4\hbar\Omega$ gave 0.179 $e$ fm$^2$ for the quadrupole moment, while the value of 0.270 $e$ fm$^2$, described by the AV8' potential, was obtained using the corresponding effective operator in the same model space. However, when the same procedure has been applied to a realistic many-body problem, the result was different. A very weak renormalized $B(E2)$ value was obtained even in small model spaces. As an example, we have looked at the $B(E2;3^+1\to1^+1)$ in $^6$Li, where a $2\hbar\Omega$ calculation gives 2.647 $e^2$ fm$^4$ when the bare operator is employed, and 2.784 $e^2$ fm$^4$ when the effective operator is used \cite{Ionel}. These results have been obtained with the Argonne V8' NN interaction. A calculation for the same observable, but with the CD-Bonn 2000 NN interaction, which is expected to give comparable results with the AV8' potential, obtained $B(E2;3^+1\to1^+1)=10.221$ $e^2$ fm$^4$ with the bare operator in $10\hbar\Omega$  model space.  Overall, the difference between the bare operator results in the $2\hbar\Omega$ and $10\hbar\Omega$ model spaces, coupled with the small renormalization at the two-body cluster level, indicate sizable effective many-body effects needed to correct the $2\hbar\Omega$  $B(E2)$ value.

The case of the kinetic energy operator is completely different from the quadrupole transitions presented above. In Refs. \cite{benchmark,Ionel}, large renormalization was obtained even at the two-body cluster for the kinetic energy. The kinetic energy is short
range, while the quadrupole is long range; at the two-body cluster level, the unitary
transformation renormalizes mainly the short-range core of the interaction, leaving
unchanged the long range part. Hence, in order to account for long-range correlations
in the two-body cluster approximation, one needs to enlarge the model space. To test this hypothesis,  a Gaussian operator of
variable range was used in Ref. \cite{Ionel}. There, by observing the variation with the model space / HO frequency of the expectation values calculated with the bare and effective operators for several different ranges, it was demonstrated that a short-range two-body operator
is renormalized accurately at the two-body cluster level, while a
long-range operator is weakly renormalized.
To further illustrate the power of the unitary transformation approach to the renormalization of short-range operators, we turn to an observable probing short-range correlations. The inclusive $(e,e')$ longitudinal data presents one of the clearest
experimental signatures for short-range correlations in the wave-function
of the ground state, at least for light nuclei. A quantitative measure of the short-range correlations is the longitudinal-longitudinal distribution function (connected to the Coulomb sum rule) \cite{Drell:1958,McVoy:1962,Schiavilla:1993tk}

\[
\rho_{LL}(q)=\frac{1}{4Z}\sum_{i\neq j}
\langle g.s.|j_{0}(q|\mathbf{r}_{i}-\mathbf{r}_{j}|)
(1+\tau_{z,i})(1+\tau_{z,j})|g.s.\rangle,
\]
where $j_0$ is the spherical Bessel function of zero order, and $q$ the momentum transfer.

\begin{figure}
\centering{\includegraphics*[scale=0.6]{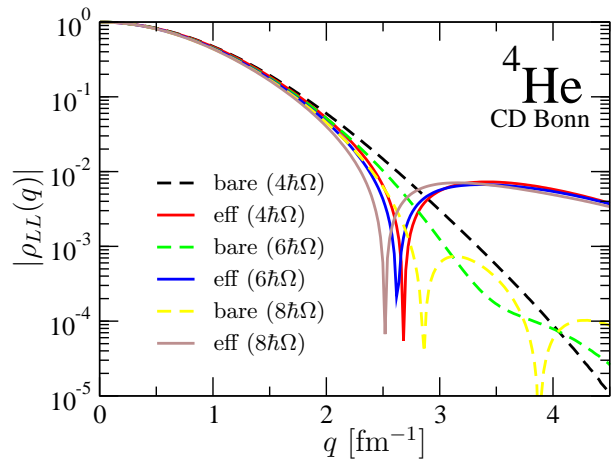}}
\caption{The longitudinal-longitudinal distribution function
in $^4$He, obtained using
bare operators (left panel) and effective operators (right panel).
The HO energy used in this calculation
was $\hbar\Omega=28$ MeV, while the NN interaction was CD Bonn.}
\label{he4rhoLL}
\end{figure}

In Figure \ref{he4rhoLL} we present the results for the
longitudinal-longitudinal distribution function for $^4$He.
At high momentum transfer, the results obtained
using bare operators depend strongly upon the model space. On
the other hand, the results obtained with effective
operators are model space invariant at high $q$, although in these model spaces 
the wave function is not fully converged, since
the energy is not converged in these very small model spaces (see Fig. 3 in
Ref.  \cite{coulomb_ren}). They
agree with the values computed in larger model spaces.
At intermediate momentum transfer, i.e., $q\approx 2.5$ fm$^{-1}$,
even the effective operator results vary. This
effect is due to the fact that the long range part of the operator
has not yet converged in these small model spaces. In larger model spaces,
where the long-range correlations are better described, the agreement is even better.

Similar results for the longitudinal-longitudinal distribution
function have been obtained for $^{12}$C, where calculations
in very large model spaces are not possible. However, even in the
smallest model space, $0\hbar\Omega$, we were able to obtain
good results for high momentum transfer, which reproduce the
values in larger model spaces \cite{coulomb_ren}.

In conclusion, short-range operators (high momentum
transfer) are very well renormalized and the results become
model-space independent even in the two-body cluster approximation,
while long-range operators, such as
the quadrupole transition operator, or the longitudinal-longitudinal
distribution function for small and intermediate momentum
transfer, are only weakly renormalized. It all comes down to the effects of the
unitary transformation, which, as discussed before, at the two-body level renormalizes only the
short-range part of the interaction, while the long-range part  is recovered in larger model
spaces.

\section{Light nuclei from chiral EFT interactions}\label{sec:chiral_NN_NNN}

Interactions among nucleons are governed by quantum chromodynamics (QCD). 
In the low-energy regime relevant to nuclear structure, 
QCD is non-perturbative, and, therefore, hard to solve. Thus, theory has 
been forced to resort to models for the interaction, which have limited physical basis. 
New theoretical developments, however, allow us connect QCD 
with low-energy nuclear physics. The chiral effective field theory 
($\chi$EFT)~\cite{Weinberg} provides a promising bridge.
Beginning with the pionic or the nucleon-pion system~\cite{bernard95} 
one works consistently with systems of increasing nucleon 
number~\cite{ORK94,Bira,bedaque02a}. 
One makes use of spontaneous breaking of chiral symmetry to systematically 
expand the strong interaction in terms of a generic small momentum
and takes the explicit breaking of chiral symmetry into account by expanding 
in the pion mass. Thereby, the NN interaction, the NNN interaction 
and also $\pi$N scattering are related to each other. 
The $\chi$EFT predicts, along with the NN interaction 
at the leading order, an NNN interaction at the third order (next-to-next-to-leading 
order or N$^2$LO)~\cite{Weinberg,vanKolck:1994,Epelbaum:2002}, 
and even an NNNN interaction at the fourth order (N$^3$LO)~\cite{Epelbaum06}.
The details of QCD dynamics are contained in parameters, 
low-energy constants (LECs), not fixed by the symmetry. These parameters 
can be constrained by experiment. At present, high-quality NN potentials 
have been determined at order N$^3$LO~\cite{N3LO}. 
A crucial feature of $\chi$EFT is the consistency between 
the NN, NNN and NNNN parts. This consistency also extends to the nuclear current.
As a consequence, at N$^2$LO and N$^3$LO, except 
for two LECs, assigned to two NNN diagrams, the potential is fully 
constrained by the parameters defining the NN interaction.

We adopt the potentials of the $\chi$EFT at the orders presently available, the NN at N$^3$LO  
of Ref.~\cite{N3LO} and the NNN interaction at N$^2$LO \cite{vanKolck:1994,Epelbaum:2002}.
Since the NN interaction is non-local, the {\it ab initio} NCSM
is the only approach currently available 
to solve the resulting many-body Schr\"odinger equation for mid-$p$-shell nuclei.
We are in a position to use the {\it ab initio} NCSM calculations in two
ways. One of them is the determination of the LECs assigned to two NNN diagrams
that must be determined in $A\geq 3$ systems. The other is testing predictions of the
chiral NN and NNN interactions for light nuclei.

\subsection{Chiral N$^2$LO three-nucleon interaction}

\begin{figure}[hbtp]
  \includegraphics*[width=0.7\columnwidth]{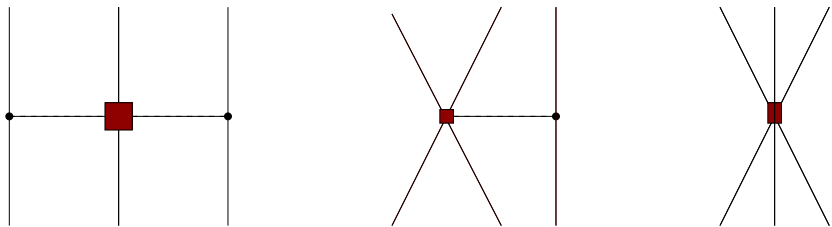}
  \caption{Terms of the N$^2$LO $\chi$EFT NNN interaction.
  \label{N2LO_NNN}}
\end{figure}

The NNN interaction at N$^2$LO of the $\chi$EFT is comprised
of three parts: (i) The two-pion exchange, (ii) the one-pion exchange plus contact
and (iii) the three-nucleon contact, see Fig.~\ref{N2LO_NNN}. In this work, we regulate the
the NNN terms with a regulator depending on the momentum transfer similarly as done, e.g., for the Tucson-Melbourne NNN interaction~\cite{Coon}, which results in a local
$\chi$EFT NNN interaction. This is advantageous for some many-body approaches, including the NCSM,
because a local NNN interaction, in particular the two-pion exchange term, is easier to implement. 
Full technical details are given in Ref.~\cite{N2LO-local}. 

The LECs associated with the 
two-pion exchange also appear in the NN interaction and are, therefore, determined
in the $A=2$ system. The one-pion exchange plus contact term (D-term)
is associated with the LEC $c_D$ and the three-nucleon contact term (E-term)
is associated with the LEC $c_E$. It is interesting to note that $c_D$ represents a renormalization of the short range interaction of two nucleons, rather than three. Consequently, it manifests itself not only in the $NN-\pi-N$ contact term of the interaction, but also in the two-nucleon contact vertex with an external probe of the exchange currents.

The $c_D$ and $c_E$ LECs, expected to be of order one, can be constrained by the $A=3$ binding energy.
One then still needs an additional observable to determine the two parameters. 
The first determination of $c_D$ and $ c_E$ was attempted using as constraints the $^3$H binding energy and $nd$ doublet scattering length, and adopting the full interaction up to N$^2$LO~\cite{Epelbaum:2002}. However, this proved to be difficult due to a correlation between these two observables, and the large experimental uncertainty on the scattering length. Later, the N$^3$LO NN potential was combined with the NNN at N$^2$LO (non-local, regulated with nucleon momenta) to study the $^7$Li structure~\cite{Nogga06}. In this work, besides the $^3$H binding energy the second constraint on the undetermined LECs was the energy of the $^4$He ground state.  As a result of  the correlation between these two observables, known as Tjon line, fitting the $^3$H ground-state energy automatically results in a  $^4$He binding energy that is within a few hundred keV of experiment.  The subsequent fine-tuning of this binding energy is then very sensitive to the structure of the adopted NNN force. Hence, small variations of the cutoff, different regularization schemes, missing terms of the interaction, etc., tend to produce large swings in the extracted values of $c_D$ and $c_E$.
A different approach that we describe and expand on it here was adopted in Ref.~\cite{NGVON07}. There, a preferred choice for the two LECs was obtained by complementing the constrain on the $A=3$ binding energies with a sensitivity study on the radius of the $^4$He and on various properties of $p$-shell nuclei.

\begin{figure}
\centerline 
{\includegraphics[width=0.7\columnwidth]{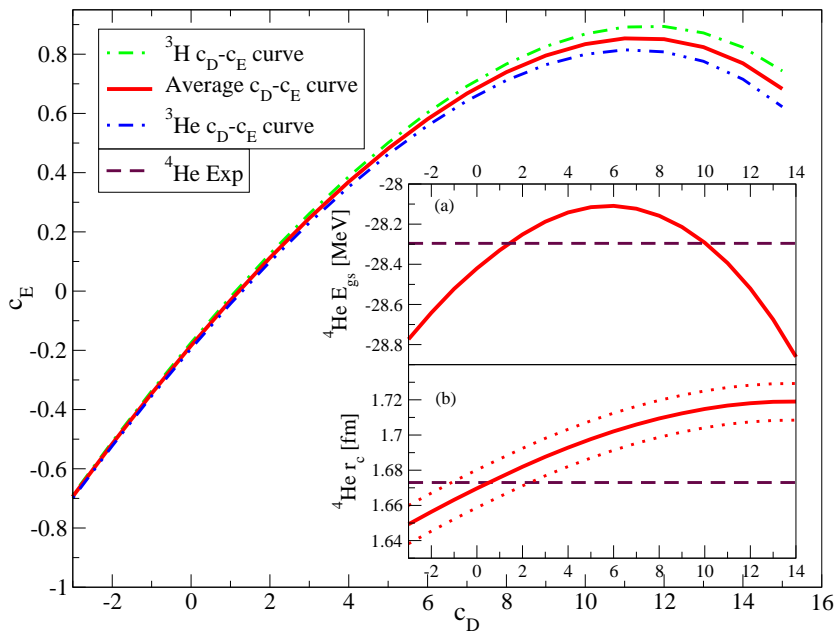}}
\caption{Relations between $c_D$ and $c_E$ for which  the 
binding energy of $^3$H ($8.482$ MeV) and  $^3$He ($7.718$ MeV) are reproduced. 
(a) $^4$He ground-state energy along the averaged curve. 
(b) $^4$He charge radius $r_c$ along the averaged curve. Dotted lines represent 
the $r_c$ uncertainty due to the uncertainties in the proton charge radius.}
\label{CDCE_curve}
\end{figure}

Fig.~\ref{CDCE_curve} shows the trajectories of the two LECs $c_D$ and $c_E$ that 
were determined in Ref.~\cite{NGVON07} from fitting the binding energies of the $A=3$ systems. 
Separate curves are shown for $^3$H and $^3$He fits, as well as their average.
We also show the calculated ground-state energy and charge radius of $^4$He 
obtained with the constrained LECs. As demonstrated in Fig.~\ref{gs_H3_He4}
in Section \ref{sec:Conv}, our $A=3$ and $A=4$ results presented in Fig.~\ref{CDCE_curve} 
are fully converged.
There are two points where the binding of $^4$He is reproduced exactly. 
One of them for $c_D\sim 1$ and the other with $c_D\sim 10$. However,
as a consequence of the correlation bewteen the triton and $^4$He binding energies, we observe 
that in the whole investigated range of $c_D$ and $c_E$, the calculated 
$^4$He  binding energy  is within a few hundred keV of experiment. 
Consequently, the determination of the LECs in this way is likely not very stringent.
By examining calculations of the $^4$He charge radius, we can see that a result consistent with 
experiment, taking into account the uncertainty of the proton charge radius, is obtained for $c_D$ values in the range from $\sim -2$ to $\sim +2$. This observation rules out 
the range of the large $c_{D}\sim 10$ values that overestimate the $^4$He radius and on top of it 
might be considered ``unnatural'' from the $\chi$EFT point of view. 

\begin{table*}[t]
\caption{$^3$H g.s. energies (in MeV), point-proton radii (in fm) and $nd$ 
 scattering lengths (in fm), obtained using the N$^3$LO NN potential~\cite{N3LO} with and without the local N$^2$LO NNN interaction~\cite{N2LO-local} with $c_D\!=\!1$ and $c_E\!=\!-0.029$, compared to experiment. Calculations performed within the NCSM and/or hyperspherical harmonics (HH) expansion approaches.}
\begin{tabular}{llcccc}
\label{predictions_3H}
&  &\multicolumn{2}{c}{$^3$H} & \multicolumn{2}{c}{$nd$} \\\cline{3-4}\cline{5-6}\\[-2mm]
&  &$E_{\rm g.s.}$ & $\langle r^2_p\rangle^{1/2}$ & $^2a$ & $^4a$ \\ [0.7mm]
\hline\\[-3mm]
NN&NCSM~\cite{N2LO-local} & $-$7.852(5) & 1.650(5) & $-$ & $-$ \\
NN&HH~\cite{Pisa} & $-$7.854$\phantom{(5)}$ & 1.655$\phantom{(5)}$ & 1.100\phantom{(8)} & 6.342\phantom{()} \\[2mm]
NN+NNN&NCSM~\cite{N2LO-local} &$-$8.473(5) & 1.608(5) & $-$ & $-$ \\
NN+NNN&HH~\cite{Pisa} &$-$8.474$\phantom{(5)}$ & 1.611$\phantom{(5)}$ & 0.675\phantom{(8)} & 6.342\phantom{()} \\[2mm]
Expt. &&$-$8.482$\phantom{(5)}$ & 1.60$\phantom{8(5)}$ & $-$ & $-$ \\
Expt.~\cite{n-d,alpharadius,n-t} &  &$-$ & $-$ &0.65(4)\phantom{8} & 6.35(2) \\
Expt.~\cite{n-d-2,n-t-2} & &$-$ & $-$&0.645(8)& $-$
\end{tabular}
\end{table*}
\begin{table*}[t]
\caption{The same as in Table ~\ref{predictions_3H} for $^4$He and $n^3$H.}
\begin{tabular}{llcccc}
\label{predictions_4He}
&  & \multicolumn{2}{c}{$^4$He} & \multicolumn{2}{c}{$n^3$H}\\\cline{3-4}\cline{5-6}\\[-2mm]
&  & $E_{\rm g.s.}$  & $\langle r^2_p\rangle^{1/2}$ & $^1a$  & $^3a$ \\ [0.7mm]
\hline\\[-3mm]
NN&NCSM~\cite{N2LO-local} &$-$25.39(1)& 1.515(2)\phantom{1} & $-$ & $-$\\
NN&HH~\cite{Pisa} &$-$25.38$\phantom{(1)}$ & 1.518\phantom{(12)} & 4.20\phantom{(29)} & 3.67\phantom{(11)}\\[2mm]
NN+NNN&NCSM~\cite{N2LO-local} &$-$28.34(2) & 1.475(2)\phantom{1} & $-$ & $-$\\
NN+NNN&HH~\cite{Pisa} &$-$28.36$\phantom{(2)}$&1.476\phantom{(12)} & 3.99\phantom{(29)} & 3.54\phantom{(11)}\\[2mm]
Expt. && $-$28.296\phantom{()} & $-$ & $-$ &$-$ \\
Expt.~\cite{n-d,alpharadius,n-t} &  & $-$& 1.467(13) &4.98(29) &3.13(11) \\
Expt.~\cite{n-d-2,n-t-2} & & $-$ &$-$ &4.45(10) &3.32(2)\phantom{1}
\end{tabular}
\end{table*}
In Tables~\ref{predictions_3H} and \ref{predictions_4He}, we present a collection of $A\!=\!3$ and 4 data, respectively, obtained with and without inclusion of the NNN force for $c_D\!=\!1$ ($c_E\!=\!-0.029$). The corresponding ground-state energy convergence within the NCSM was shown in Figs.~\ref{gs_H3_He4}
in Section \ref{sec:Conv}. Besides the triton ground-state energy, which is by construction within a few keV of experiment, the NN+NNN results for the $nd$ doublet and quartet scattering lengths and $^4$He ground-state energy and point-proton radius are in perfect agreement with measurement. While for the $n^3$H singlet scattering length the inclusion of the $NNN$ force worsens the disagreement with respect to experiment to some extent, the $n^3$H triplet scattering length improves with the NNN included. We also note a perfect agreement between the two theoretical approaches, the {\it ab initio} NCSM and the variational HH method of Ref.~\cite{Pisa}, for the bound-state results.

As the $c_D$ LEC enters also the nuclear current, it is possible to utilize, e.g., the triton half life,
as another observable in addition to the $A=3$ binding energy constraint for the determination of the NNN LECs. This was done recently in Ref.~\cite{Doron}. Taking into account experimental errors, a narrow range of $c_D$ values around $c_D\sim -0.2$ (the corresponding $c_E=-0.205$ from the $A=3$ binding energy) was found to agree with the measurements. An interesting observation was made in Ref.~\cite{Doron}, namely,
the triton half life is quite insensitive, unlike most other observables, to the NNN terms in the Hamiltonian. This is makes the LEC determination of Ref.~\cite{Doron} rather robust.

\subsection{Results for $p$-shell nuclei}

Sensitivity of the $p$-shell nuclear properties to the choice of the $c_D$ and $c_E$ LECs 
was investigated in Ref.~\cite{NGVON07}. First, the $A=3$ binding energy 
constraint was maintained. Second, the sensitivity study was limited to the $c_D$ values in the vicinity of the point $c_{D}\sim 1$, in particular values from $-3$ to $+2$ that include the range compatible with the $^4$He radius. 

\begin{figure}
\begin{minipage}{8cm}
{\includegraphics[width=0.9\columnwidth]{cD_dep_Li613.eps}}
\end{minipage}
\begin{minipage}{8cm}
{\includegraphics[width=0.9\columnwidth]{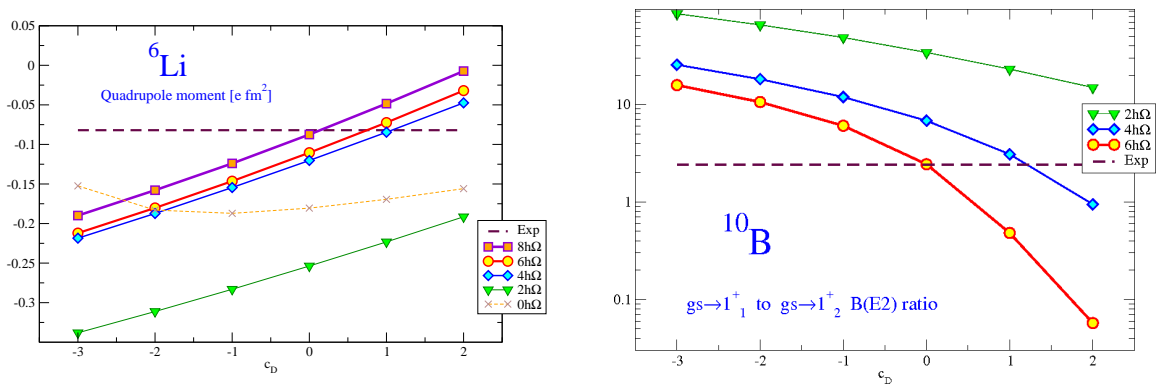}}
\end{minipage}
\caption{Dependence  of the $^{6}$Li quadrupole moment (left) and the $^{10}$B 
B(E2;$3^+_1 0 \rightarrow 1^+_1 0$)/B(E2;$3^+_1 0 \rightarrow 1^+_2 0$) ratio (right)
on the value of $c_D$ with the $c_E$ constrained by the $A=3$ 
binding energy fit for different basis sizes.
The HO frequencies of $\hbar\Omega=13$ MeV (left) and 14 MeV (right) were employed.}
\label{cD_dep_6Li}
\end{figure}

While most of the $p$-shell nuclear properties, e.g., excitation spectra,
are not very sensitive to variations of $c_D$ in the vicinity of the $c_{D}\sim 1$ point,
we were able to identify several observables that do demonstrate strong dependence on $c_D$. 
For example, the $^6$Li quadrupole moment changes sign 
depending on the choice of $c_D$, as can be seen in Fig.~\ref{cD_dep_6Li}. 
In the right of Fig.~\ref{cD_dep_6Li}, we display the ratio 
of the B(E2) transitions from the $^{10}$B ground state to the first and the second $1^+ 0$ state.
This ratio changes by several orders of magnitude 
depending on the $c_D$ variation. This is due to the fact
that the structure of the two $1^+ 0$ states is exchanged depending on $c_D$. 
%
%
In addition, $c_D$ dependence of the $^{12}$C B(M1) transition
from the ground state to the $1^+ 1$ state was discussed in Ref.~\cite{NGVON07}. 
Also, the importance of the NNN interaction in reproducing 
the experimental value was illustrated \cite{NGVON07,Hayes03}. 
Overall the results show that for $c_D<-2$ the $^4$He radius 
and the $^6$Li quadrupole moment underestimate experiment, while for $c_D>0$ the lowest
two $1^+$ states of $^{10}$B are reversed and the $^{12}$C B(M1;$0^+0\rightarrow 1^+1$)
is overestimated. Therefore, the value of $c_D=-1$ was chosen in Ref.~\cite{NGVON07}
as globally the best choice. We note that the triton half-life study suggests a $c_D$ 
value in the range around $c_D\sim -0.2$. This result is not inconsistent with the study of the $p$-shell nuclei.
It is straightforward to reconcile these findings.
First, one may consider the $p$-shell calculations less reliable than the much-less-involved 
$A=3$ calculations. However, it is quite plausible that the re-normalization of the $c_D$ value for $p$-shell nuclei mimics the effect of (neglected) higher-order $NNN$ force terms, 
which are irrelevant for the calculation of the triton half life. In fact, a closer look at 
the $^4$He results shown in Fig.~\ref{CDCE_curve} and Table~\ref{predictions_4He}, the $^6$Li results from Fig.~\ref{cD_dep_6Li}, the $^{10}$B results from the right of Fig.~\ref{cD_dep_6Li} and the $^{12}$C B(M1) results, suggests a drift of the optimal $c_D$ value towards smaller (increasingly more negative) values with nuclear mass. It is natural to expect that an effect of the higher order NNN terms will become more important with increasing mass of the nucleus. Another
issue that deserves attention is the determination of the $c_3$ and $c_4$ LECs from the NN data
and the assessment of the extent to which they would influence the determination of $c_D$ from the triton half life and the $p$-shell nuclei calculations. The $c_4$ LEC is, in particular, poorly constrained by the NN data fit~\cite{Machleidt-private}.

\begin{figure}
\begin{minipage}{8cm}
{\includegraphics[width=1.0\columnwidth]{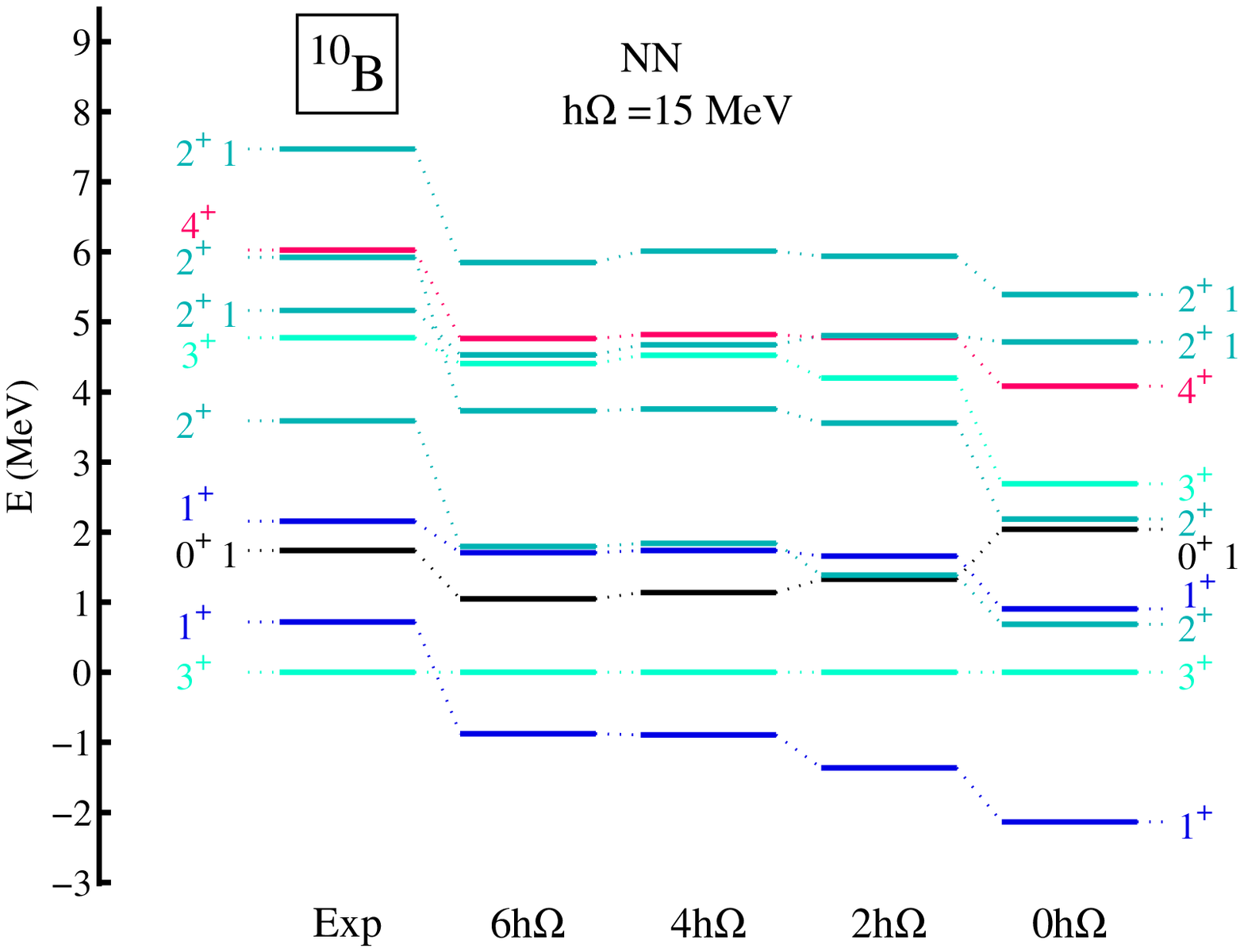}}
\end{minipage}
\begin{minipage}{8cm}
{\includegraphics[width=1.0\columnwidth]{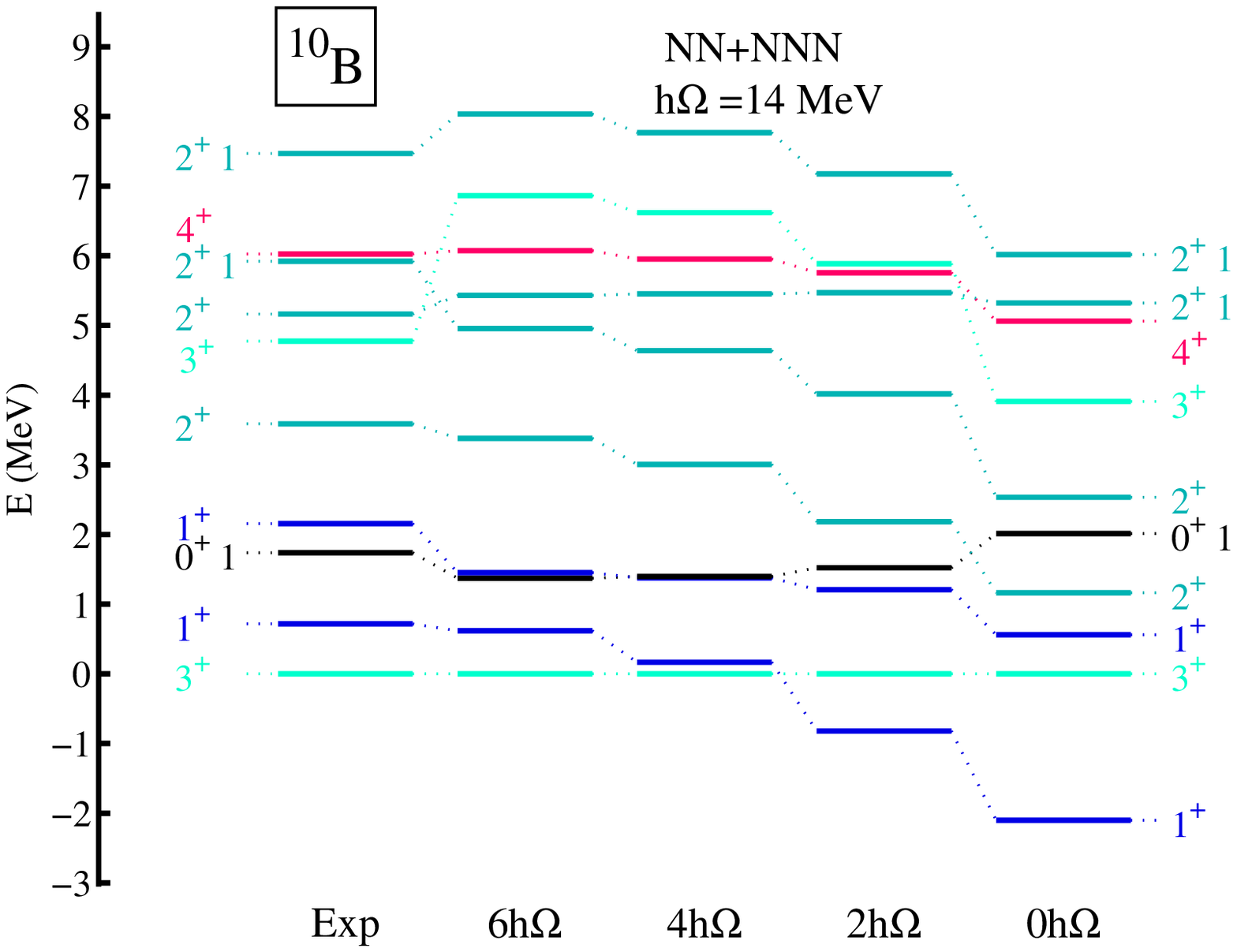}}
\end{minipage}
\caption{$^{10}$B excitation spectra as function of the basis-space size $N_{\rm max}$ 
with the chiral NN (left) and NN+NNN (right) 
interaction compared to experiment. The isospin of the states 
not explicitly depicted is $T$=0. The HO frequencies of  $\hbar\Omega=14$ MeV (left) and 15 MeV (right) were used.}\label{B10_NN_NNN}
\end{figure}

In Figs.~\ref{B10_NN_NNN}, we present the excitation spectra of $^{10}$B, as 
a function of $N_{\rm max}$, for both the chiral NN+NNN, as well as with 
the chiral NN interaction alone. In both cases, the convergence 
with increasing $N_{\rm max}$ is quite reasonable for the low-lying states. 
Similar convergence rates are obtained for our other $p-$shell nuclei calculations.

A remarkable feature of the $^{10}$B results is the observation that the chiral NN
interaction alone predicts the incorrect ground-state spin of $^{10}$B. The experimental
value is $3^+ 0$, while the calculated one is $1^+ 0$. On the other hand, once we also 
include the chiral NNN interaction in the Hamiltonian, which is actually required by 
the $\chi$EFT, the correct ground-state spin is predicted. Further, once we select the $c_D$
value, as discussed above, i.e., $c_D=-1$, we also obtain the two lowest $1^+ 0$ states
in the experimental order.  

\begin{figure}[t]
{\includegraphics[width=1.0\columnwidth]{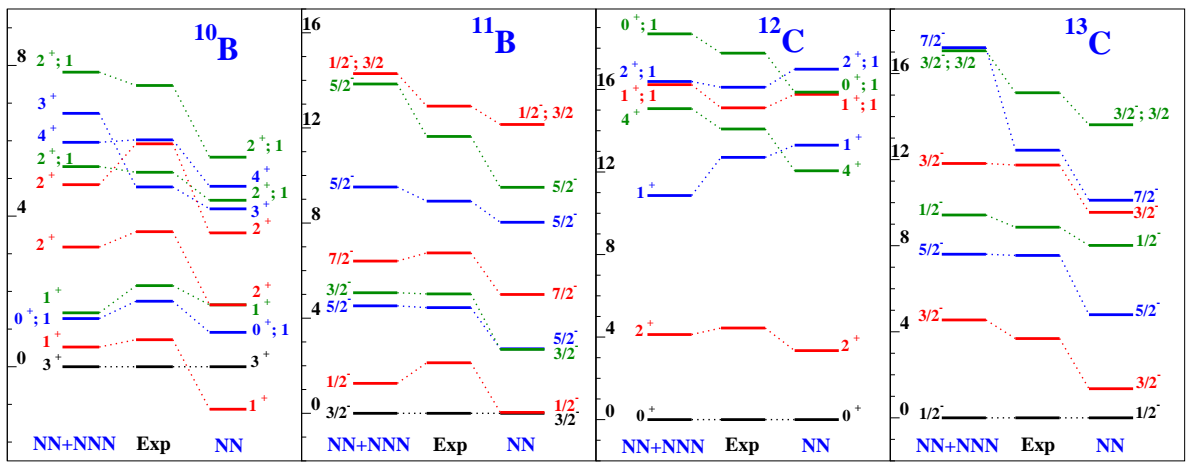}}
\caption{States dominated by $p$-shell configurations for $^{10}$B, $^{11}$B, $^{12}$C, 
and $^{13}$C calculated at $N_{\rm max}=6$ using  $\hbar\Omega=15$ MeV 
(14 MeV for $^{10}$B). Most of the eigenstates are isospin $T$=0 or 1/2, 
the isospin label is explicitly shown only for states with $T$=1 or 3/2.
The excitation energy scales are in MeV.}
\label{B10B11C12C13}
\end{figure}

We display in Fig.~\ref{B10B11C12C13} the natural-parity excitation spectra of four nuclei 
in the middle of the $p-$shell with both the NN and the NN+NNN effective interactions 
from $\chi$EFT. The results shown are obtained 
in the largest basis spaces achieved to date for these nuclei with the NNN interactions, 
$N_{\rm max}=6$ ($6\hbar\Omega$). Overall, the NNN interaction contributes 
significantly to improve theory in comparison with experiment. 
This is especially well-demonstrated in the odd mass nuclei for the lowest, few excited states. 
The case of the ground-state spin of $^{10}$B and its sensitivity to the presence 
of the NNN interaction, discussed also in Fig.~\ref{B10_NN_NNN},
is clearly evident. We note that the $^{10}$B results in the left panel of Fig.~\ref{B10_NN_NNN},
only with the NN interaction, were obtained with the HO frequency of $\hbar\Omega=15$ MeV, 
while those
in Fig.~\ref{B10B11C12C13} are with $\hbar\Omega=14$ MeV. A weak HO frequency dependence
of the $N_{\rm max}=6$ results is evident. The $^{10}$B results with the NN+NNN interaction,
presented in Figs.~\ref{B10_NN_NNN}
and \ref{B10B11C12C13}, were obtained using the same HO frequency. Still, one may notice
small differences of the $N_{\rm max}=6$ results. The reason behind those differences
is the use of two alternative D-term regularizations, discussed in Ref.~\cite{N2LO-local}.
As the dependence on the regulator is a higher-order
effect than the $\chi$EFT expansion order used to derive the NNN interaction, 
these differences should have only minor effect. It is satisfying that the present
$^{10}$B results appear to support this expectation.

Concerning the $^{12}$C results, there is an initial indication 
that the chiral NNN interaction is somewhat over-correcting the inadequacies of the NN interaction 
since, e.g., the $1^+ 0$ and $4^+ 0$ states in $^{12}$C
are not only interchanged, but they are also spread apart 
more than the experimentally observed separation. 
In the $^{13}$C results, we can also identify an indication of 
an overly strong correction arising from the chiral NNN interaction,
as seen in the upward shift of the $\frac{7}{2}^-$ state.
However, the experimental $\frac{7}{2}^-$ may have significant intruder components 
and is not well-matched with our state. 
In addition, convergence for some higher-lying states is affected by incomplete treatment 
of clustering in the NCSM. This point will be elaborated upon later. 
These results required substantial computer resources. 
A typical $N_{\rm max}=6$ spectrum, shown in Fig. \ref{B10B11C12C13}, 
and a set of additional experimental observables 
took 4 hours on 3500 processors of the LLNL's Thunder machine. 
The $A$-nucleon calculations were performed in the Slater determinant
HO basis using the shell model code MFD \cite{MFD}. 

The calculations presented in this section demonstrate that the chiral NNN 
interaction makes substantial contributions 
to improving the spectra and other observables. However, there is 
room for further improvement in comparison with experiment. 
In these calculations we used a strength of the 2$\pi$-exchange piece of the NNN interaction,
which is consistent with the NN interaction that we employed (i.e. from Ref.~\cite{N3LO}). 
As we already discussed, this strength is somewhat uncertain (see, e.g., Ref.~\cite{Nogga06}). 
Therefore, it will be important to study the sensitivity of our results 
with respect to this strength.
Further on, it will be interesting to incorporate sub-leading NNN interaction terms 
that are currently under development~\cite{Be08} and also
four-nucleon interactions, which are also order N$^3$LO \cite{Epelbaum06}.
Finally, it is useful and currently feasible
to extend the basis spaces to $N_{\rm max}=8$ ($8\hbar\Omega$) also for $A>6$ 
to further improve convergence.

\subsection{Photodisintegation of $^4$He}

Over the years, measurements of the $\alpha$-particle photodisintegration in the near-threshold region have been controversial, particularly concerning the height of the cross section at the peak, for which one can find differences up to a factor of two between different experiments (see, e.g., Ref.~\cite{chiral_LIT_4He} and references therein). On the other hand, theoretical calculations of the $^4$He photoabsorption cross section have reached an unprecedented level of accuracy in the recent past~\cite{chiral_LIT_4He,Gazit}. Predictions obtained using high-precision NN and NNN interactions models lie in a rather contained band, which is remarkable compared to the large discrepancies still present among the different experimental data.  In this section, we review the results obtained for the $^4$He total photoabsorption cross section using the above presented chiral NN+NNN interactions (for $c_D=-1$).

Nuclear photoabsorption processes can be described in good approximation by the cross section
\begin{equation}
\label{sigma}
\sigma_\gamma(\omega)=4\pi^2\frac{e^2}{\hbar c}\omega R(\omega)\;,
\end{equation}
where $\omega$ is the incident photon energy and $R(\omega)$ is the inclusive response function in the long wavelength approximation
\begin{equation}
\label{response}
R(\omega)=\int d\Psi_{f}\left|\left\langle \Psi_{f}\right|\hat D\left|\Psi_{0}\right\rangle \right|^{2}\delta(E_{f}-E_{0}-\omega)\,.
\end{equation}
This  is the sum of all  the transitions from the ground state $|\Psi_0\rangle$ (of energy $E_0$) to the various allowed final states $|\Psi_f\rangle$ (of energy $E_f$) induced by the electric dipole operator $\hat D$.
Here, a fully {\em ab initio} result for the response function (hence for the cross section) was obtained by means of the Lorentz integral transform (LIT) method~\cite{EBLO07} implemented in the framework of the NCSM approach~\cite{LIT_NCSM}. More specifically, we first performed an accurate NCSM calculation of the $^4$He ground state reaching the same level of convergence as shown in Fig.~\ref{gs_H3_He4}, by means of effective interactions at the three-body cluster level. We then evaluated the LIT of the response~(\ref{response}) by applying the Lanczos algorithm to the chiral Hamiltonian, using as starting vector the transition operator acting on the ground state, $\hat D|\psi_0\rangle$. Indeed the LIT can be expressed in terms of a continued fraction of the so-called Lanczos coefficients (i.e., the elements of the tridiagonal Hamiltonian in the Lanczos basis)~\cite{Marchisio}. After inversion of the LIT~\cite{inversion}, the cross section is obtained from Eq.~(\ref{sigma}).  
\begin{figure}
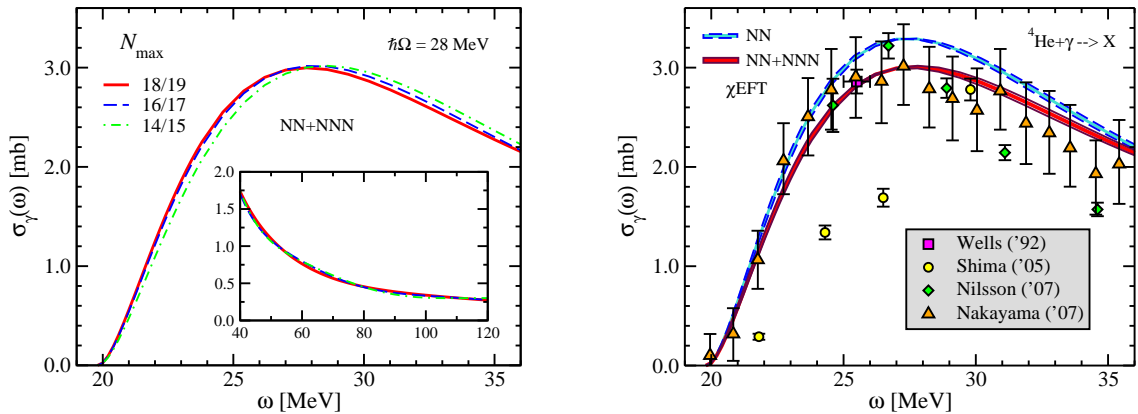

\begin{minipage}{8cm}
{ \includegraphics[height=.23\textheight]{conv_cs_ChEFT_Review.eps}}
\end{minipage}
\begin{minipage}{8cm}
{\includegraphics[height=.23\textheight]{comp_cs_ChEFT_Review.eps}}
\end{minipage}
  \caption{The $^4$He photo-absorption cross section as a function of the excitation energy $\omega$. Present results are for $\chi$EFT interactions, and in particular: (left panel) convergence pattern of the NN+NNN calculation with respect to the model-space truncation $N_{\rm max}$ for $\hbar\Omega=28$ MeV; (right panel) comparison to the most recent experiments~\cite{Shima:2005,Nilsson:2005,nakayama:021305}.}
  \label{csexp}
\end{figure}
The use of three-body effective interaction allows a stable and accurate convergence not only of the $^4$He ground state, but also of the photoabsorption cross section. This is presented in the left panel of Fig.~\ref{csexp}, where the inset shows the tail of the cross section.
Due to the selection rules induced by the dipole operator, the Lanczos vectors have parity opposite to the $^4$He ground state. Therefore the convergence of our results is studied as a function of  $N_{\rm max}/N_{\rm max}+1$, the truncations adopted for the $0^+0/1^-1$ model spaces, respectively. 
Our results, presented in the right panel of Fig.~\ref{csexp} together with the most recent experiments, show a peak around the excitation energy of $\omega=27.8$ MeV, with a peak height mildly sensitive to the NNN force. The experimental situation in the near-threshold region is controversial:  two direct measurements performed using quasi-mono-energetic photons~\cite{Shima:2005,Nilsson:2005} show discrepancies up to a factor of two on the absolute height of the cross-section peak.  We find an overall good agreement with the photo-disintegration data from bremsstrahlung photons of Nilsson {\em et al}.~\cite{Nilsson:2005}, while we reach only the last of the experimental points of Ref.~\cite{Shima:2005}. In particular, the   confused experimental situation drawn by these two data sets does not allow to assess the role of the NNN force effect. Recently Nakayama {\em et al}. performed an indirect measurements of the $\alpha$-particle total photo-absorption cross section~\cite{nakayama:021305} by observing its analog via the $^4$He($^7$Li,$^7$Be) reaction at an incident energy of $455$ MeV and at forward scattering angles. Although the uncertainty on this extracted absolute cross section is 20\% or more, the inclusion of the NNN terms of the interaction appear to improve the agreement of the calculated cross section with the latter indirect measurement.  

\section{Radii, moments and transitions in light nuclei}\label{sec:Mom}

\subsection{Electric dipole moment of $^3$He}

Massive (and expensive) experimental efforts are directed toward high precision tests of the Standard Model.
While obviously the experimental confirmation of the Higgs boson is of extreme importance, somewhat smaller scale
experiments can be set up in search for physics beyond the standard model. Presently, several
experimental programs are pushing the limits on the detection of electric dipole moments (EDMs) of the
nucleon, nuclei and atoms. Thus, a permanent EDM of a system requires a direct violation of the time-reversal (T) and
parity (P) and hence CP violation through the CPT theorem. The standard model allows for a very little CP-violation, at levels that
are too small to be observed currently, and thus any non-zero measurement of EDMs would be a clear
signal of sources of CP-violation beyond the Standard Model. 

Motivated by the recent proposal of a new scheme for measuring EDMs of light nuclei stripped of their electrons (e.g., deuteron, $^3$He) in magnetic storage rings, the first comprehensive \textit{ab initio} calculation of the EDM of $^3$He has been reported in Ref.  \cite{Stetcu_edm}. In this investigation, two distinct contributions have been considered: (i) the intrinsic EDMs of each nucleon, $d_p$ and $d_n$ respectively, and (ii) polatization effects induced by the CP violation in the NN interaction.

The three-body problem was solved with high accuracy in relative coordinates, using several potential models. Thus, we have obtained results for the CD Bonn potential \cite{cdb2k}, Argonne V18 interaction \cite{AV18}, as well as the latest generation chiral two- \cite{N3LO} and three-body interactions~\cite{N2LO-local}. When three-body forces are included, an excellent description of the ground-state properties is achieved, as discussed above. The polarization effects, which give the largest contribution to the EDM, were obtained in perturbation theory, using the Podolsky method \cite{podolsky}, similar to the LIT application to exclusive processes~\cite{Marchisio}. However, in the absence of a EFT derivation of the CP-violating interaction, a one-boson exchange model had to be used in this case. Thus, we have considered all possible $\pi$, $\rho$ and $\omega$ CP-violating exchanges (for a detailed expression of the CP-violating interaction used, see Ref. \cite{Stetcu_edm}), the final expression of the EDM being expressed as a superposition of CP-violating coupling constants $\bar G_x^T$, where $x$ stands for the meson exchanged, and $T$ for the isospin. The coupling constants are unknown, and only limits exist.

A consistent approach would require that the same transformation used to derive the effective interaction be used for any observable calculated with the respective wave function. In this case, this means that the dipole-moment operator and the CP-violating interaction should be also renormalized, as discussed in Sec. \ref{sec:effOp}. However, as shown in Sec. \ref{sec:Conv}, the long-range operators, like the dipole transition operator, are insensitive to the renormalization. The long-range correlations are built by increasing the model space, in this case by increasing the number of HO shells used to construct the many-body basis. Hence, we observe the convergence of the EDM with the model space, and, thus, we find that in large-sized model spaces, the results become independent of the parameters used (HO frequency and number of HO shells).

\begin{figure}
\centering\includegraphics*[width=0.8\columnwidth]{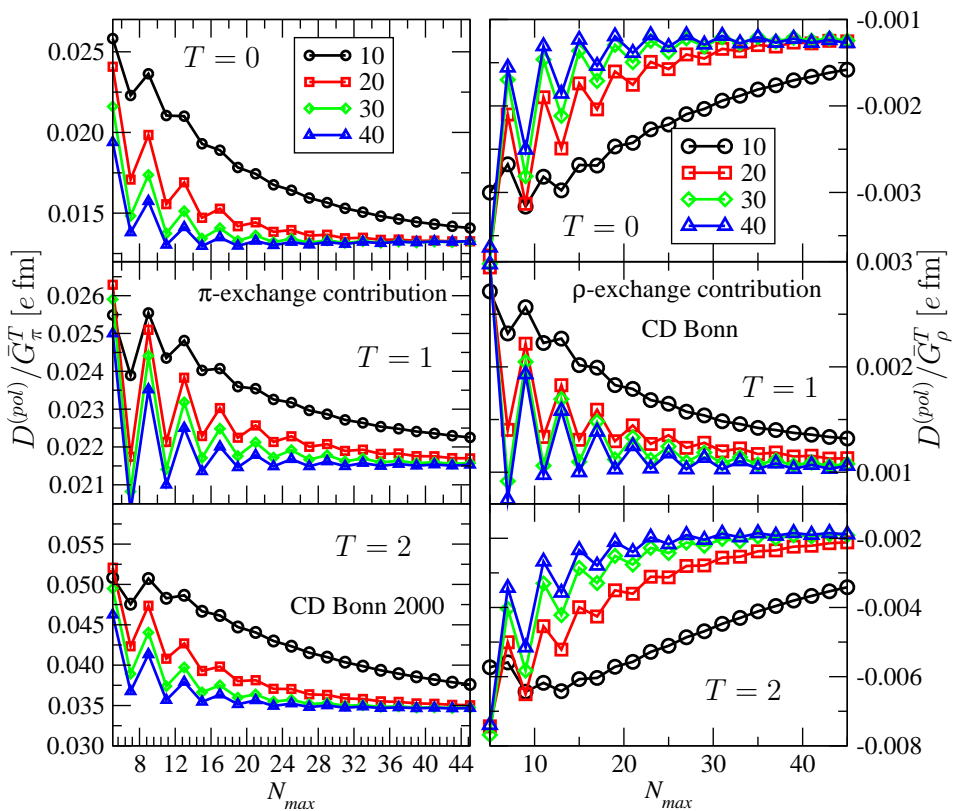}
  \caption{The polarization contribution to the EDM of $^3$He, decomposed into isoscalar, isovector and isotensor components. We present contributions for both $\pi$ and $\rho$ exchanges; the $^3$He ground state has been obtained using the CD Bonn NN interaction. Convergence with the model-space size for different HO frequencies is shown.}
  \label{fig:edm3He}
\end{figure}

Complete results for the $\pi$, $\rho$ and $\omega$ exchanges have been reported in Ref. \cite{Stetcu_edm}. In Fig. \ref{fig:edm3He}, we present the convergence of the isoscalar, isovector and isotensor components of the EDM for the $\pi$ and $\rho$ exchanges ($\omega$ exchange has the same order of magnitude as the $\rho$). In the case presented here, the ground-state wave function was obtained by the diagonalization of the effective interaction, obtained from the non-local CD Bonn NN interaction, while three-body forces were neglected. In the absence of an effective theory, which would achieve consistent description of CP-conserving and CP-violating observables, the model dependence cannot be completely removed from our results. However, the model dependence for the pion exchange is small, since the long-range part of the wave function shows little model dependence. Not surprisingly, the short-range contributions to the EDM, i.e., $\rho$ and $\omega$ meson exchanges, present a fairly strong dependence on the choice of the CP-conserving NN potential model. Nevertheless,  because $^3$He is mainly an $S$-state nucleus, and the CP-violating Hamiltonian involves $S$- to $P$-wave transitions, the effect is two fold: (i) the long range contribution, i.e., $\pi$, is enhanced, and (ii) the short-range contribution ($\rho$ and $\omega$ exchanges) are suppressed. Hence, if the CP-coupling constants $\bar G^T_x$ are of similar magnitude as expected, the $\pi$ contribution dominates. A quick comparison in Fig. \ref{fig:edm3He} shows that indeed the $\rho$ (and similarly the $\omega$) contribution to the $^3$He EDM is only about $10\%$ of the $\pi$ exchange. In the hypothesis that the unknown CP-violating coupling constants $\bar{G}^T_x$ are of the same order of magnitude for $\pi$, $\rho$ and $\omega$ exchanges, we, thus, conclude that the $\pi$ contribution dominates the EDM, and its value is

\begin{equation}
D=(0.024\, \bar G_\pi^0+0.023\, \bar G_\pi^1+0.027\,\bar G_\pi^2)\:e\,\mathrm{fm}.
\end{equation}
This value was obtained after a compilation of all the potential models we have used (see Ref. \cite{Stetcu_edm}).
(Note that in the absence of isospin violation in the Hamiltonian, there is a trivial relationship between the EDM of $^3$He and that of the triton, i.e., the isoscalar and isotensor contributions change sign; the small isospin violation slightly breaks this symmetry.) If we compare this result with the neutron dipole moment $d_n=(0.010\,\bar G_\pi^0-0.010\,\bar G_\pi^2)\: e$ fm, and deuteron EDM, $D_{deut}=0.015\, \bar G_\pi^1\:e$ fm, we see immediately that the three EDMs are complementary. Consequently, we can conclude that a measurement of these three systems would provide a valuable constraint for the theoretical models of CP-violating NN interactions.

\subsection{Charge radii of He isotopes}

Recent advances in the theory of the atomic structure of helium 
as well as in the techniques of isotopic shift measurement made it possible
to determine accurately the charge radius of $^6$He \cite{6He_chrad} and $^8$He \cite{8He_chrad}. 
Precision laser spectroscopy on individual $^6$He and  $^8$He atoms confined and cooled in a 
magneto-optical trap was performed and measured the isotope shift between
$^6$He, $^8$He and $^4$He. With the help of precise quantum-mechanical calculations
with relativistic and QED corrections \cite{Drake} and from the knowledge of 
the charge radius of $^4$He (1.673(1) \cite{4He_chrad}), 
it was possible to determine the charge radius of $^6$He to be $2.054\pm0.014$ fm \cite{6He_chrad} 
and the charge radius of  $^8$He to be $1.93\pm3$ fm~\cite{8He_chrad}. 
The large differences between the $^4$He and $^6$He and between the $^4$He and $^8$He 
charge radii is due to the extra loosely-bound neutrons in $^6$He and $^8$He that form a halo \cite{Tanihata}. 
The reduction in charge radius from $^6$He to $^8$He indicates a change in the correlations 
of the excess neutrons.

It is a challenge for {\it ab initio} many-body methods to calculate 
the nuclear radii with an accuracy comparable to current experimental 
accuracy and, thereby, test the nuclear Hamiltonians used as 
the input of {\it ab initio} calculations. In Ref.~\cite{NCSM_He_rad}, the ground-state properties 
of $^4$He, $^6$He and $^8$He were calculated within the NCSM using two different high-precision NN potentials:
the CD-Bonn~\cite{cdb2k} and the INOY\cite{INOY}. 

The $^4$He calculations were performed both in the Slater determinant basis
using the Antoine code~\cite{Antoine} with model spaces up to $N_{\rm max}=22$ 
within the two-body effective interaction approximation and the
Jacobi-coordinate HO basis using the Manyeff code~\cite{Jacobi_NCSM} with model spaces
up to $N_{\rm max}=20$ within either the two-body 
or the three-body effective interaction approximation (with both approximations 
converging to the same result). The ground-state energy and radius
convergences are good for both NN potentials. The NCSM calculations for $^6$He and $^8$He nuclei 
were performed within the two-body effective interaction approximation that allows one to reach much larger model-space sizes than within the three-body effective interaction approximation. As the radius operator is a long-range operator, it is essential to use as large an HO basis as possible. Using the Antoine code, we were able to reach model spaces up to $N_{\rm max}=16$ and $N_{\rm max}=12$ for $^6$He and $^8$He, respectively, for a wide range of HO frequencies. 

\begin{figure}
\begin{minipage}{8cm}
{\includegraphics[width=0.9\columnwidth]{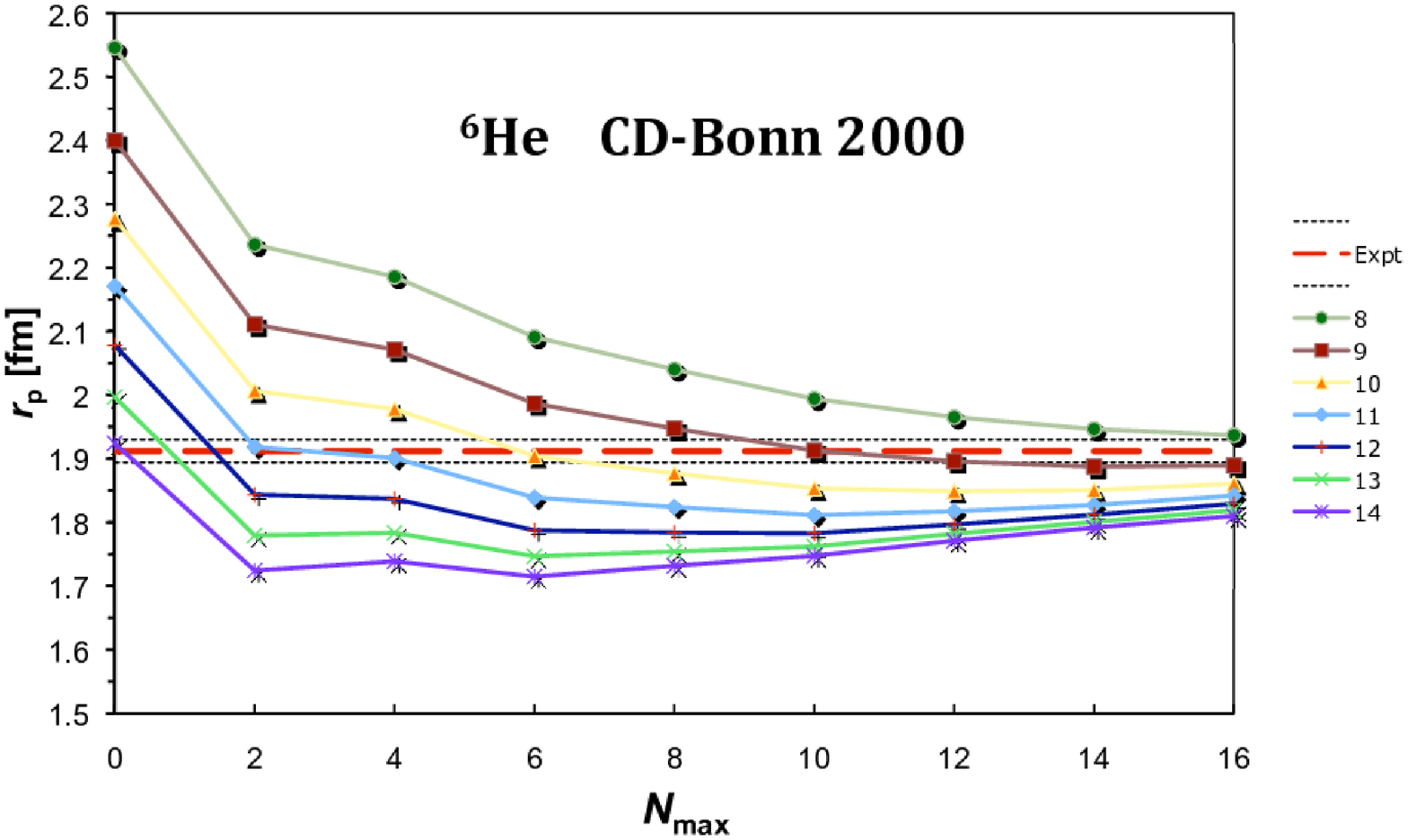}}
\end{minipage}
\begin{minipage}{8cm}
{\includegraphics[width=0.9\columnwidth]{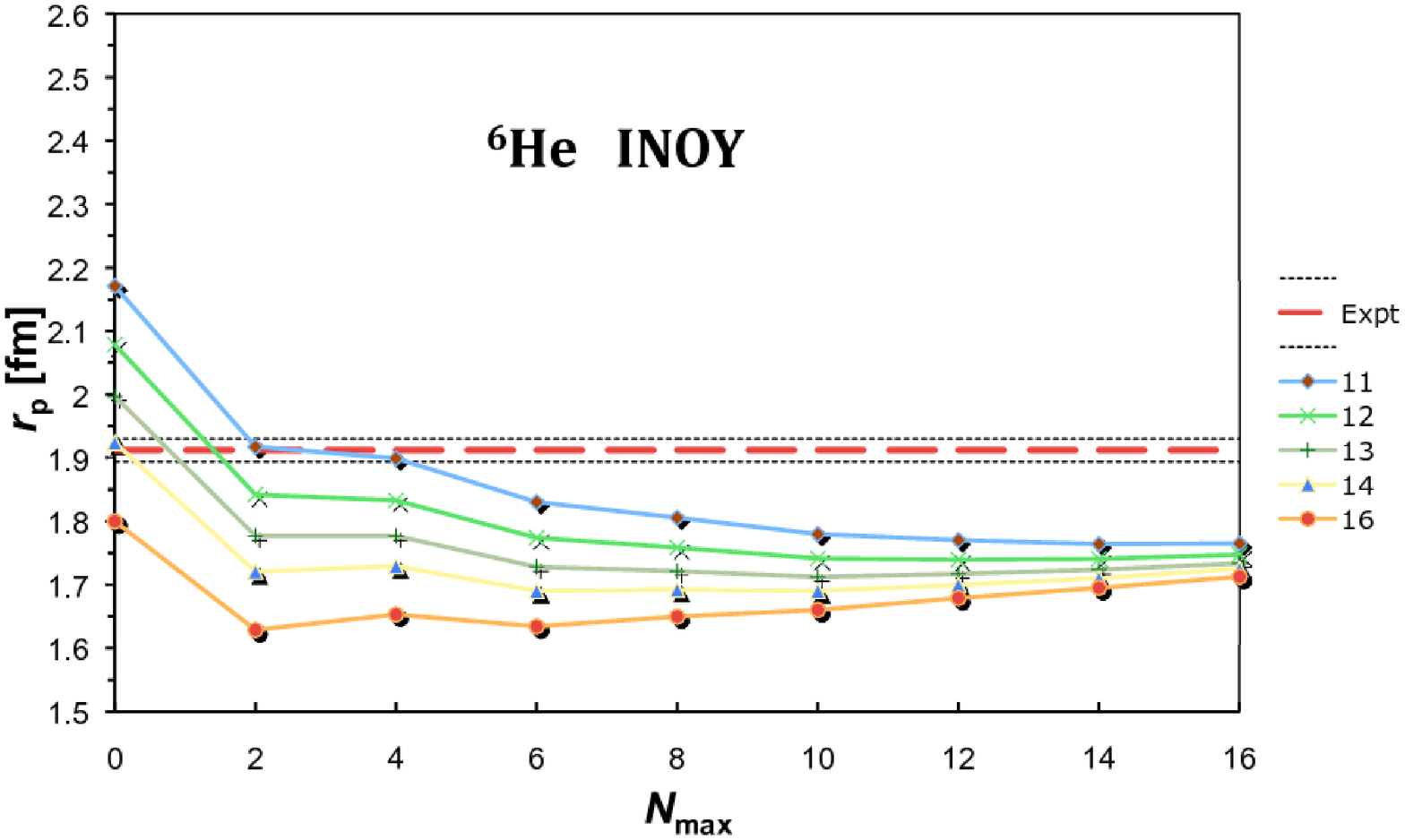}}
\end{minipage}
\begin{minipage}{8cm}
{\includegraphics[width=0.9\columnwidth]{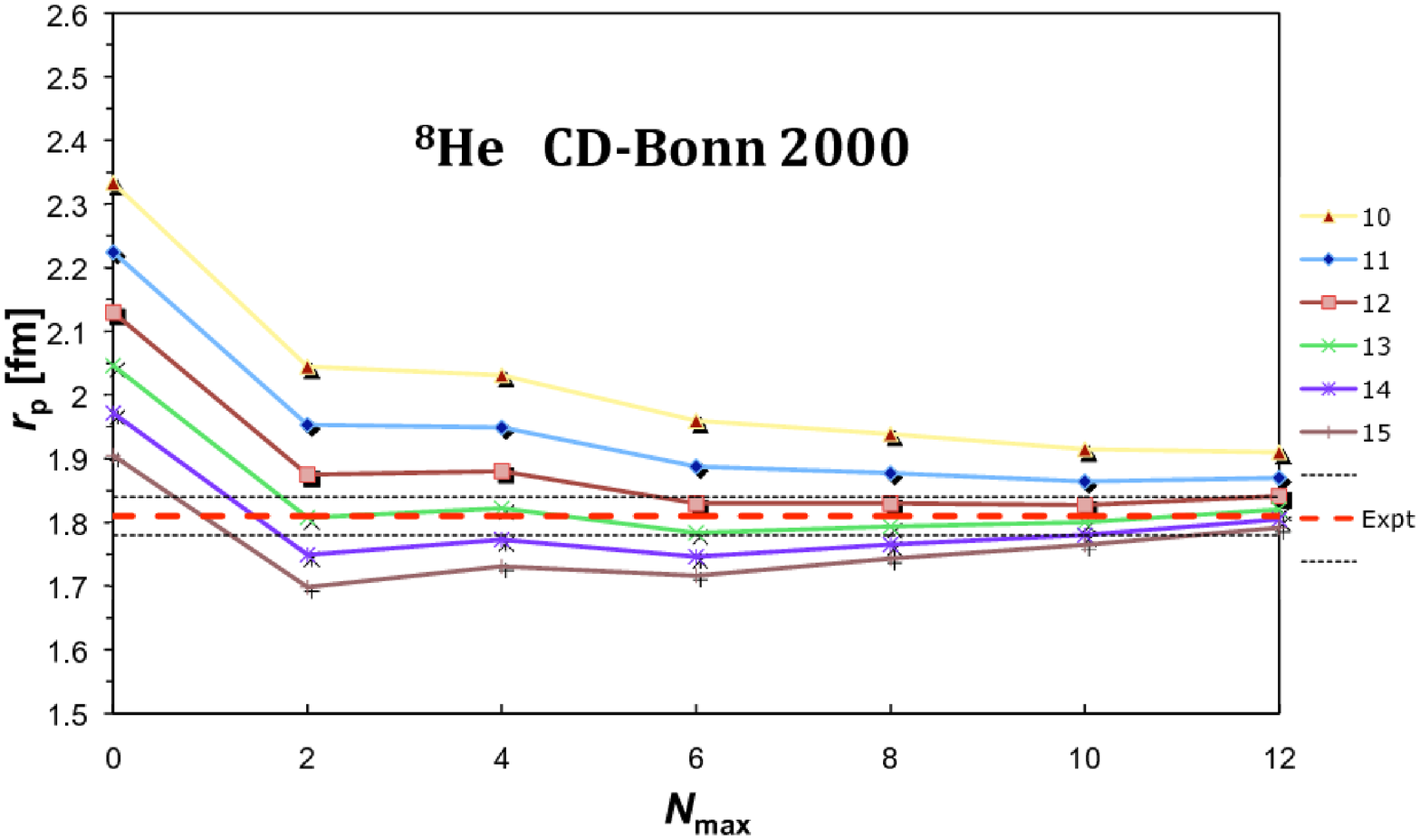}}
\end{minipage}
\begin{minipage}{8cm}
{\includegraphics[width=0.9\columnwidth]{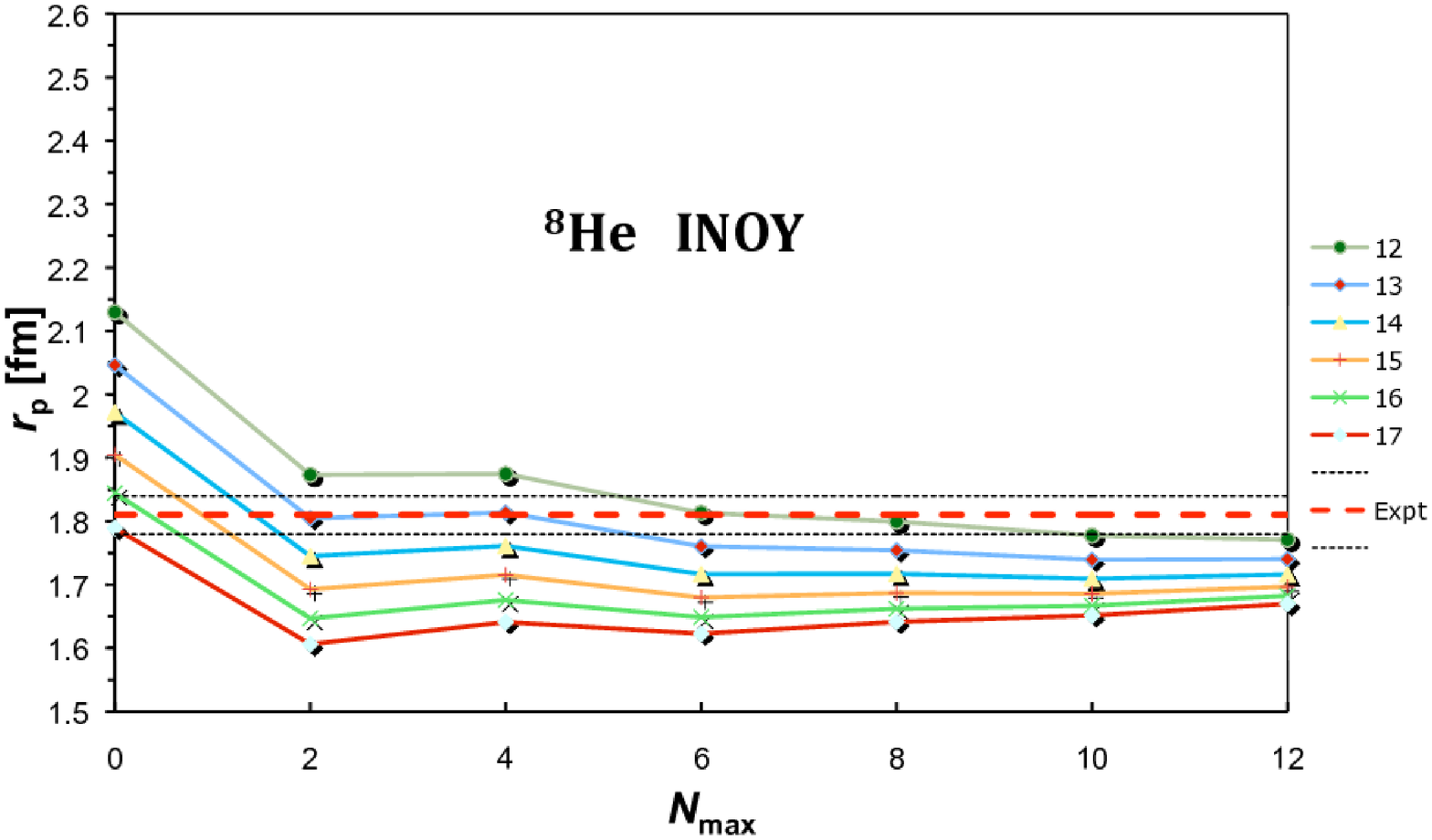}}
\end{minipage}
\caption{Dependence of the $^6$He (top panels) and $^8$He (bottom panels) point-proton rms radius on the model-space size
for different HO frequencies, obtained using the CD-Bonn 2000 (left panels) 
and the INOY (right panels) NN potentials. 
The experimental values are from Refs. \protect\cite{6He_chrad} and \protect\cite{8He_chrad}.}
\label{He6_rp}
\end{figure}

The point-proton root-mean-square (rms) radii results are summarized in Table~\ref{tab:r_p}. We note that the point-proton rms radius is related to the proton charge rms radius as follows \cite{6He_chrad}
$\langle r_p^2\rangle = \langle r_c^2 \rangle - \langle R_p^2\rangle 
- \langle R_n^2\rangle (N/Z)$, with $(\langle R_p^2\rangle)^{1/2}=0.895(18)$ fm 
\cite{R_p} (the charge radius of the proton) and $\langle R_n^2\rangle=-0.120(5)$ fm$^2$
\cite{R_n} (the mean-square-charge radius of the neutron).
\begin{table}[t]
  \caption{Point-proton ($r_p$) rms radii of $^{4,6,8}$He isotopes.
The calculated values were obtained within the {\it ab initio} NCSM~\protect\cite{NCSM_He_rad}.
The experimental values are from Refs.~\protect\cite{4He_chrad,6He_chrad,8He_chrad}.
\label{tab:r_p}}
     \begin{tabular}{cccc}
$r_p$ [fm]  & Expt.    & CD-Bonn 2000 & INOY \\
\hline
$^4$He & 1.455(1)   & 1.45(1)      & 1.37(1) \\ 
$^6$He & 1.912(18) & 1.89(4)      & 1.76(3) \\
$^8$He & 1.81(3)     & 1.88(6)      & 1.74(6) \\
    \end{tabular}
 \end{table}
In Fig.~\ref{He6_rp}, we show the model-space size dependence 
of the rms radii for different HO frequencies. A general feature is a decrease
of the HO frequency dependence with increasing model-space size defined by $N_{\rm max}$. 
In all cases, the rms radii exhibit convergence. The $^6$He point-proton rms
radius experimental value is shown as a dashed line in top panels of Fig.~\ref{He6_rp}, 
with the dotted lines indicating the experimental error.
The CD-Bonn 2000 $^6$He point-proton rms radius stabilizes
at $N_{\rm max}=16$ for the HO frequencies of $\hbar\Omega=9$ and 10 MeV, while
it is still decreasing for $\hbar\Omega=8$ MeV but is
increasing for the HO frequencies higher than $\hbar\Omega=10$ MeV. Clearly,
the stable result is very close to the experimental value. The error was estimated from the
HO frequency dependence of the $N_{\rm max}=16$ calculations. Based on these results, 
we arrive at the CD-Bonn 2000 point-proton rms radius of 1.89(4) fm that, taking 
into account the error bars, agrees with the experimental value of 1.912(18) fm.
We observe a better convergence for the INOY NN potential not only for the binding
energies, discussed in detail in Ref.~\cite{NCSM_He_rad}, but also for the radii. 
This is apparent from Fig.~\ref{He6_rp}. For this NN potential, we find the $^6$He 
point-proton rms radius to be 1.76(3) fm. This is significantly less than in experiment. 
Clearly, the INOY NN potential underpredicts both the $^4$He and $^6$He point-proton rms radii.  

The NCSM $^8$He point-proton rms radius results are shown in bottom panels of Fig.~\ref{He6_rp} 
for the CD-Bonn 2000 and INOY potentials. From the basis size and the HO frequency dependence, 
a prediction was made in Ref.~\cite{NCSM_He_rad} that the $^8$He point-proton rms radius is 1.88(6) fm based on the NCSM CD-Bonn results.
The INOY NN potential gives a smaller value, 1.74(6) fm, consistent with the 
smaller $^4$He and $^6$He results. In both cases, the $^8$He point-proton radius is 
slightly smaller then the corresponding one in $^6$He. The subsequent experimental 
measurement~\cite{8He_chrad} found the $^8$He radius of $1.93(3)$ fm, which translates to a point-proton rms radius of $1.81(3)$ fm, which is slightly lower than the NCSM CD-Bonn result but still consistent with it, taking into account the theoretical and experimental error bars.

\subsection{Electromagnetic moments of Li and Be isotopes}

Recent developments of both experimental and theoretical techniques have
allowed for very precise measurements of charge radii and ground-state
electromagnetic moments of exotic
isotopes~\cite{san06:96,nor08:0809.2607,bor05:72,neu08:101}.
Electric quadrupole and magnetic dipole moments in particular can
be determined using an experimental method that is based on the nuclear
magnetic resonance technique~\cite{bor05:72,neu08:101}. 
These observables reflect, in different ways, the evolving nuclear structure along the isotopic chains.

In response to these and forthcoming experimental programs, extensive {\it ab initio} NCSM calculations were performed for charge radii and electromagnetic moments of Li and Be isostopes \cite{FCN09} in a similar way as for the charge radii of He isotopes discussed in the previous subsection. In particular, the CD-Bonn and the INOY NN potentials were employed. The calculations were performed up to very large model spaces in a wide range of HO frequencies. Efforts were made to quantify the rates of convergence of observables. In order to maximize the size of the basis, a two-body effective interaction approximation was employed. 

\begin{figure}
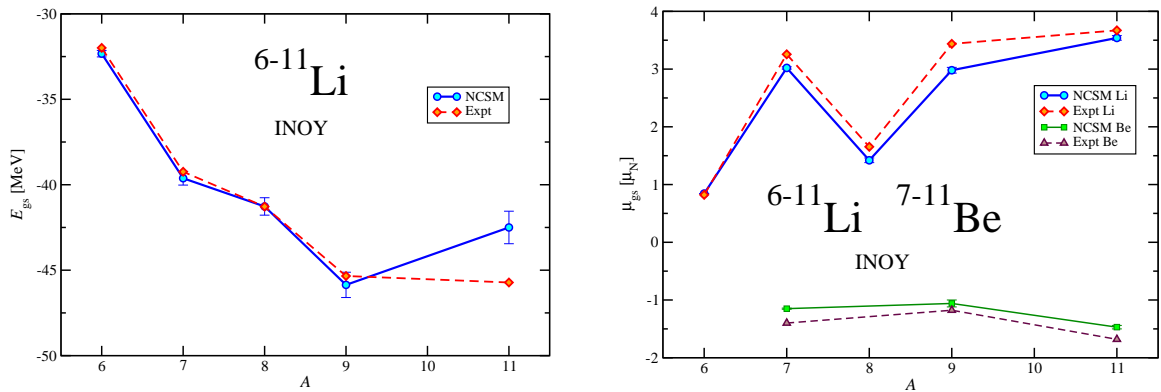

\vspace{0.5cm}
\begin{minipage}{8cm}
{\includegraphics[width=0.9\columnwidth]{Li_gs_inoy.eps}}
\end{minipage}
\begin{minipage}{8cm}
{\includegraphics[width=0.9\columnwidth]{Li_Be_mu_inoy.eps}}
\end{minipage}
\caption{The Li isotope ground-state energies (left panel) and Li and Be isotope magnetic moments (right panel) obtained using the INOY NN potential. 
The magnetic moment experimental values are from Refs.~\protect\cite{bor05:72,neu08:101,til02:708,til04:745}.}
\label{gs_INOY_Li_Be}
\end{figure}
Detailed results for the ground-state energies, charge radii, quadrupole and magnetic moments of $^{6,7,8,9,11}$Li and $^{7,9,10,11}$Be were published in Ref.~\cite{FCN09}. Here we compare in Fig.~\ref{gs_INOY_Li_Be} the calculated ground-state energies of Li isotopes and magnetic moments of Li and Be isotopes obtained using the INOY NN potential with experimental trends. As already discussed in Subsection~\ref{sec:Conv}, convergence of eigenenergies with this potential is very good and a reliable exponential extrapolation, i.e. using $E(N_{\rm max})=E_\infty+a \;{\rm exp}(-bN_{\rm max})$, can be performed. The ground-state energies of the Li isotopes are nicely reproduced by the INOY potential except for $^{11}$Li where lower $N_{\rm max}$ truncation is used compared to lighter Li isotopes as dimensions grow steeply with $A$. Consequently, the ground-state energy extrapolation is more uncertain. Overall, with the exception
of the radius of the $^{11}$Li halo ground-state we find a very good
agreement between NCSM results and recent experiments. The overall
trends of all observables are well reproduced. 
Magnetic dipole moments characterized by good convergence properties with the NCSM
are found in agreement with the experimental trend. Another success is the tiny
quadrupole moment of $^{6}$Li that is known to pose a difficult task for
most theoretical calculations. In particular, the general failure of
three-body models for this observable has been blamed on missing
antisymmetrization of the valence nucleons and the nucleons in the
alpha-core. The NCSM correctly reproduces the very small
value, but with CD-Bonn and INOY giving different signs. Simultaneously,
the trend for the much larger moments of $A=7-11$ is nicely
reproduced. We note that the ratio $Q(^{11}{\rm Li})/Q(^9{\rm Li})$ is found to be
very close to unity, as confirmed recently by very precise experimental
data~\cite{neu08:101}. This finding is obtained without a very accurate
description of the dilute halo structure of $^{11}$Li; a structural feature
that we find would require an extension of the HO basis used in the
standard NCSM. Still, the decrease of the charge radius of $A=6-9$
isotopes is reproduced, although the INOY interaction gives too high
nuclear densities. 

\subsection{Natural and unnatural parity states of $^{9}$Be and $^{11}$Be}\label{sec:berillium}

It is a challenge for nuclear theory to describe odd-$A$ beryllium isotopes from first principles.
The $A=11$ isobar is of particular interest in this respect since it exhibits some anomalous
features that are not easily explained in a simple shell-model
framework. Most importantly, the parity-inverted $1/2^+$ ground state of
$^{11}$Be was noticed by Talmi and Unna~\cite{tal60:4} already in the early
1960s, and it still remains one of the best examples of the
disappearance of the $N=8$ magic number. 

Large-basis {\it ab initio} NCSM calculations for $^9$Be and $^{11}$Be were reported in Ref.~\cite{Fo05}.
Calculations were performed for both natural-parity and unnatural-parity states in model spaces up to $N_{\rm max}=9$ using four different accurate NN potentials: CD-Bonn 2000, Argonne V8$^\prime$~\cite{GFMC,AV18}, INOY and chiral EFT N$^3$LO.  To maximize the model-space size, the NNN forces were not included and the two-body effective interaction approximation was used. The investigation included a $^{11}$B $N_{\rm max}=9$ calculation with a basis dimension of $1.1\times 10^9$, the largest NCSM diagonalization at that time. 

\begin{figure}
\begin{minipage}{8cm}
{\includegraphics[width=0.9\columnwidth]{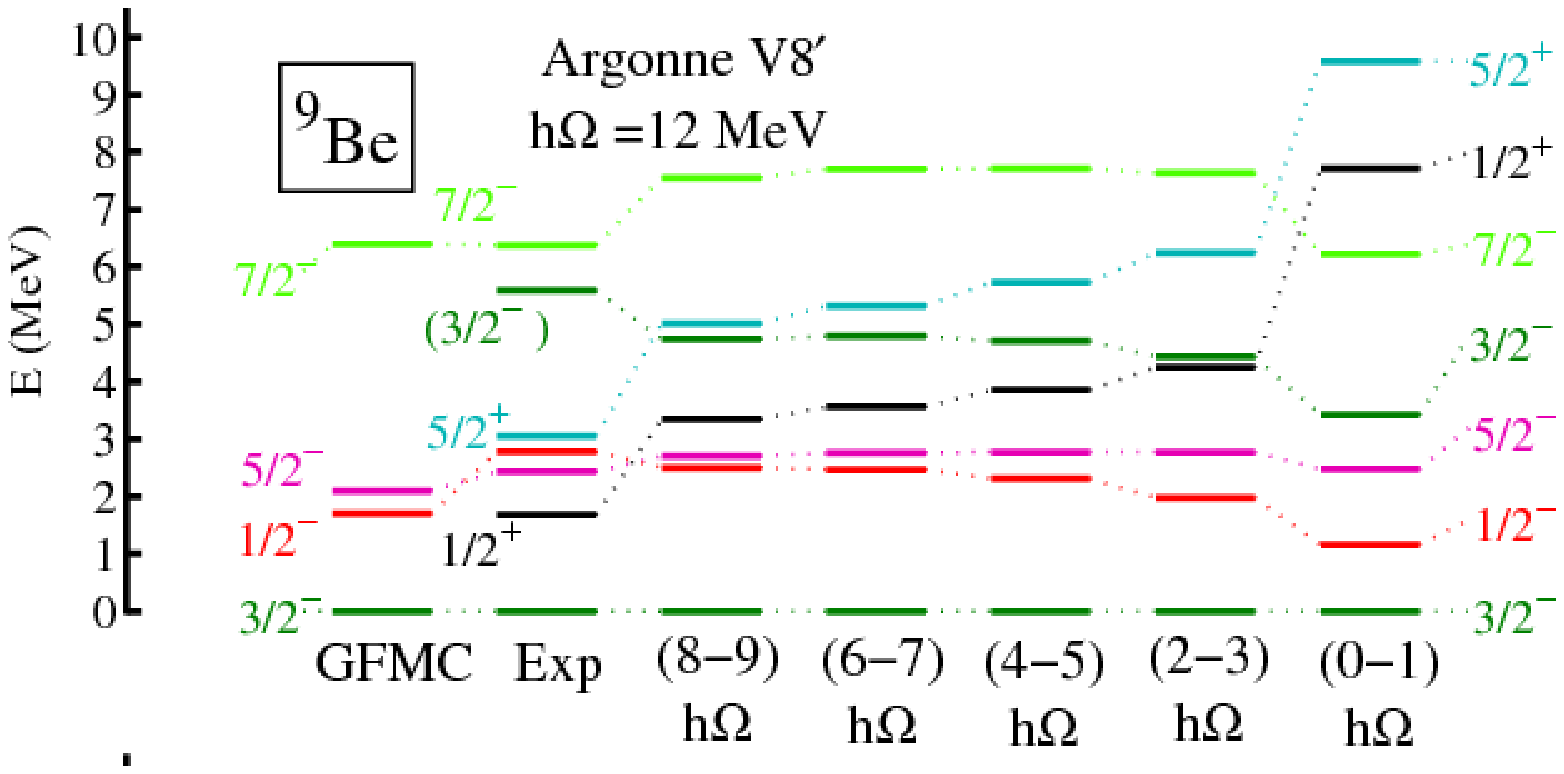}}
\end{minipage}
\begin{minipage}{8cm}
{\includegraphics[width=0.9\columnwidth]{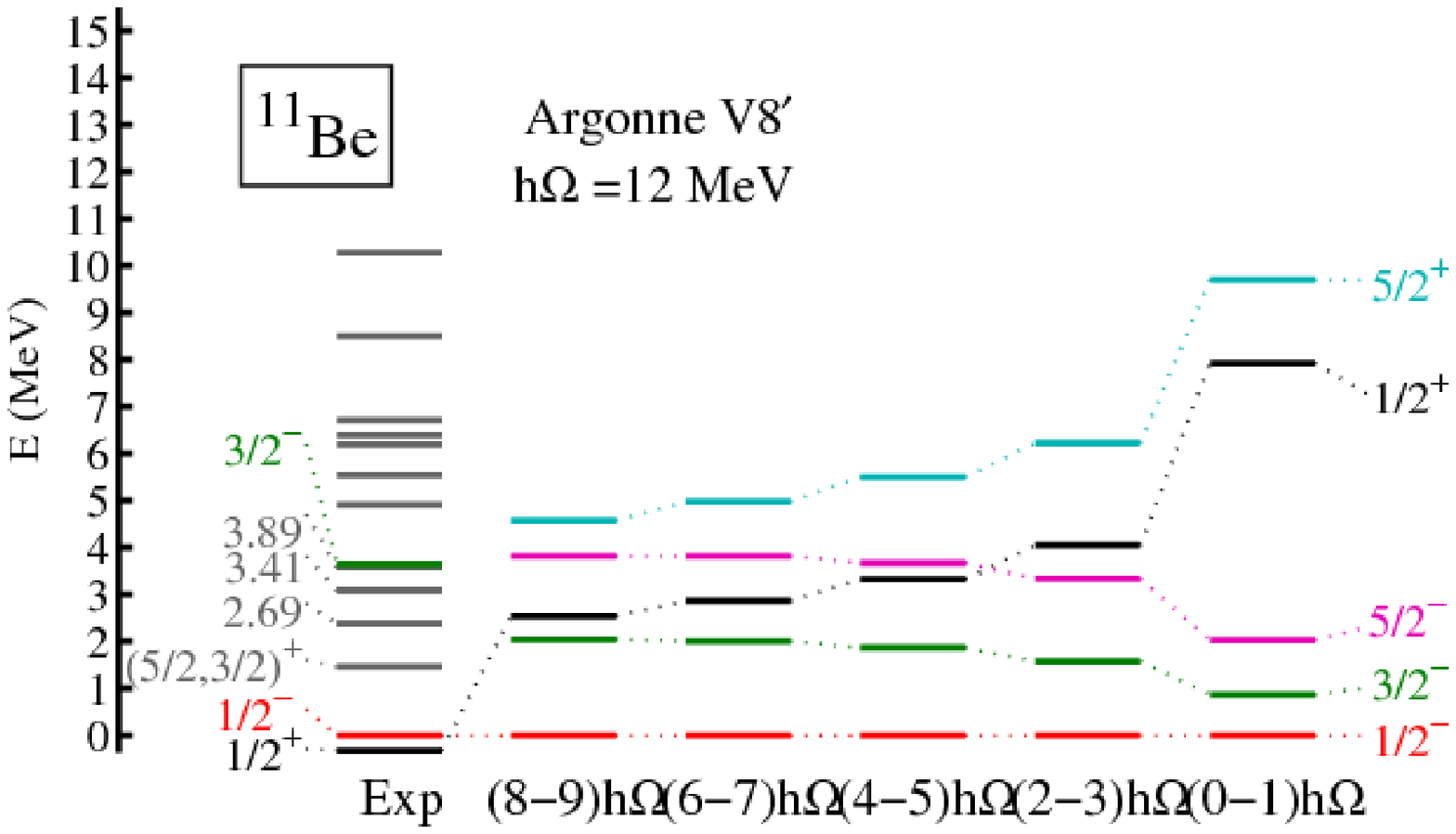}}
\end{minipage}
\caption{Excitation spectrum for $^{9}$Be (left panel) and $^{11}$Be (right panel) 
calculated using the AV8$^\prime$ interaction in $N_{\rm max}=0-9$ model spaces with a fixed HO
  frequency of $\hbar\Omega = 12$~MeV. The experimental values are from
  Refs.~\cite{til04:745} and \cite{ajz90:506rev}, respectively. 
The  AV8$^\prime$ $^9$Be results obtained by the GFMC method~\cite{GFMC} 
are shown for comparison.}
\label{Be9_Be11_levels}
\end{figure}
General features of the excitation energy results are represented in Fig.~\ref{Be9_Be11_levels}. 
We observe a very reasonable agreement with experimental
levels of natural parity, while the unnatural-parity states are
consistently high in excitation energy. For both parities, there is a
general trend of convergence with increasing model space. 
It is clear, however, that the relative position of the
negative- versus positive-parity states is still not
converged. Furthermore, when studying the AV8$^\prime$ convergence pattern, 
it seems as if this interaction will predict the positive-parity states at too high
excitation energies, in particular for $^{11}$Be, even when the calculations will be converged.  
For $^9$Be, it was found that calculations with the AV8$^\prime$ and N$^3$LO interactions
predict the first-excited negative-parity state to be a $1/2^-$, while experiments
show that it is a $5/2^-$. This level reversal was also found in the GFMC calculations
using AV8$^\prime$. The INOY interaction, on the other hand, gives the correct
level ordering, but instead overpredicts the spin-orbit splitting.

The experimental ground state of $^{11}$Be is an intruder $1/2^+$ level,
while the first $p$-shell state is a $1/2^-$ situated at
$E_x=320$~keV. The neutron separation energy is only 503~keV, and there
are no additional bound states. This level-ordering anomaly constitutes
the famous parity-inversion problem. 
An important topic of this work has been the investigation of the parity
inversion found in $^{11}$Be. The large-scale {\it ab initio} NCSM calculations of Ref.~\cite{Fo05} 
did not reproduce the anomalous $1/2^+$ ground state, but did observe a
dramatic drop of the positive-parity excitation energies with increasing
model space. Furthermore, the behavior of the INOY results suggested
that a realistic NNN force might have an important influence on the
parity inversion. However, as we discuss in the last part of this review, one cannot draw any conclusions without first including the extended $n-^{10}$Be configurations. 

\begin{table}[t]
  \caption{B(E1) values (in [$e^2$fm$^2$]) for the strong ground-state transitions 
     in $^9$Be and $^{11}$Be.
    The NCSM calculations were performed in the $N_{\rm max}=8(9)$ model
    space for negative-(positive-)parity states using the AV8$^\prime$
    interaction. Experimental values are
    from~\cite{til04:745,ajz90:506rev,tan88:206,fri95:60}.
\label{tab:BE1}}
     \begin{tabular}{cccc}
          & $^9$Be($\frac{1}{2}_{_1}^+ \rightarrow \frac{3}{2}_{_1}^-$)   & 
              $^9$Be($\frac{5}{2}_{_1}^+ \rightarrow \frac{3}{2}_{_1}^-$)   &
              $^{11}$Be($\frac{1}{2}_{_1}^- \rightarrow \frac{1}{2}_{_1}^+$) \\
\hline
Expt.   & 0.061(25) & 0.0100(84) & 0.116(12) \\
NCSM  & 0.033       & 0.0057       & 0.0065     \\
    \end{tabular}
 \end{table}
This conclusion is further re-inforced by the E1 transition calculations for the strong ground-state transitions in  $^9$Be and $^{11}$Be, reported in Ref.~\cite{Fo05} and summarized in Table~\ref{tab:BE1}.
The strength of the electric dipole transition between the two bound
states in $^{11}$Be is of fundamental importance. This is an observable,
which has attracted much attention, since it was first measured in
1971~\cite{han71:3}, and again in 1983~\cite{mil83:28}. The cited value
of 0.36 W.u. is still the strongest known transition between low-lying
states, and it has been attributed to the halo character of the
bound-state wave functions. Unfortunately, by working in a HO basis, we
suffer from an incorrect description of the long-range asymptotics, and
we would need an extremely large number of basis states in order to
reproduce the correct form. This shortcoming of the HO basis is
illustrated by the fact that we obtain a value for the E1 strength which
is 20 times too small. When studying the
dependence of this value on the size of the model space, we observe an
almost linear increase, indicating that our result is far from
converged. A similar increase is observed for $^9$Be. However, for this nucleus we note that,
in the largest model space, our calculated E1 strength is only off by a
factor of two compared to experiment. In addition, a consistent result
is found for the much weaker $\frac{5}{2}_{_1}^+ \rightarrow \frac{3}{2}_{_1}^-
$ E1 transition in $^9$Be, where we also obtain a factor of two smaller
$B(\mathrm{E}1)$ than experiment. These results accentuates the
anomalous strength observed for $^{11}$Be. As argued in the last part of this review, 
a proper asymptotic behavior of the $n-^{10}$Be $S$-wave configurations that cannot be achieved withou extending the NCSM basis by cluster configurations is critical for explanation of the  B(E1) strength between the two bound states of $^{11}$Be.

\section{Extension to heavier nuclei and larger model spaces}\label{sec:Ext}

No-core shell model calculations become computationally intractable for heavier nuclei. At least
$N_{\rm max}=6$(8) model spaces are required to obtain stable excitation energies of the lowest states in the three-body (two-body) cluster approximation (i.e. using the three-body (two-body) effective interaction). Still larger spaces are needed for a full convergence of binding energies. The $m$-scheme dimension grows exponentially with the number of nucleons $A$ and the truncation level $N_{\max}$. For $^{16}$O this limits the presently tractable model space to $N_{\max}=8$ \cite{CaMa05}, corresponding to an effective $m$-scheme dimension of $6\times10^{8}$. For $^{40}$Ca the dimension of the 8$\hbar\Omega$ model space is $2 \times10^{12}$---well beyond the capabilities of current shell model codes. In order to extend applicability of the {\it ab initio} NCSM to heavier nuclei and larger model spaces, we must resort to approximation schemes. In this section we discuss two such schemes that were developed recently.

\subsection{Importance truncated no-core shell model}\label{sec:IT}

It turns out that many of the $m$-scheme basis states used in the NCSM calculations are irrelevant for the description of any particular eigenstate, e.g., the ground state. Therefore, if one were able to identify the important basis states beforehand, one could reduce the dimension of the matrix eigenvalue problem without losing predictive power. This can be done using an importance truncation scheme based on many-body perturbation theory~\cite{RothNa07}.

The general concept of the importance truncation is as follows: One starts with a reference state $|\Psi_{\rm ref}\rangle$, which in the simplest case of closed-shell nuclei could be a single harmonic-oscillator Slater-determinant, but, in a general case, a reference state of any complexity can be used. Also, excited states can be included, yielding a set of reference states. Starting from $|\Psi_{\rm ref}\rangle$, one can build a many-body space by generating all possible $n$-particle--$n$-hole ($npnh$) excitations up to the excitation energy $N_{\max}\hbar\Omega$. By increasing $n$ ($\leq A$), we eventually recover the translationally invariant $N_{\max}\hbar\Omega$ model-space of the NCSM. We now estimate the contribution of a given basis state $|\Phi_{\nu}\rangle$ to the exact eigenstate after the diagonalization via many-body perturbation theory. In first-order the amplitude of the state $|\Phi_{\nu}\rangle$ is given by 
\begin{equation}\label{eq:importanceweight}
  \kappa_{\nu} = -\frac{\langle\Phi_{\nu}|H^\prime|\Psi_{\rm ref}\rangle}{\epsilon_{\nu}-\epsilon_{\rm ref}} \;,
\end{equation}
where $H^\prime$ is the Hamiltonian of the perturbation and $\epsilon_{\nu}, \epsilon_{\rm ref}$ are the unperturbed energies of the two configurations. In the case of closed-shell nuclei with a single-Slater-determinant reference state, the unperturbed Hamiltonian is just the one-body harmonic oscillator Hamiltonian, $H_0 = H_{\rm HO}$, given, e.g., as the first term on the right-hand side of Eq.~(\ref{hamomega}), with the Slater determinants $|\Phi_{\nu}\rangle$ as eigenstates, i.e., $H_{\rm HO}|\Phi_{\nu}\rangle=\epsilon_{\nu}|\Phi_{\nu}\rangle$ and $H_{\rm HO}|\Psi_{\rm ref}\rangle=\epsilon_{\rm ref}|\Psi_{\rm ref}\rangle$. The perturbation is given by $H^\prime=H_A - H_{\rm HO}$, with the internal Hamiltonian $H_A$ given by Eq.~(\ref{ham}).
  
If we restrict ourselves to two-body interactions in $H_A$ ~(\ref{ham}), then $H^\prime$ contains only one- and two-body terms such that  $\kappa_{\nu}$ vanishes for $3p3h$ and higher-order configurations, when starting with a $0p0h$ reference state. In principle, higher orders of perturbation theory are required to directly generate states beyond the $2p2h$ level. Since this becomes computationally inefficient, one can resort to an iterative scheme. In a first iteration we generate $1p1h$ and $2p2h$, states starting from a Slater determinant as the reference state; retain those with an importance weight $|\kappa_{\nu}| \geq \kappa_{\min}$; and solve the eigenvalue problem in this space. In the next iteration, the dominant components of the ground state $|\Psi_{0}\rangle=\sum_{\nu} C_{\nu} |\Phi_{\nu}\rangle$ obtained from the diagonalization in the previous step are used as reference state, i.e., $|\Psi_{\rm ref}\rangle=\sum_{\nu}^{|C_{\nu}|\geq C_{\rm ref}} C_{\nu} |\Phi_{\nu}\rangle$ with $C_{\rm ref}\approx0.0005$. Since this state already contains up to $2p2h$ admixtures, one obtains nonvanishing importance weights~(\ref{eq:importanceweight}) for states up to the $4p4h$ level. After applying the importance truncation $|\kappa_{\nu}| \geq \kappa_{\min}$, we solve the eigenvalue problem in the extended space. This cycle can be repeated until the full $N_{\max}\hbar\Omega$ model-space is generated in the limit $\kappa_{\min}=0$. 

Alternative schemes to the above can be developed~\cite{Roth09}. For example, one can start from a more complex reference state (or a set of reference states) obtained in an $N_{\rm max}\hbar\Omega$ space. Then by application of Eq.~(\ref{eq:importanceweight}) with an appropriate re-definition of $H_0$ and $H^\prime$, we generate a new basis with components up to $2p2h$ above the highest $npnh$ configurations present in the reference state(s). We can retain only the configurations up to, e.g., $(N_{\rm max}+2)\hbar\Omega$, diagonalize the Hamiltonian in this importance-truncated basis, and obtain a new reference state (or set of reference states). This procedure can be repeated up to a desired $N_{\rm max}\hbar\Omega$ space, optimally until convergence is reached.

To remove a dependence of the results on the cutoff $\kappa_{\min}$, a series of calculations can be performed with $\kappa_{\min}$ varied. It then becomes feasible to extrapolate to the desired  $\kappa_{\min}=0$ case ~\cite{RothNa07}.

\begin{figure}
\centering\includegraphics*[width=0.7\columnwidth]{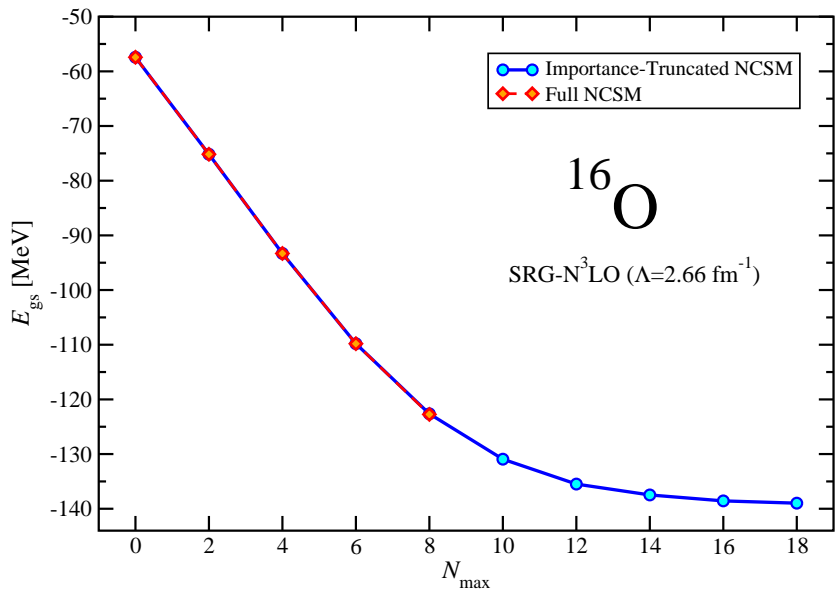}
  \caption{Ground-state energy dependence on the model-space size $N_{\rm max}$ for $^{16}$O, obtained within the importance-truncated NCSM, using the SRG-N$^3$LO NN potential of Ref.~\cite{Roth_SRG} with a cutoff of 2.66 fm$^{-1}$. The HO frequency $\hbar\Omega=24$ MeV was employed. The calculation is variational. No NCSM effective interaction was used. The full NCSM results were obtained with the code Antoine~\cite{Antoine}.}
  \label{fig:O16_ITNCSM}
\end{figure}

In Fig.~\ref{fig:O16_ITNCSM}, we show an example of $^{16}$O ground-state convergence obtained by the application of the importance-truncated NCSM using the above iterative procedure~\cite{NaRoQua09}. More results for $^4$He, $^{16}$O and $^{40}$Ca were presented in Ref.~\cite{RothNa07}.

It should be noted that the reduction of the basis dimension that can be achieved by this method is quite dramatic. For example, importance-truncated NCSM calculations for $^{16}$O in the $N_{\rm max}=16$ model space reach dimensions of the order of $10^7$, which can be managed easily. This is one order of magnitude less than the dimension of full $N_{\rm max}=8$ NCSM calculations for the same nucleus.  

The importance-truncated NCSM calculations are completely variational, i.e., they provide an upper bound for the ground-state energy of the system, if no NCSM model-space effective interaction is used, as is done in the calculations with the soft potentials used in Fig.~\ref{fig:O16_ITNCSM} and also in Ref.~\cite{RothNa07}. If an NCSM effective interaction is used for a specific $N_{\rm max}$ model space, the importance-truncated NCSM calculation is variational only with respect to the full NCSM calculation for the same $N_{\max}$. 

The {\it ab initio} NCSM preserves translational invariance, when working with a Slater determinant basis, because it uses the HO single-particle states and truncates the model space according to the total HO excitation energy $N_{\rm max}\hbar\Omega$. This is important not only for obtaining proper binding or excitation energies, but also for a correct extraction of physical wavefunctions, which is crucial, e.g., in the {\it ab initio} reaction calculations, discussed later in the paper. The importance-truncated NCSM violates the translational invariance in a minimal way that can be monitored by looking at the invariance of intrinsic properties (e.g., energy, radii, density~\cite{trinvde}) with the value of $\beta$ in a Lawson projection term ( discussed, e.g., after Eq.~(\ref{Ham_A_Omega_2eff})). The variational nature and the minimal violation of the translational invariance distinguish the importance-truncated NCSM from other truncation schemes, such as those used in the coupled-cluster method. This was pointed out in the discussion \cite{Dean_comm,Repl}, following the publication of Ref.~\cite{RothNa07}.

\subsection{{\it Ab initio} shell model with a core}\label{sec:core}

The two-body and three-body cluster approximations presented in subsections \ref{sec:tbeff} and \ref{sec:thrbeff} can be generalized further 
 for a larger size of clusters \cite{navratil1997,Lis08}. For example, if we take a six-nucleon system,  we may construct a
 six-body effective Hamiltonian that takes into account the full-space six-nucleon correlations.
 Assuming that it is possible to solve the six-body problem in a very large space using two- or three-body interactions,
 and to achieve reasonable eigenenergy convergence, we may treat the obtained solutions as the full space results.
 In fact, the NCSM calculation for the $A=6$ system in the $N_{\rm max}=12$ space yields nearly converged energies for the
 lowest states dominated by the $N=0$ components.   
 
Moreover, if it is possible to solve the six-body problem for $A=6$, then it is possible to solve the six-body problem
for arbitrary $A$, using the corresponding effective Hamiltonian $H^{N_{\rm max},\Omega}_{A,2}$ obtained in
the two-body cluster approximation. This means that we can determine for any $A$-body system the secondary effective 
six-body Hamiltonian which accounts for six-body dynamics in the large $N_{\rm max}=12$ space. The six-body effective 
Hamiltonian can be constructed for an arbitrary $P_1$ space with $N_{1,\rm max} < N_{\rm max}$; however, the simplest and most 
instructive case is  $N_{1,\rm max}=0$ (single $p$-shell), {\it i.e.}, when the secondary six-body effective 
Hamiltonian has the dimension of the two-body Hamiltonian in the $p$-shell. Thus, we can construct an effective two-body
 Hamiltonian for the $p$-shell, which also contains information about 3-,4-, 5- and 6-body correlations in the 
large $N_{\rm max}=12$ space for the $A$-body system. Because the $P_1$ space has $N_{1,\rm max}=0$, the projection into this 
space "freezes" four of the $A$ nucleons into fixed single-particle configurations, which can be thought of as the "inert core" 
states in the Standard Shell Model (SSM) approach. Consequently, we are dealing with a two-body valence cluster approximation
 for $A-4$ valence nucleons in the $p$-shell. Remember that we have solved the six-body problem for $6 \le A \le 16$.

To calculate the matrix elements of the secondary effective Hamiltonian,
${\cal H}_{\rm eff}^{A}$, for two-body HO basis states,
$|\alpha_{P_1} \rangle$, we employ the transformation given 
by Eq.~(\ref{effham}), where the two-body eigenstates $|k\rangle$ and eigenvalues $E_k$ have to be replaced with six-body eigenstates and eigenvalues, respectively.   

In the general case of a doubly magic closed shell with two extra nucleons i.e., $A=6,18,42$, {\em etc.},
the dimension of the effective Hamiltonian ${\cal H}_{\rm eff}^{A}$ is a 2-body
 Hamiltonian in the $p-$, $sd-$, $pf-$spaces, {\em etc.}, respectively. This means that the secondary effective
 Hamiltonian  does not contain 3- and higher-body terms, even after the {\it exact} $A$-body cluster
transformation. This effective Hamiltonian, which now contains the correlation energy of \underline{all}
 $A$ nucleons, is the correct two-body Hamiltonian to use in a SSM calculation 
with an inert core.
The $A_c=A-2$ nucleon-spectators fully occupy the shells below the valence shell and the total $A$-body wave-function can
 be exactly factorized as the $A_c$-body ''core'' and the valence 2-body wave functions.
 This considerably simplifies the calculation of the effective Hamiltonian, because only the
$0\hbar \Omega$ part (P$_1$-space part) of the complete $N_{\rm max}\hbar \Omega$ wave function needs to be specified.

 Utilizing the approach outlined above, we have calculated effective $p$-shell Hamiltonians for $^{6}$Li,
using the 6-body Hamiltonians with $N_{\rm max}=2,4,..,12$ and $\Omega=14$ MeV, constructed from the INOY interaction \cite{INOY}. 
The corresponding excitation energies of $p$-shell dominated states and
the binding energy of $^{6}$Li are shown as a function of $N_{\rm max}$ on the left of Fig.\ref{li6_exc_12} discussed already in Sect.~\ref{sec:Conv}.

In the SSM an effective two-body Hamiltonian for a nucleus with mass number A is represented
in terms of three components:
\begin{equation}
\label{hssm}
H_{\rm SSM}^A= H_0 + H_1 + V_2^A,
\end{equation}
where $H_0$ is the inert core part associated with the interaction of the nucleons occupying closed shells,
$H_1$ is the one-body part corresponding to the interaction of valence nucleons with core nucleons, and
$V_2^A$ is the two-body part referring to the interaction between valence particles.
  It is usually assumed that  the core and one-body parts are constant for an arbitrary number of valence particles
 and that only the two-body part $V_2^A$ may contain mass dependence that includes effects of three-body and 
higher-body interactions.

To represent the ${\cal H}_{\rm eff}^{A}$ Hamiltonian in the SSM format,
we develop a valence cluster expansion (VCE),
\begin{equation}
\label{hexp}
{\cal H}_{\rm eff}^{A,a_{\rm v}}=H_0^{A,4}+H_1^{A,5}+\sum_{k=2}^{a_{\rm v}}V_k^{A,k+4},
\end{equation}
where the lower index, $k$, stands for the $k$-body interaction in the $a_{\rm v}$-body valence cluster;
 the first upper index $A$ for the mass dependence; and the second upper index, $k+4$,  for the number of particles
 contributing to the corresponding $k$-body part. Thus, we consider the more general case of allowing the core ($k=0$),
 one-body ($k=1$) and other $k$-body parts to vary with the mass number $A$. This appears necessary to include 
 the $A$-dependence of the excitations of the four core nucleons treated in the original $N_{\rm max}$ basis space.

For the $A=6$ case the two-body valence cluster (2BVC) approximation is exact:
 \begin{equation}
\label{hncsm}
{\cal H}^{A=6,2}_{\rm eff}= H_{0}^{6,4} + H_{1}^{6,5} + V_{2}^{6,6},
\end{equation}
where the core part, $H_{0}^{6,4}={\cal H}^{A=6,0}_{\rm eff}$, is defined as the ground state $J^\pi=0^+$  
energy of $^4$He calculated in the $N_{\rm max} \hbar \Omega$ space with the TBMEs of the primary effective 
Hamiltonian, $H^{N_{\rm max},\Omega}_{6,2}$ for $A=6$. Then the one-body part, $H_{1}^{6,5}$,is determined as
\begin{equation}
\label{v1}
 H_{1}^{6,5}={\cal H}^{6,1}_{\rm eff}-H_{0}^{6,4}.
\end{equation}
The TBMEs of the one-body part, $H_{1}^{6,5}$,
\begin{equation}
\label{v1me}
 \langle ab| H_1^{6,5}| cd \rangle_{JT} = ( \epsilon_a + \epsilon_b ) \delta_{a,c} \delta_{b,d}
\end{equation}
may be represented in terms of single particle energies (SPE) , $\epsilon_a$:
\begin{equation}
\label{v1spe}
\epsilon_a^p=  E(^5\mbox{Li},j_a) - H_0^{6,4}, \\
\epsilon_a^n=  E(^5\mbox{He},j_a) - H_0^{6,4}.
\end{equation}
where the index $a$ (as well as $b,c$, and $d$) denotes the set of single particle HO quantum numbers $(n_a,l_a,j_a)$, 
the upper index stands for proton (p) and neutron (n), and
the $E(^5$Li$,J$) and $E(^5$He$,J$) are the NCSM energies of the lowest $J^\pi_i=3/2^-_1$ and $J^\pi_i=1/2^-_1$ states calculated
in the $N_{\rm max}\hbar \Omega$ space for the $5$-body system using the TBMEs of the $A=6$ effective
Hamiltonian, $H^{N_{\rm max},\Omega}_{A=6,2}$, which includes Coulomb energy.
Finally, the two-body part $V_2^{6,6}$ is obtained by subtracting the two Hamiltonians:
\begin{equation}
\label{v2}
 V_{2}^{6,6}={\cal H}^{6,2}_{\rm eff}-{\cal H}^{6,1}_{\rm eff}.
\end{equation}
It is worth noting that since the Coulomb energy is included in the original Hamiltonian, the proton-proton (pp),
neutron-neutron (nn) and proton-neutron (pn) $T=1$ TBMEs of the two-body part, $V_{2}^{6,6}$, have different values.

\begin{figure}[!ht]
 \includegraphics[scale = 0.40]{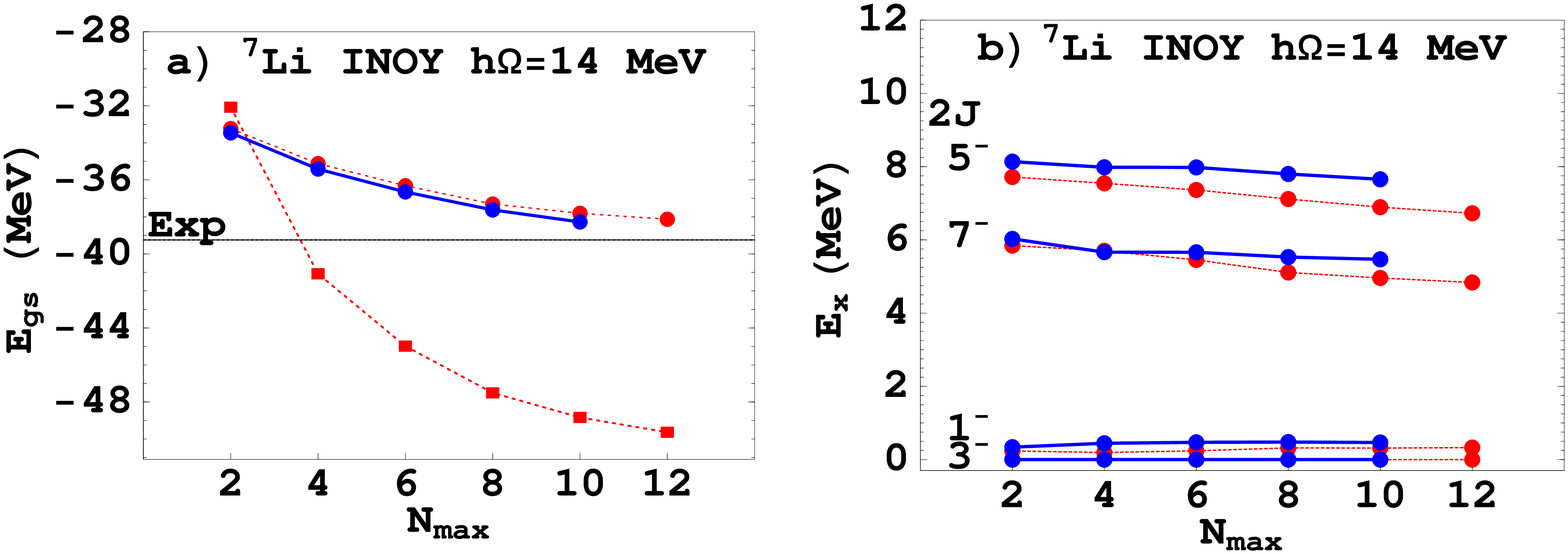}%
 \caption{\label{li7x} a) The ground-state energy, $E_{\rm gs}$,  of $^7$Li as a function of $N_{\rm max}$ for the INOY 
interaction. The NCSM results with the $H^{N_{\rm max},\Omega}_{A=7,2}$ Hamiltonian are shown by filled circles connected 
with the solid line. The SSM results with the effective ${\cal H}^{7,2}_{\rm eff}$ Hamiltonian decomposed
according to Eq.(\ref{hdec1}) are shown by filled circles connected with a dashed line.
The SSM results with the effective ${\cal H}^{7,6}_{\rm eff}$ Hamiltonian decomposed
according to Eq.(\ref{hdec2}) are shown by filled squares connected with a dashed line.
b) NCSM (solid line) and SSM (using Eq.(\ref{hdec1}), dashed line) spectra for $^7$Li.
The states with spin J are marked by 2J. }
 \end{figure}

The VCE given by the Eq.(\ref{hexp}) would require a three-body part
V$_3^{7,7}$ of the $p$-shell effective interaction ${\cal H}^{7,3}_{\rm eff}$ to reproduce exactly the NCSM results for $A=7$ nuclei:
\begin{equation}
\label{hncsm7}
{\cal H}^{A=7,a_{\rm v}=3}_{\rm eff} = H_{0}^{7,4} + H_{1}^{7,5} + V_{2}^{7,6} + V_{3}^{7,7}.
\end{equation}
Therefore, it is worth knowing how good the 2BVC approximation is
for $A=7$, as well as for $A>7$. To test the  2BVC approximation, we have constructed
the ${\cal H}^{A=7,2}_{\rm eff}$ Hamiltonian, using an expansion in
terms of zero-, one- and two-body valence clusters, {\em i.e.}, omitting the three-body part:
\begin{equation}
\label{hdec1}
{\cal H}^{A=7,2}_{\rm eff} = H_{0}^{7,4} + H_{1}^{7,5} + V_{2}^{7,6}.
\end{equation}
In other words, we have first performed NCSM calculations for the $k+4$-body  systems ($k=0,1,2$)
with the $H^{N_{\rm max},\Omega}_{A=7,2}$ Hamiltonian. Thus, $H_{0}^{7,4}$ is the $^4$He ``core'' energy and $H_{1}^{7,5}$ is the one-body part determined, as in Eqs.~(\ref{v1})-(\ref{v1spe}), but with $A=7$; and
$V_{2}^{7,6}$ is obtained by subtracting $H_{0}^{7,4} + H_{1}^{7,5}$
from ${\cal H}^{A=7,2}_{\rm eff}$. The resulting parts of the ${\cal H}^{A=7,2}_{\rm eff}$ Hamiltonian are given in
Table \ref{V6dec}. 
\begin{center}
 \begin{table}[tbp]
 \caption{\label{V6dec} The core energy, SPEs and pn TBMEs of the effective $p$-shell Hamiltonian 
  ${\cal H}^{A,2}_{\rm eff}$ for the $A=6$ and the $A=7$ systems in the 2BVC approximation.}
 \begin{tabular}{c|c|c|c}
 \hline \hline 
                      & Eq.(\ref{hncsm}) & Eq.(\ref{hdec1}) & Eq.(\ref{hdec2})  \\
\cline{2-4}
                      &  A=6     &  A=7   &   A=7   \\
\hline
 Core energy          &  -54.830    & -63.336   & -30.500  \\
 $\epsilon(p_{3/2})$  &   13.922    &  10.637   &   2.784  \\
 $\epsilon(p_{1/2})$  &   17.964    &  16.355   &   5.355  \\
 $\langle p_{3/2}^2|V_{2}^{A,2}| p_{3/2}^2 \rangle_{J=3,T=0}$ 
                      &   -2.181    & -2.457    & -19.586  \\
 $\langle p_{3/2}^2|V_{2}^{A,2}| p_{3/2}^2 \rangle_{J=2,T=1}$ 
                      &    2.094    &  2.875    & -14.245  \\
$\langle p_{3/2}^2|V_{2}^{A,2}| p_{1/2}^2 \rangle_{J=0,T=1}$ 
                      &   -2.823    & -3.104    &  -3.104  \\ 
\hline
\end{tabular}
\end{table}
\end{center}
Comparing several TBMEs for $A=6$ and $A=7$ (Table \ref{V6dec}), we find that they differ considerably.
There is a big change separately for the core and one-body parts, but weaker changes for the
two-body parts, which tend to become larger in magnitude with increasing $A$.  
We have then performed SSM calculations for the ground state energy of  $^7$Li,
using the zero-, one- and
two-body parts in Eq.(\ref{hdec1}). Namely, the one- and two-body parts were employed in a SSM calculation of the
 ground and excited-state energies of the valence nucleons in the $p$-shell, {\em i.e.}, $0\hbar\Omega$ space, to which
 the $^4$He core energy, $H_{0}^{7,4}$, was added, in order to yield the total energies. These calculations were
 repeated for $N_{\rm max}=0,2,...10$. Next we carried out NCSM calculations for $^7$Li with
 $H^{N_{\rm max},\Omega}_{A=7,2}$ for the same values of $N_{\rm max}$. The SSM and NCSM results for the
 ground-state energy are shown in Fig.\ref{li7x}.

It is also of interest to find out what would be the result if we
take the fixed core and one-body parts at values, which are appropriate for
 the $A=4$ and $A=5$ systems, respectively, because this is analogous to what is done in the SSM to determine
energies relative to an inert core.  To do this we adopt an alternative two-body VCE,
 which assumes that the core and one-body parts are A-independent, {\em i.e.},
\begin{equation}
\label{hdec2}
{\cal H}^{A,2}_{\rm eff} = H_{0}^{4,4} + H_1^{5,5} + W_2^{A,6},
\end{equation}
similar to the SSM convention given by Eq.(\ref{hssm}). We have then performed another set of SSM calculations
for $A=7$ in the same manner as described previously, but using the decomposition given in Eq.(\ref{hdec2}). 
To distinguish between the two-body parts of the VCE given
by the Eqs.(\ref{hexp}) and (\ref{hdec2}), we have introduced the new notation, $W_2^{A,6}$,  in Eq.(\ref{hdec2}).
The Hamiltonian ${\cal H}^{7,2}_{\rm eff}$ expanded according to the
Eq.({\ref{hdec2}}) is shown in last column of Table \ref{V6dec} and the corresponding results are depicted in
Fig.\ref{li7x}a) by the solid squares connected with a dashed line.
Figure \ref{li7x} indicates that for light systems a realistic balance of core, one-body and two-body parts of
the effective interaction may be achieved only when both the core and one-body parts are calculated according to the prescription given above, {\em i.e.}, Eq.(\ref{hdec1}).
 Adoption of the decomposition procedure with $A$-independent core and one-body parts leads 
to a very strong diagonal two-body part for the valence nucleons and, subsequently, to drastic overbinding.
It is obvious, that, in order to compensate for such an effect, one would need to introduce a strongly repulsive 
three-body effective interaction with an unrealistic strength of about 10 MeV. 
The VCE with the $A$-dependent core and one-body parts also yields
 better agreement with the exact NCSM results for the excited states. The corresponding low-energy spectra
of $^7$Li obtained with the NCSM and the $A$-dependent SSM are shown in Fig.\ref{li7x}b). 
The differences observed in Fig.\ref{li7x}a) and b) for the ground state and excited states,
respectively, may be attributed to the neglected three-body part of the effective interaction  at the two-body
valence cluster level.

We have generalized the 2BVC expansion procedure of Eq.(\ref{hdec1}) for arbitrary mass number $A$,
\begin{equation}
\label{hdecg}
{\cal H}^{A,a_{\rm v}=2}_{\rm eff} = H_{0}^{A,4} + H_{1}^{A,5} + V_{2}^{A,6},
\end{equation}
 and applied it to the $A=7,8,9$, and 10 isobars for $N_{\rm max}=6$ in \cite{Lis08}.
We have found that the three-body and higher-body correlations become more important with
increasing mass number. There is also a very strong isospin dependence of the obtained results. 
For the highest isospin values the SSM systematically underbinds nuclei in comparison to the 
NCSM and higher-body correlations appear to be small for systems
 containing only valence neutrons. However, there is an opposite effect in the vicinity
of the $N=Z$ line, where SSM yields considerably more binding energy than the NCSM.

The analysis of the $A=7$ systems  allowed us to derive an effective three-body Hamiltonian for the $p$-shell 
and to give an idea about the strength of the effective three-body interaction \cite{Lis08}.

\section{Cluster overlap functions and spectroscopic factors}\label{sec:overlap}

In the {\it ab initio} NCSM calculations, we are able to obtain wave functions of
low-lying states of light nuclei in large model spaces. An interesting and important
question is, what is the cluster structure of these wave functions. That is we
want to understand, how much, e.g., an $^6$Li eigenstate looks like $^4$He plus deuteron,
an $^7$Be eigenstate looks like $^4$He plus $^3$He, an $^8$B eigenstate looks like
$^7$Be plus proton, and so on. This information
is important for the description of low-energy nuclear reactions.
To gain insight, one introduces channel-cluster form factors
(or overlap integrals, overlap functions). The formalism for calculating
the channel-cluster form factors from the NCSM wave functions was developed in Ref.~\cite{cluster}.
Here we just briefly repeat a part of the formalism relevant to the simplest case,  
when the lighter of the two clusters is a single nucleon.   

We consider a composite system of $A$ nucleons, e.g., $^8$B, a nucleon projectile, e.g., a proton,
and an $A-1$-nucleon target, e.g., $^7$Be.
Both nuclei are assumed to be described by eigenstates of the NCSM effective Hamiltonians
expanded in the HO basis with identical HO frequency and the same
(for the eigenstates of the same parity) or differing by one unit
of the HO excitation (for the eigenstates of opposite parity)
definitions of the model space. The target and the composite system is assumed to be described
by wave functions expanded in a Slater determinant single-particle HO basis (that is obtained from
a calculation using a shell-model code like Antoine).

Let us introduce a projectile-target wave function
\begin{eqnarray}\label{proj-targ_state_delta}
\langle\vec{\xi}_1 \ldots \vec{\xi}_{A-2} r^\prime \hat{r}
|\Phi_{(l\frac{1}{2})j;\alpha I_1}^{(A-1,1)J M};\delta_{r}\rangle
&=&\sum (j m I_1 M_1 | J M) (l m_l \textstyle{\frac{1}{2}} m_s | j m)
\frac{\delta(r-r^\prime)}{r r^\prime}
\nonumber \\
&\times &
Y_{l m_l}(\hat{r}) \chi_{m_s}
\langle \vec{\xi}_1 \ldots \vec{\xi}_{A-2} | A-1 \alpha I_1 M_1\rangle \; ,
\end{eqnarray}
where
$\langle \vec{\xi}_1 \ldots \vec{\xi}_{A-2} | A-1 \alpha I_1 M_1\rangle$
and
$\chi_{m_s}$ are the target and the nucleon wave functions, respectively.
Here, $l$ is the channel relative orbital angular momentum, $\vec{\xi}_i$ 
are the target Jacobi coordinates defined in Eq.~(\ref{jacobiam11}) and
$\vec{r}=\left[\frac{1}{A-1}
      \left(\vec{r}_1+\vec{r}_2 + \ldots+ \vec{r}_{A-1}\right)-\vec{r}_{A}\right]$
describes the relative distance between the nucleon and the center of mass of the target.
The spin and isospin coordinates were omitted for simplicity.

The channel cluster form factor is then defined by
\begin{equation}\label{cluster_form_factor}
{\mathfrak g}^{A\lambda J}_{(l\frac{1}{2})j;A-1 \alpha I_1}(r)=
\langle A \lambda J |{\cal A}\Phi_{(l\frac{1}{2})j;\alpha I_1}^{(A-1,1)J};
\delta_{r}\rangle \; ,
\end{equation}
with ${\cal A}$ the antisymmetrizer and $|A\lambda J\rangle$ an eigenstate 
of the $A$-nucleon composite system (here $^8$B). It can be calculated from the
NCSM eigenstates obtained in the Slater-determinant basis from a
reduced matrix element of the creation operator. The derivation is as follows.
First, we use the relation (\ref{state_relation}) for both
the composite $A$-nucleon and the target $A-1$-nucleon
eigenstate. With the help of HO transformations:
\begin{eqnarray}\label{ho_tr}
&&\sum_{M m} (L M l m|Q q) \varphi_{N L M}(\vec{R}^{A-1}_{\rm CM}) 
\varphi_{n l m}(\vec{r}_A) = 
\nonumber \\
&&\sum_{n' l' m' N' L' M'} \langle n'l' N'L' Q|N L n l Q\rangle_{\frac{1}{A-1}}
(l' m' L' M'|Q q) 
\nonumber \\
&&\times \varphi_{n'l'm'}(\vec{\xi}_{A-1}) \varphi_{N'L'M'}(\vec{\xi}_0)
\; ,
\end{eqnarray}
we obtain
\begin{equation}\label{SD_Jacobi_overlap}
 _{\rm SD}\langle A\lambda J | {\cal A} 
\Phi_{(l\frac{1}{2})j;\alpha I_1}^{(A-1,1)J};nl\rangle_{\rm SD} \;
= \langle nl00l|00nll\rangle_{\frac{1}{A-1}} \;
\langle A\lambda J | {\cal A} 
\Phi_{(l\frac{1}{2})j;\alpha I_1}^{(A-1,1)J};nl\rangle
\; ,  
\end{equation}
with a general HO bracket due to the CM motion. The $nl$ in 
(\ref{SD_Jacobi_overlap}) refers to a replacement of $\delta_{r}$ by the HO 
$R_{nl}(r)$ radial wave function. 
Second, we relate the SD overlap to a linear combination of matrix elements 
of a creation operator 
between the target and the composite eigenstates
$_{\rm SD}\langle A\lambda J |a^\dagger_{nlj}| A-1 \alpha I_1 \rangle_{\rm SD}$. 
The subscript SD refers to the fact that these states were obtained
in the Slater determinant basis.
Such matrix elements are easily calculated by shell model codes.
The result is
\begin{eqnarray}\label{single-nucleon}
\langle A \lambda J|{\cal A} \Phi_{(l \textstyle{\frac{1}{2}},j);\alpha I_1}^{(A-1,1) J};
\delta_{r}\rangle
&=& \sum_n R_{nl}(r)
\frac{1}{\langle nl00l|00nll\rangle_{\frac{1}{A-1}}} \frac{1}{\hat{J}}
(-1)^{I_1-J-j}
\nonumber \\
&&\times
\; _{\rm SD}\langle A\lambda J||a^\dagger_{nlj}||A-1\alpha I_1\rangle_{\rm SD} \;
\; .
\end{eqnarray}
The eigenstates expanded in the Slater determinant basis contain CM components.
A general HO bracket, whose value is simply given by
\begin{equation}\label{cm_ho_br}
\langle nl00l|00nll\rangle_{\frac{1}{A-1}} = (-1)^l
\left(\frac{A-1}{A}\right)^{\frac{2n+l}{2}}
\; ,
\end{equation}
then appears in Eq. (\ref{single-nucleon}) in order to remove these components.
The $R_{nl}(r)$ in Eq. (\ref{single-nucleon}) is the radial HO
wave function with the oscillator length parameter $b=\sqrt{\frac{\hbar}{\frac{A-1}{A}m\Omega}}$,
where $m$ is the nucleon mass.

A conventional spectroscopic factor is obtained by integrating the square of the cluster form
factor:
\begin{equation}\label{spec_fac}
S^{A\lambda J}_{(l\frac{1}{2})j;A-1 \alpha I_1}=
\int dr r^2
|{\mathfrak g}^{A\lambda J}_{(l\frac{1}{2})j;A-1 \alpha I_1}(r)|^2
\; .
\end{equation}

A generalization for projectiles (= the lighter of the two clusters) with 2, 3 or 4
nucleons is straightforward, although the expressions become more involved. In all cases,
the projectile is described by a wave function expanded in the Jacobi coordinate HO basis,
while the composite and the target eigenstates are expanded in the Slater determinant
HO basis. Full details are given in Ref.~\cite{cluster}.

\section{S-factors of capture reactions}\label{sec:cap_reac}

The overlap functions introduced in the previous section can be used as a strating point for description of low-energy $\gamma$-capture reactions important for nuclear astrophysics.

\subsection{$^7$Be(p,$\gamma$)$^8$B}

The $^7$Be(p,$\gamma$)$^8$B capture reaction serves as an important
input for understanding the solar neutrino flux \cite{Adelberger}.
Recent experiments have determined the neutrino flux emitted from
$^8$B with a precision of ~9\% \cite{SNO}. On the other hand,
theoretical predictions have uncertainties of the order of 20\%
\cite{CTK03,BP04}. The theoretical neutrino flux depends on the
$^7$Be(p,$\gamma$)$^8$B S-factor. Many experimental and theoretical
investigations studied this reaction.

In this subsection, we discuss a calculation of the
$^7$Be(p,$\gamma$)$^8$B S-factor starting from {\it ab initio}
wave functions of $^8$B and $^7$Be.
It should be noted that the aim of {\it ab initio} approaches is to predict
correctly absolute cross sections (S-factors), not only relative cross sections.
The full details
of our $^7$Be(p,$\gamma$)$^8$B investigation were published in 
Refs.~\cite{Be7_p_NCSM_lett,Be7_p_NCSM}.

\begin{figure}[t]
\vspace{0.5cm}
\begin{minipage}{8cm}
  \includegraphics[width=0.9\columnwidth]
{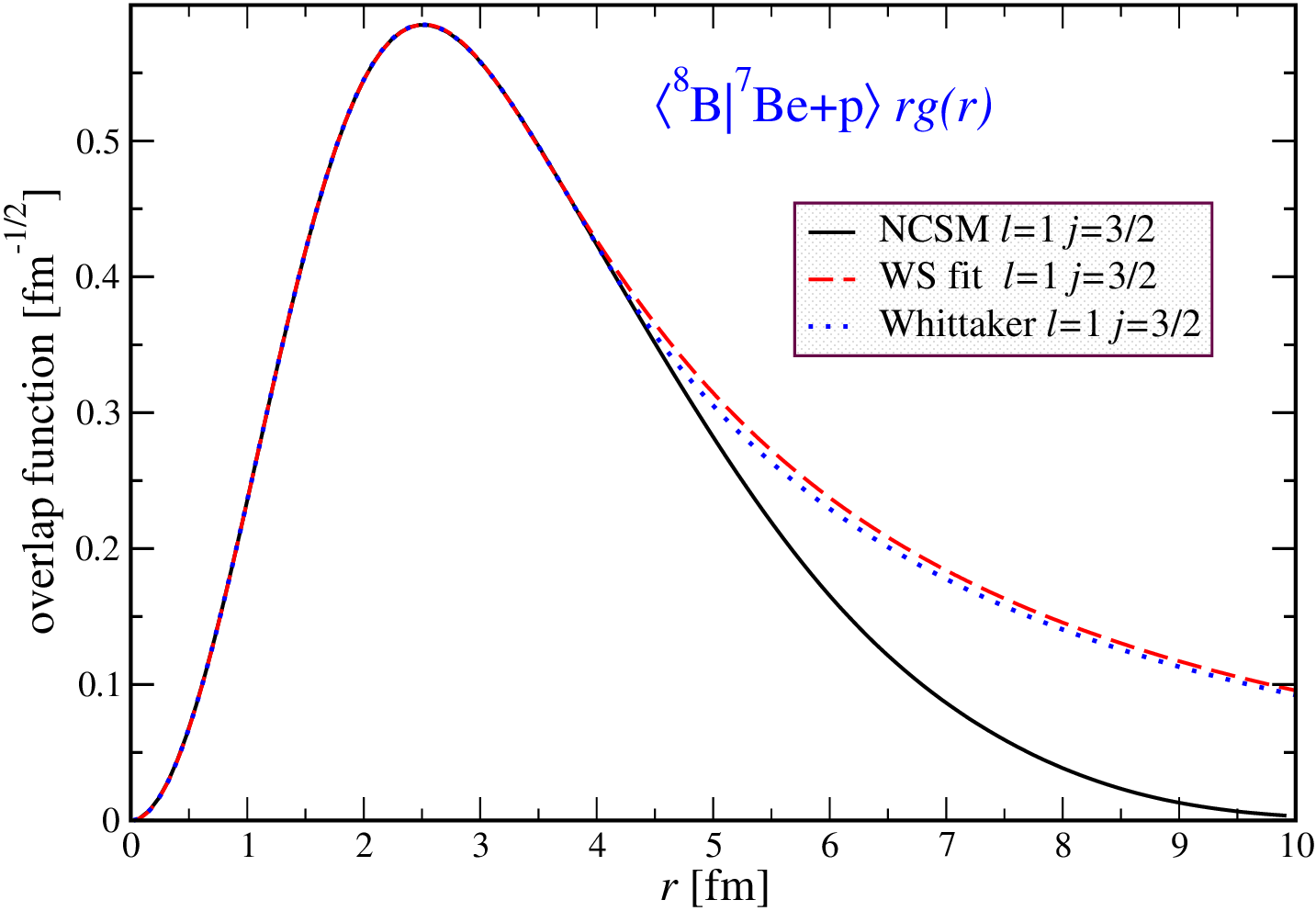}
\caption{\label{B8_Be7+p_overlap} 
Overlap function, $r{\mathfrak g}(r)$, for the ground state of $^8$B with the ground 
state of $^7$Be plus proton.
See the text for details.
}
\end{minipage}
\begin{minipage}{8cm}
  \includegraphics[width=0.9\columnwidth]{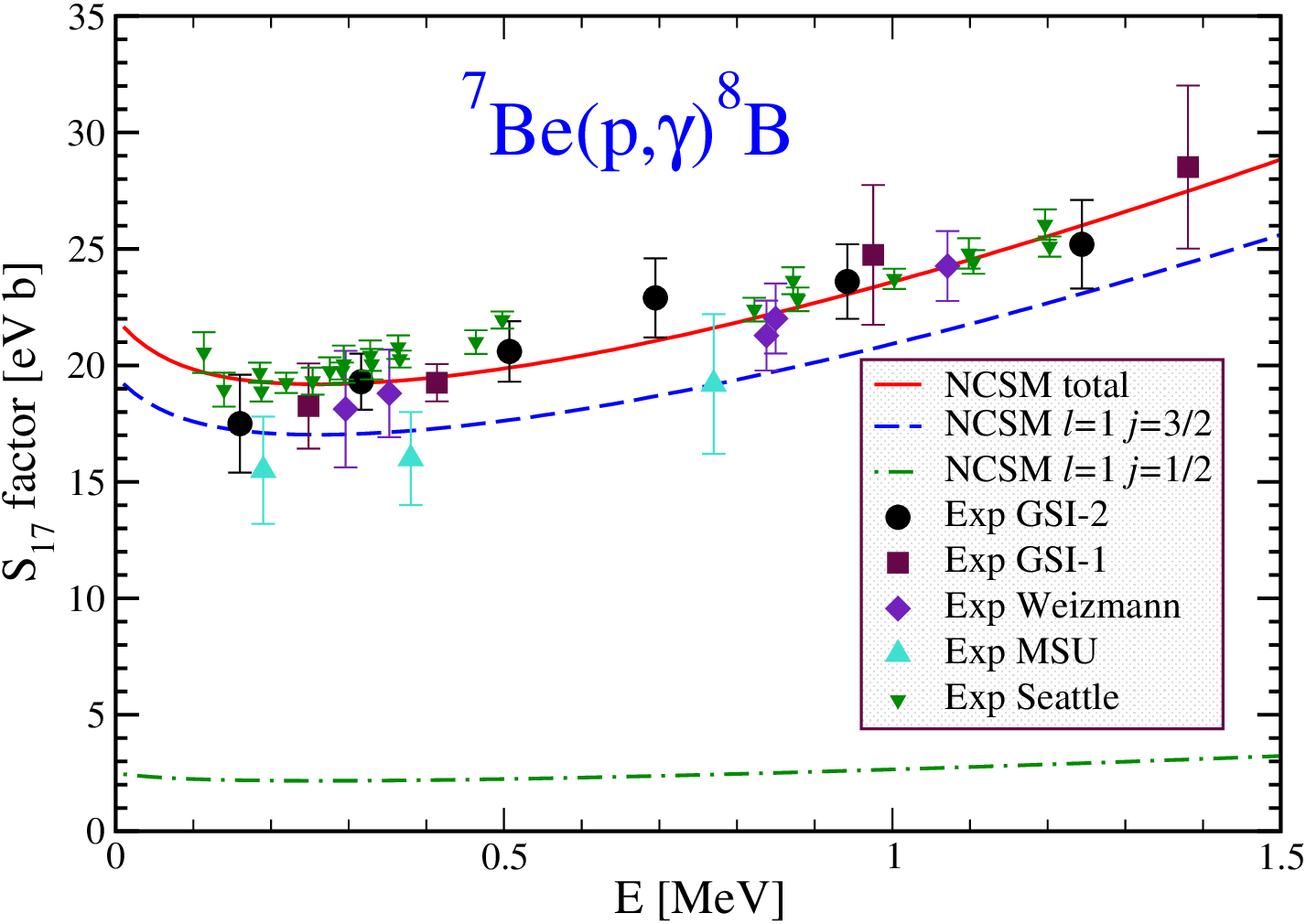}
  \caption{\label{S-factor_12_Nmax} 
The $^7$Be(p,$\gamma$)$^8$B S-factor 
from the corrected NCSM overlap functions. 
Experimental values are from Refs. \protect\cite{Seattle,Be7pgamm_exp}.
}
\end{minipage}
\end{figure}

Our calculations for both $^7$Be and $^8$B nuclei were performed
using the high-precision CD-Bonn 2000 NN potential \cite{cdb2k}
in model spaces up to $10\hbar\Omega$ ($N_{\rm max}=10$) for a wide range of HO frequencies.
From the obtained $^8$B and $^7$Be wave functions, we calculate the channel cluster
form factors (overlap functions, overlap integrals)
${\mathfrak g}^{A\lambda J}_{(l\frac{1}{2})j;A-1 \alpha I_1}(r)$, as discussed in the previous section.
Here, $A=8$, $l$ is the channel
relative orbital angular momentum and
$\vec{r}=\left[\frac{1}{A-1}
      \left(\vec{r}_1+\vec{r}_2 + \ldots+ \vec{r}_{A-1}\right)-\vec{r}_{A}\right]$
describes the relative distance between the proton and the center of mass of
$^7$Be.
The two most important channels are the $p$-waves, $l=1$, with the proton
in the $j=3/2$ and $j=1/2$ states, $\vec{j}=\vec{l}+\vec{s}, s=1/2$. In these channels,
we obtain the spectroscopic factors of $0.96$ and $0.10$, respectively. The dominant
$j=3/2$ overlap integral is presented in 
Fig. \ref{B8_Be7+p_overlap} by the full line. The $10\hbar\Omega$ model space 
and the HO frequency of $\hbar\Omega=12$ MeV were used.
Despite the fact that a very large basis was employed in the present calculation, 
it is apparent that the overlap function is nearly zero at about 10 fm. This is
a consequence of the HO basis asymptotic behavior. As already discussed, 
in the {\it ab initio} NCSM, the short-range correlations are taken into account
by means of the effective interaction. The medium-range correlations are then included
by using a large, multi-$\hbar\Omega$ HO basis. The long-range behavior is not treated
correctly, however. 
The proton capture on $^7$Be to the 
weakly bound ground state of $^8$B associated dominantly by the $E1$ radiation
is a peripheral process. In order to calculate the S-factor of this process we need
to go beyond the {\it ab initio} NCSM, as done up to this point.
We expect, however, that the interior part of the overlap function
is realistic. It is then straightforward to find a simple correction 
to the asymptotic behavior of the overlap functions, which should be
proportional to the Whittaker function. 

One possibility we explored utilizes solutions of a Woods-Saxon (WS) potential.
In particular, we performed
a least-square fit of a WS potential solution to the interior of the
NCSM overlap in the range of $0-4$ fm. The WS potential parameters
were varied in the fit under the constraint that the experimental
separation energy of $^7$Be+p, $E_0=0.137$~MeV, was reproduced. In this way we obtain a perfect
fit to the interior of the overlap integral and a correct asymptotic behavior
at the same time. The result is shown in Fig. \ref{B8_Be7+p_overlap}
by the dashed line.

Another possibility is a direct matching of logarithmic derivatives of the NCSM overlap integral
and the Whittaker function:
$\frac{d}{dr}ln(r{\mathfrak g}_{lj}(r))=\frac{d}{dr}ln(C_{lj} W_{-\eta,l+1/2}(2k_0r))$,
where $\eta$ is the Sommerfeld parameter, $k_0=\sqrt{2\mu E_0}/\hbar$ with $\mu$ the reduced mass
and $E_0$ the separation energy.
Since asymptotic normalization constant (ANC) $C_{lj}$ cancels out, there 
is a unique solution at $r=R_m$.
For the discussed overlap presented in Fig.~\ref{B8_Be7+p_overlap}, 
we found $R_m=4.05$~fm.
The corrected overlap using the Whittaker function matching is shown 
in Fig.~\ref{B8_Be7+p_overlap}
by a dotted line. In general, we observe that the approach using the WS fit leads 
to deviations from the
original NCSM overlap starting at a smaller radius. In addition, the WS solution fit introduces
an intermediate range from about 4 fm to about 6 fm, where the corrected overlap deviates
from both the original NCSM overlap and the Whittaker function. Perhaps, this is a more realistic
approach compared to the direct Whittaker function matching. In any case, 
by considering the two alternative
procedures we are in a better position to estimate uncertainties in our S-factor results.

In the end, we re-scale the corrected overlap functions to preserve the original
NCSM spectroscopic factors (Table 2 of Ref.~\cite{Be7_p_NCSM_lett}).
In general, we observe a faster convergence of the spectroscopic
factors than that of the overlap functions. The corrected
overlap function should represent the infinite space result. By re-scaling
a corrected overlap function obtained at a finite $N_{\rm max}$, we approach
faster the infinite space result. At the same time, by re-scaling we preserve the
spectroscopic factor sum rules.

The S-factor for the reaction  $^7{\rm Be(p},\gamma)^8{\rm B}$
also depends on the continuum wave function,
$R_{lj}^{(c)}$. As the capture reaction calculations were first performed before
the extension of the NCSM to describe continuum wave functions
introduced in Sect.~\ref{sec:NCSMC},
the continuum wave functions $R_{lj}^{(c)}$ were obtained for $s$ and $d$ waves from
a WS potential model.
Since the largest part of the integrand stays outside the
nuclear interior, one expects that the continuum wave functions are
well-described in this way.
In order to have the same scattering wave function in all the calculations,
we chose a WS potential from Ref.~\cite{Esbensen} that was fitted to
reproduce the $p$-wave $1^+$ resonance in $^8$B.
It was argued \cite{Robertson}
that such a potential is also suitable for the description of $s$- and $d$-waves.
We note that the S-factor is very weakly dependent on the choice
of the scattering-state potential (using our fitted potential for the scattering state
instead changes the S-factor by less than 1.5 eV b at 1.6 MeV with no change at 0 MeV).

Our obtained S-factor is presented in Fig. \ref{S-factor_12_Nmax}, 
where the contribution
from the two partial waves are shown together with the total result. 
It is interesting
to note a good agreement of our calculated S-factor with the recent Seattle direct
measurement \cite{Seattle}.

In order to judge the convergence of our S-factor calculation, we performed
a detailed investigation of the model-space-size and the HO frequency dependencies.
We used the HO frequencies in the range from
$\hbar\Omega=11$ MeV  to $\hbar\Omega=15$ MeV and the model spaces from 
$6\hbar\Omega$ to $10\hbar\Omega$.
By analysing these results, we arrived at the S-factor value of
$S_{17}(10\;{\rm keV})=22.1\pm 1.0$ eV b.

\subsection{$^3$He($\alpha$,$\gamma$)$^7$Be and $^3$H($\alpha$,$\gamma$)$^7$Li}

\begin{figure}[t]
\vspace{1cm}
\begin{center}
  \includegraphics[width=0.6\columnwidth]{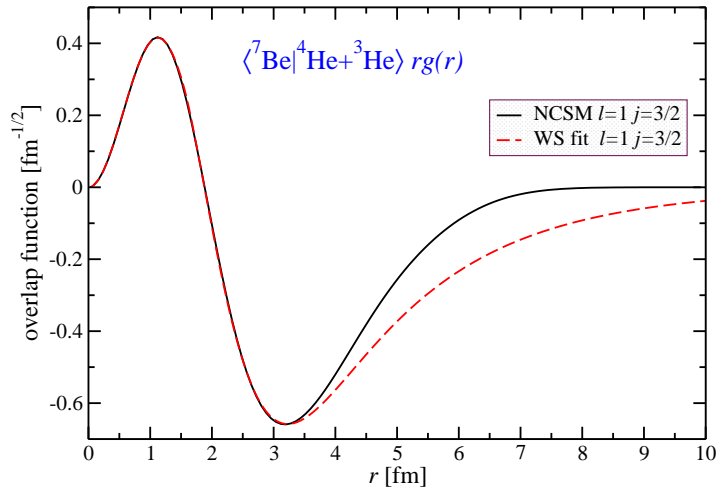}
  \caption{\label{overlap_34} 
The overlap function, $r{\mathfrak g}(r)$, for the ground state of $^7$Be with 
$^3$He plus $\alpha$. 
See the text for details.
}
\end{center}
\end{figure}

The $^3$He($\alpha$,$\gamma$)$^7$Be capture reaction cross section has been
identified as the most important uncertainty in the solar model predictions 
of the neutrino fluxes in the p-p chain \cite{BP04}. 
We investigated the bound states
of $^7$Be, $^3$He and $^4$He within the {\it ab initio} NCSM and calculated the overlap
functions of $^7$Be bound states with the ground 
states of $^3$He plus $^4$He as a function of separation between the $^3$He 
and the $\alpha$ particle.
The obtained $p$-wave overlap function of the $^7$Be
$3/2^-$ ground state is presented in Fig.~\ref{overlap_34} by the full line.
The dashed lines show the corrected overlap function obtained by the least-square fits
of the WS parameters done in the same way as in the $^8$B$\leftrightarrow ^7$Be+p case.
The corresponding NCSM spectroscopic factors obtained using the CD-Bonn 2000 in the
$10\hbar\Omega$ model space for $^7$Be ($12\hbar\Omega$ for $^{3,4}$He) 
and HO frequency of $\hbar\Omega=13$ MeV are 0.93 and 0.91
for the ground state and the first excited state of $^7$Be, respectively. We note 
that contrary to
the $^8$B$\leftrightarrow ^7$Be+p case, the $^7$Be$\leftrightarrow ^3$He+$\alpha$ 
$p$-wave overlap functions have a node.

\begin{figure}[t]
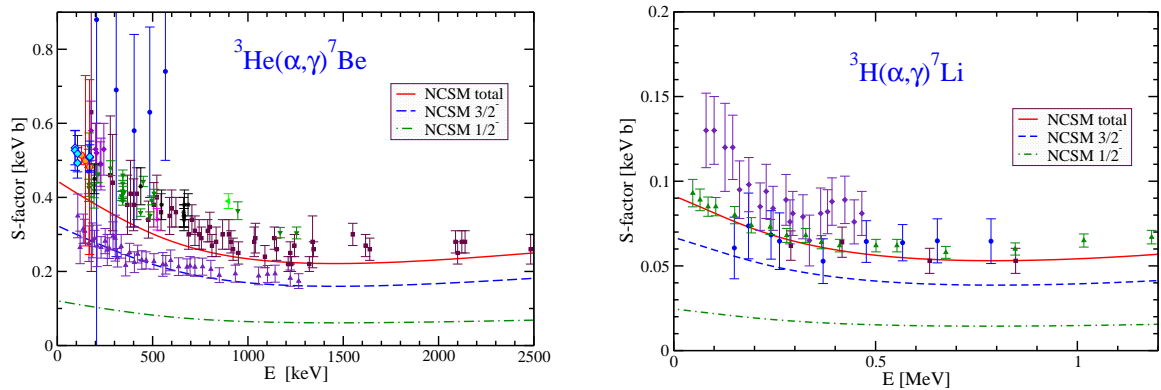

\vspace{0.5cm}
\begin{minipage}{8cm}
  \includegraphics[width=0.9\columnwidth]{S_factor.Be7_10_He4cdb2k12_13.eps}
\end{minipage}
\begin{minipage}{8cm}
  \includegraphics[width=0.9\columnwidth]{S_factor_Li7_10_He4cdb2k_12_13.eps}
\end{minipage}
  \caption{\label{S-factor_34_Nmax}
The full line shows the $^3$He($\alpha$,$\gamma$)$^7$Be (left) and 
$^3$H($\alpha$,$\gamma$)$^7$Li (right) S-factors obtained
using the NCSM overlap functions with corrected asymptotics. The dashed lines
show the $^7$Be (left) and $^7$Li (right) ground- and the first excited state contributions.
The calculations were done using the CD-Bonn 2000 NN 
potential and the $10\hbar\Omega$ model space for $^7$Be and $^7$Li 
($12\hbar\Omega$ for $^3$H, $^3$He and $^{4}$He) 
with the HO frequency of $\hbar\Omega=13$ MeV.  
}
\end{figure}

Using the corrected overlap functions and a $^3$He+$\alpha$ 
scattering state obtained using the potential
model of Ref.~\cite{Kim}, we calculated the $^3$He($\alpha$,$\gamma$)$^7$Be S-factor.
Our $10\hbar\Omega$ result is presented in the left panel of Fig.~\ref{S-factor_34_Nmax}.
We show the total S-factor as well as the contributions from the capture to the ground
state and the first excited state of $^7$Be.
By investigating the model space dependence for $8\hbar\Omega$ and $10\hbar\Omega$
spaces, 
we estimate the $^3$He($\alpha$,$\gamma$)$^7$Be S-factor at zero energy 
to be higher than 0.44 keV b, the value that we obtained in the discussed
case is shown in the left panel of Fig.~\ref{S-factor_34_Nmax}.
Our results are similar to those obtained by K. Nollett \cite{Nollett},
using the variational Monte Carlo wave functions for the bound states and 
potential-model wave functions for the scattering state. We note that the recent evaluation ~\cite{Cy08} that took into account precise LUNA measurements at low energies~\cite{Be06,Co07} found the S-factor of $0.580\pm0.054$ keV b at zero energy.

An important check on the consistency of the $^3$He($\alpha$,$\gamma$)$^7$Be S-factor
calculation is the investigation of the mirror reaction $^3$H($\alpha$,$\gamma$)$^7$Li,
for which more accurate data exist \cite{Brune}. Our results obtained using the CD-Bonn 2000
NN potential are shown on the right of Fig.~\ref{S-factor_34_Nmax}. It is apparent that
our $^3$H($\alpha$,$\gamma$)$^7$Li results are consistent with our 
$^3$He($\alpha$,$\gamma$)$^7$Be calculation. We are on the lower side of the data
and find an increase of the S-factor as we increase the size of our basis.


More details on the {\it ab initio} NCSM investigation of the $^3$He($\alpha$,$\gamma$)$^7$Be
and $^3$H($\alpha$,$\gamma$)$^7$Li S-factors are given in Ref.~\cite{S_fact_NPA}.

\section{Light nuclei as open systems}\label{sec:NCSMC}

Nuclei are open quantum systems with bound states, unbound resonances, and scattering states.
A realistic {\it ab initio} description of light nuclei with predictive power must
have the capability to describe all the above classes of states within a unified framework. Over the past decade, significant progress has been made in our understanding of the properties of the bound states of light nuclei starting from realistic NN and NNN interactions. This progress was also in part due to the development of the {\it ab initio} NCSM, as described in the previous sections of this review. The solution of the nuclear many-body problem is even more complex, when scattering or nuclear reactions are considered. {\em Ab initio} calculations for scattering processes involving more than four nucleons overall are challenging and still a rare exception~\cite{GFMC_nHe4}. Even calculations of resonant states are quite complicated~\cite{Ha07}.
The development of an {\em ab initio} theory of low-energy nuclear 
reactions on light nuclei is key to further refining our understanding of the fundamental nuclear interactions among the constituent nucleons and providing, at the same time, 
accurate predictions of crucial reaction rates for nuclear astrophysics.  

The use of the harmonic oscillator (HO) basis in the NCSM results in an incorrect description of 
the wave-function asymptotic and a lack of coupling to the continuum.
The first applications of the NCSM to the calculation of nuclear reactions that were described in Sections~\ref{sec:cap_reac} required a phenomenological correction of the asymptotic behavior of the overlap functions. A fully {\it ab initio} approach to nuclear reactions based on the NCSM requires an extension or a modification of the NCSM basis.

\subsection{Coupling of {\it ab initio} no-core shell model with the
  resonating group method}

The resonating-group method (RGM) is a microscopic cluster technique in which the many-body Hilbert space is spanned by cluster wave functions describing a system of two or more clusters in relative motion. Here, we will limit our discussion to the two-cluster RGM, which is based on binary-cluster channel states of total angular momentum $J$, parity $\pi$, and isospin $T$,
\begin{eqnarray}
|\Phi^{J^\pi T}_{\nu r}\rangle &=& \Big [ \big ( \left|A{-}a\, \alpha_1 I_1^{\,\pi_1} T_1\right\rangle \left |a\,\alpha_2 I_2^{\,\pi_2} T_2\right\rangle\big ) ^{(s T)}\nonumber\\
&&\times\,Y_{\ell}\left(\hat r_{A-a,a}\right)\Big ]^{(J^\pi T)}\,\frac{\delta(r-r_{A-a,a})}{rr_{A-a,a}}\,.\label{basis}
\end{eqnarray}
In the above expression, $\big ( \left|A{-}a\, \alpha_1 I_1^{\,\pi_1} T_1\right\rangle$ and $\left |a\,\alpha_2 I_2^{\,\pi_2} T_2\right\rangle$ are the internal (antisymmetric) wave functions of the first and second clusters, containing $A{-}a$ and $a$ nucleons ($a{<}A$), respectively. They are characterized by angular momentum quantum numbers $I_1$ and $I_2$ coupled together to form channel spin $s$. For their parity, isospin and additional quantum numbers we use, respectively, the notations $\pi_i, T_i$, and $\alpha_i$, with $i=1,2$. The cluster centers of mass are separated by the relative coordinate 
\begin{equation}
\vec r_{A-a,a} = r_{A-a,a}\hat r_{A-a,a}= \frac{1}{A - a}\sum_{i = 1}^{A - a} \vec r_i - \frac{1}{a}\sum_{j = A - a + 1}^{A} \vec r_j\,,
\end{equation}
where $\{\vec{r}_i, i=1,2,\cdots,A\}$ are the $A$ single-particle coordinates.
The channel states~(\ref{basis}) have relative angular momentum $\ell$. It is convenient to group all relevant quantum numbers into a cumulative index $\nu=\{A{-}a\,\alpha_1I_1^{\,\pi_1} T_1;\, a\, \alpha_2 I_2^{\,\pi_2} T_2;\, s\ell\}$.
 
The former basis states can be used to expand the many-body wave function according to
\begin{equation}
|\Psi^{J^\pi T}\rangle = \sum_{\nu} \int dr \,r^2\frac{g^{J^\pi T}_\nu(r)}{r}\,\hat{\mathcal A}_{\nu}\,|\Phi^{J^\pi T}_{\nu r}\rangle\,. \label{trial}
\end{equation}
As the basis states~(\ref{basis}) are not anti-symmetric under exchange of nucleons belonging to different clusters, in order to preserve the Pauli principle one has to introduce the appropriate inter-cluster anti-symmetrizer, schematically
\begin{equation}
\hat{\mathcal A}_{\nu}=\sqrt{\frac{(A{-}a)!a!}{A!}}\sum_{P}(-)^pP\,.
\end{equation}   
Here the sum runs over all possible permutations $P$ that can be carried out 
among nucleons pertaining to different clusters, and $p$ is the number of interchanges characterizing them. The coefficients of the expansion are the relative-motion wave functions $g^{J^\pi T}_\nu(r)$, which represent the only unknowns of the problem. To determine them one has to solve the non-local integro-differential coupled-channel equations 
\begin{equation}
\sum_{\nu}\int dr \,r^2\left[{\mathcal H}^{J^\pi T}_{\nu^\prime\nu}(r^\prime, r)-E\,{\mathcal N}^{J^\pi T}_{\nu^\prime\nu}(r^\prime,r)\right] \frac{g^{J^\pi T}_\nu(r)}{r} = 0\,,\label{RGMeq}
\end{equation}
where the two integration kernels, the Hamiltonian kernel,
\begin{equation}
{\mathcal H}^{J^\pi T}_{\nu^\prime\nu}(r^\prime, r) = \left\langle\Phi^{J^\pi T}_{\nu^\prime r^\prime}\right|\hat{\mathcal A}_{\nu^\prime}H\hat{\mathcal A}_{\nu}\left|\Phi^{J^\pi T}_{\nu r}\right\rangle\,,\label{H-kernel}
\end {equation}
and the norm kernel,
\begin{equation}
{\mathcal N}^{J^\pi T}_{\nu^\prime\nu}(r^\prime, r) = \left\langle\Phi^{J^\pi T}_{\nu^\prime r^\prime}\right|\hat{\mathcal A}_{\nu^\prime}\hat{\mathcal A}_{\nu}\left|\Phi^{J^\pi T}_{\nu r}\right\rangle\,,\label{N-kernel}
\end{equation}
contain all the nuclear structure and anti-symmetrization properties of the problem. In particular, the non-locality of the kernels is a direct consequence of the exchanges of nucleons between the clusters. We have used the notation $E$ and $H$ to denote the total energy in the center-of-mass frame, and the intrinsic $A$-nucleon microscopic Hamiltonian, respectively.

The formalism presented above can be combined with the {\em ab initio} NCSM as follows. First, we note that the Hamiltonian can be written as
\begin{equation}\label{Hamiltonian}
H=T_{\rm rel}(r)+ {\mathcal V}_{\rm rel} +\bar{V}_{\rm C}(r)+H_{(A-a)}+H_{(a)}\,,
\end{equation}
where $H_{(A-a)}$ and $H_{(a)}$, the ($A{-}a$)- and $a$-nucleon intrinsic Hamiltonians, respectively, $T_{\rm rel}(r)$ is the relative kinetic energy 
and ${\mathcal V}_{\rm rel}$ is the sum of all interactions between nucleons belonging to different clusters after subtraction of the average Coulomb interaction between them, explicitly singled out in the term $\bar{V}_{\rm C}(r)=Z_{1\nu}Z_{2\nu}e^2/r$ ($Z_{1\nu}$ and $Z_{2\nu}$ being the charge numbers of the clusters in channel $\nu$).   We use identical realistic potentials in both the cluster's Hamiltonians and inter-cluster interaction ${\mathcal V}_{\rm rel}$. Accordingly, $\left|A{-}a\, \alpha_1 I_1^{\,\pi_1} T_1\right\rangle$ and $\left |a\,\alpha_2 I_2^{\,\pi_2} T_2\right\rangle$ are obtained by diagonalizing $H_{(A-a)}$ and $H_{(a)}$, respectively, in the model space spanned by the NCSM basis.  If the adopted potential generates strong short-range correlations, we derive consistent NCSM effective interactions. While the cluster eigenstates are obtained by employing the usual NCSM effective interaction, in place of the bare potential entering ${\mathcal V}_{\rm rel}$ we adopt a modified effective interaction, which avoids renormalizations related to the kinetic energy.  At the two-body cluster level this is given by $V^\prime_{2\rm{eff}}=\bar{H}_{2\rm{eff}}-\bar{H}^\prime_{2\rm{eff}}$, where $\bar{H}^\prime_{2\rm{eff}}$ is the effective Hamiltonian derived from $H^{\Omega\,\prime}_2 = H_{02}+V^\prime_{12}$, with $V^\prime_{12} = - m\Omega^2\vec{r}^{\,2}/A$, compare to Eq.~(\ref{hamomega2}). 
Note that $V^\prime_{2\rm{eff}}\rightarrow V_{N}$ in the limit $N_{\rm max}\rightarrow\infty$. 

As indicated by the presence of the norm kernel ${\mathcal N}^{J^\pi T}_{\nu^\prime\nu}(r^\prime, r)$, Eq.~(\ref{RGMeq}) does not represent a system of multichannel Schr\"odinger equations, and $g^{J^\pi T}_\nu(r)$ do not represent Schr\"odinger wave functions. This short-range non-orthogonality, induced by the non-identical permutations in the inter-cluster anti-symmetrizers, can be removed by introducing normalized Schr\"odinger wave functions
\begin{equation}
\frac{\chi^{J^\pi T}_\nu(r)}{r} = \sum_{\gamma}\int dy\, y^2 {\mathcal N}^{\frac12}_{\nu\gamma}(r,y)\,\frac{g^{J^\pi T}_\gamma(y)}{y}\,,
\end{equation}
where ${\mathcal N}^{\frac12}$ is the square root of the norm kernel, and applying the inverse-square root of the norm kernel, ${\mathcal N}^{-\frac12}$, to both left and right-hand side of the square brackets in Eq.~(\ref{RGMeq}).  This procedure, explaned in more detail in Ref.~\cite{NCSM_RGM_long}, leads to the system of multichannel Schr\"odinger equations:
\begin{eqnarray}
&&[\hat T_{\rm rel}(r) + \bar V_{\rm C}(r) -(E - E_{\alpha_1}^{I_1^{\pi_1} T_1} - E_{\alpha_2}^{I_2^{\pi_2} T_2})]\frac{\chi^{J^\pi T}_{\nu} (r)}{r} \nonumber\\[2mm]
&&+ \sum_{\nu^\prime}\int dr^\prime\,r^{\prime\,2} \,W^{J^\pi T}_{\nu \nu^\prime}(r,r^\prime)\,\frac{\chi^{J^\pi T}_{\nu^\prime}(r^\prime)}{r^\prime} = 0,\label{r-matrix-eq}
\end{eqnarray} 
where $E_{\alpha_i}^{I_i^{\pi_i} T_i}$ is the energy eigenvalue of the $i$-th cluster ($i=1,2$), and $W^{J^\pi T}_{\nu^\prime \nu}(r^\prime,r)$ is the overall non-local potential between the two clusters, which depends upon the channel of relative motion, while it does not depend upon the energy.     

\begin{figure}
\begin{minipage}{6.5cm}
\rotatebox{-90}{ \includegraphics[height=.24\textheight]{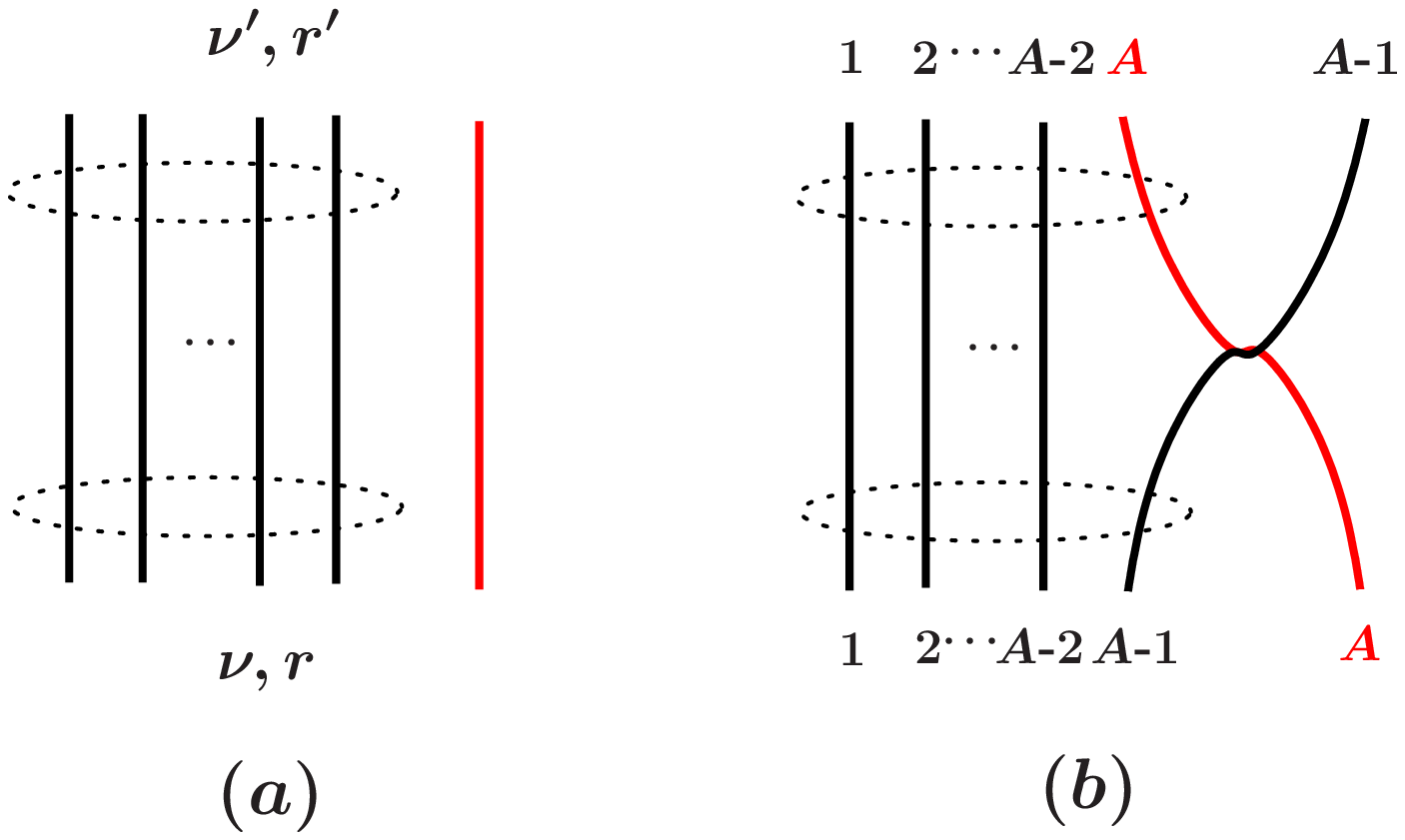}}
\end{minipage}
\begin{minipage}{9.5cm}
\rotatebox{-90}{\includegraphics[height=.39\textheight]{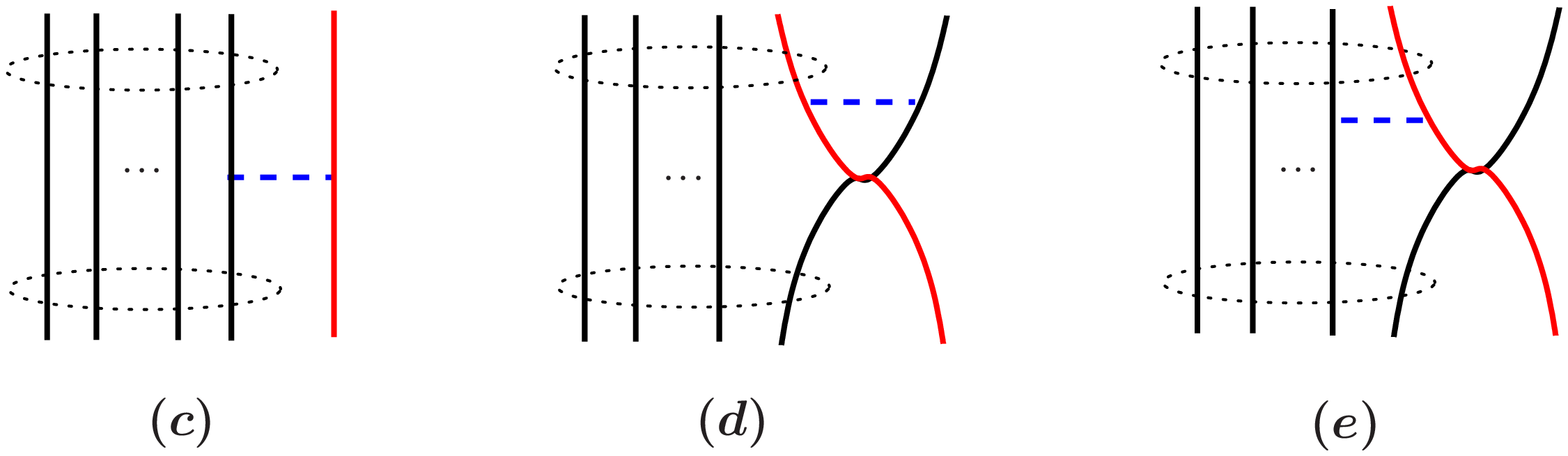}}
\end{minipage}
\caption{Diagrammatic representation of: ($a$) ``direct" and ($b$) ``exchange"  components of the norm kernel; ($c$ and $d$) ``direct"  and ($e$) ``exchange"   components of the potential kernel. The first group of circled black lines represents the first cluster, the bound state of $A{-}1$ nucleons. The separate red line represents the second cluster, in the specific case  a single nucleon. Bottom and upper part of the diagram represent initial and final states, respectively.}\label{diagram-norm-pot}
\end{figure}

So far we have fully developed and tested this formalism in the single-nucleon projectile basis, i.e., for binary-cluster channel states~(\ref{basis}) with $a=1$ (with channel index $\nu = \{ A{-}1 \, \alpha_1 I_1^{\pi_1} T_1; \, 1\, \frac 1 2 \frac 1 2;\, s\ell\}$). In this model space, the norm kernel is rather simple and is given by 
\begin{eqnarray}
{\mathcal N}^{J^\pi T}_{\nu^\prime\nu}(r^\prime, r)& = &\left\langle\Phi^{J^\pi T}_{\nu^\prime r^\prime}\right|1-\sum_{i=1}^{A-1}\hat P_{iA} \left|\Phi^{J^\pi T}_{\nu r}\right\rangle
\\
&=&\delta_{\nu^\prime\,\nu}\,\frac{\delta(r^\prime-r)}{r^\prime\,r}-(A-1)\sum_{n^\prime n}R_{n^\prime\ell^\prime}(r^\prime) R_{n\ell}(r)\nonumber\\
&&\times \left\langle\Phi^{J^\pi T}_{\nu^\prime n^\prime}\right|\hat P_{A-1,A} \left|\Phi^{J^\pi T}_{\nu n}\right\rangle\,,\label{norm}
\end{eqnarray}
where it is easy to recognize a direct term, in which initial and final state are identical (corresponding to diagram $(a)$ of Fig.~\ref{diagram-norm-pot}), and a many-body correction due to the exchange part of the inter-cluster anti-symmetrizer (corresponding to diagram $(b)$ of Fig.~\ref{diagram-norm-pot}). We note that in calculating the matrix elements of the exchange operator $\hat P_{A-1,A}$ we replaced the delta function of eq.~(\ref{basis}) with its representation in the HO model space. This is appropriate whenever (as it is the case of $\hat P_{A-1,A}$) the operator is short-to-medium range. 
The presence of the inter-cluster anti-symmetrizer affects also the Hamiltonian kernel, and, in particular, the matrix elements of the interaction. For a NN potential one obtains a direct term involving interaction and exchange of two nucleons only (see diagrams ($c$) and ($d$) of Fig.~\ref{diagram-norm-pot}), and an exchange term involving three-nucleons. Diagram ($e$) of Fig.~\ref{diagram-norm-pot} describes this latter term, in which  the last nucleon is exchanged with one of the nucleons of the first clusters, and interacts with yet another nucleon. For more details on the integration kernels in the single-nucleon projectile basis we refer the readers to Ref.~\cite{NCSM_RGM_long}.  

Being translationally-invariant  quantities, the norm and Hamiltonian kernels can be ``naturally" derived working within the NCSM Jacobi-coordinate basis. However, by introducing Slater-determinant channel states of the type 
\begin{eqnarray}
|\Phi^{J^\pi T}_{\nu n}\rangle_{\rm SD}   &=&    \Big [\big (\left|A{-}a\, \alpha_1 I_1 T_1\right\rangle_{\rm SD} 
\left |a\,\alpha_2 I_2 T_2\right\rangle\big )^{(s T)}\nonumber\\
&&\times Y_{\ell}(\hat R^{(a)}_{\rm c.m.})\Big ]^{(J^\pi T)} R_{n\ell}(R^{(a)}_{\rm c.m.})\,,
\label{SD-basis}
\end{eqnarray}
in which the eigenstates of the $(A{-}a)$-nucleon fragment are obtained in the SD basis (while the second cluster is still a NCSM Jacobi-coordinate eigenstate), it can be easily demonstrated that translationally invariant matrix elements can be extracted from those calculated in the SD basis of Eq.~(\ref{SD-basis}) by inverting the following expression:
 \begin{eqnarray}
&& {}_{\rm SD}\!\left\langle\Phi^{J^\pi T}_{\nu^\prime n^\prime}\right|\hat{\mathcal O}_{\rm t.i.}\left|\Phi^{J^\pi T}_{\nu n}\right\rangle\!{}_{\rm SD} = \nonumber\\
&&\nonumber\\
&&\sum_{n^\prime_r \ell^\prime_r, n_r\ell_r, J_r}
 \left\langle\Phi^{J_r^{\pi_r} T}_{\nu^\prime_r n^\prime_r}\right|\hat{\mathcal O}_{\rm t.i.}\left|\Phi^{J_r^{\pi_r} T}_{\nu_r n_r}\right\rangle\nonumber\\
&&  \times \sum_{NL} \hat \ell \hat \ell^\prime \hat J_r^2 (-1)^{(s+\ell-s^\prime-\ell^\prime)}
  \left\{\begin{array}{ccc}
 s &\ell_r&  J_r\\
  L& J & \ell
 \end{array}\right\}
 \left\{\begin{array}{ccc}
 s^\prime &\ell^\prime_r&  J_r\\
  L& J & \ell^\prime
 \end{array}\right\}\nonumber\\
 &&\nonumber\\
&& \times\langle  n_r\ell_rNL\ell | 00n\ell\ell \rangle_{\frac{a}{A-a}} 
 \;\langle  n^\prime_r\ell^\prime_rNL\ell | 00n^\prime\ell^\prime\ell^\prime \rangle_{\frac{a}{A-a}} \,.\label{Oti}
 \end{eqnarray}
Here $\hat {\mathcal O}_{\rm t.i.}$ represents any scalar and parity-conserving translational-invariant operator ($\hat {\mathcal O}_{\rm t.i.} = \hat{\mathcal A}$, $\hat{\mathcal A} H \hat{\mathcal A}$, etc.).
We exploited both Jacobi-coordinate and SD channel states to verify our results.  The use of the SD basis is computationally advantageous and allows us to explore reactions involving $p$-shell nuclei.

\subsection{{\it Ab initio} many-body calculations of nucleon-nucleus scattering}
The two-cluster NCSM/RGM formalism outlined in the previous section, can be used to calculate nucleon-nucleus phase shifts below three-body break threshold, by solving the 
system of multi-channel Schr{\"o}dinger equations~(\ref{r-matrix-eq}) with scattering boundary conditions.  Results for neutrons scattering on $^3$H, $^4$He and $^{10}$Be and protons scattering on $^{3,4}$He, using realistic NN potentials, were presented in Refs.~\cite{NCSM_RGM} and~\cite{NCSM_RGM_long}. In the following we review part of these calculations.  

\begin{figure}
\begin{minipage}{8cm}
{ \includegraphics[height=.21\textheight]{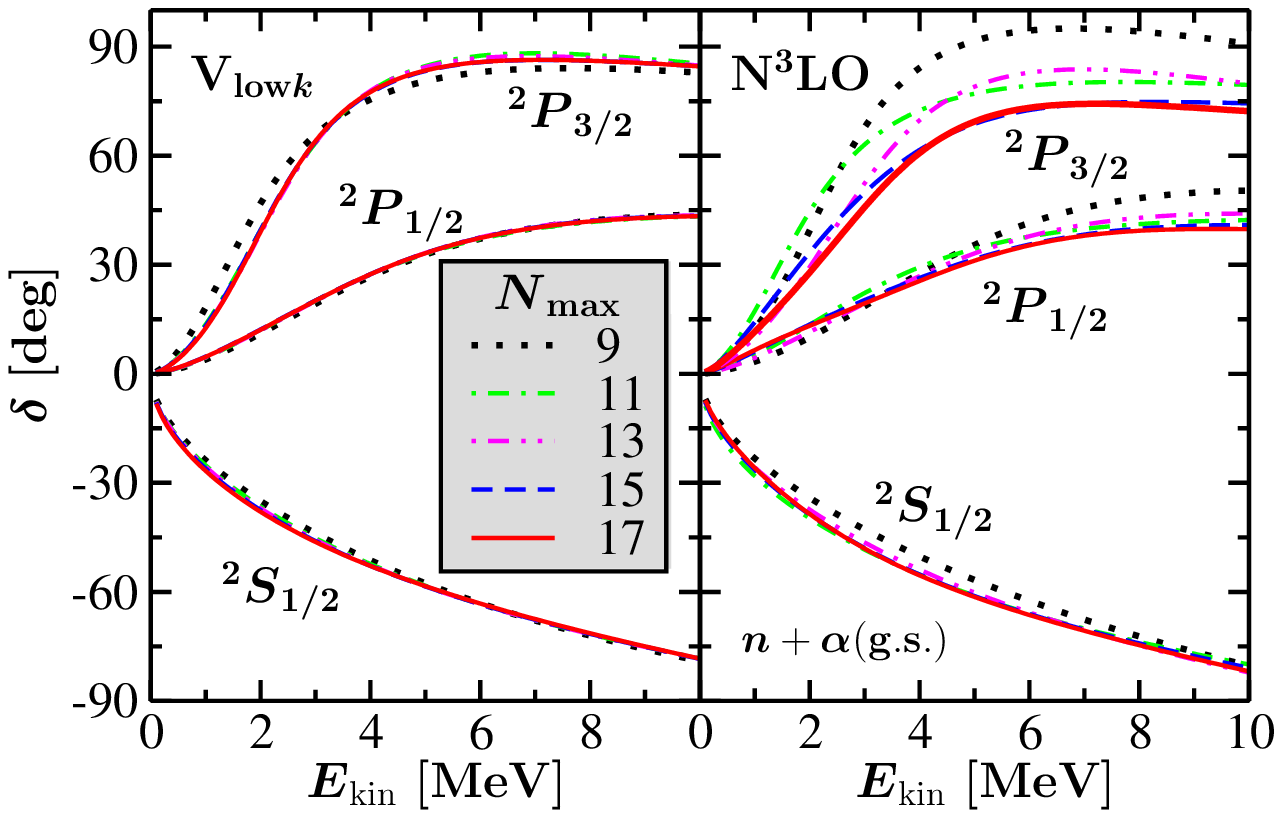}}
\end{minipage}
\begin{minipage}{8cm}
{\includegraphics[height=.20\textheight]{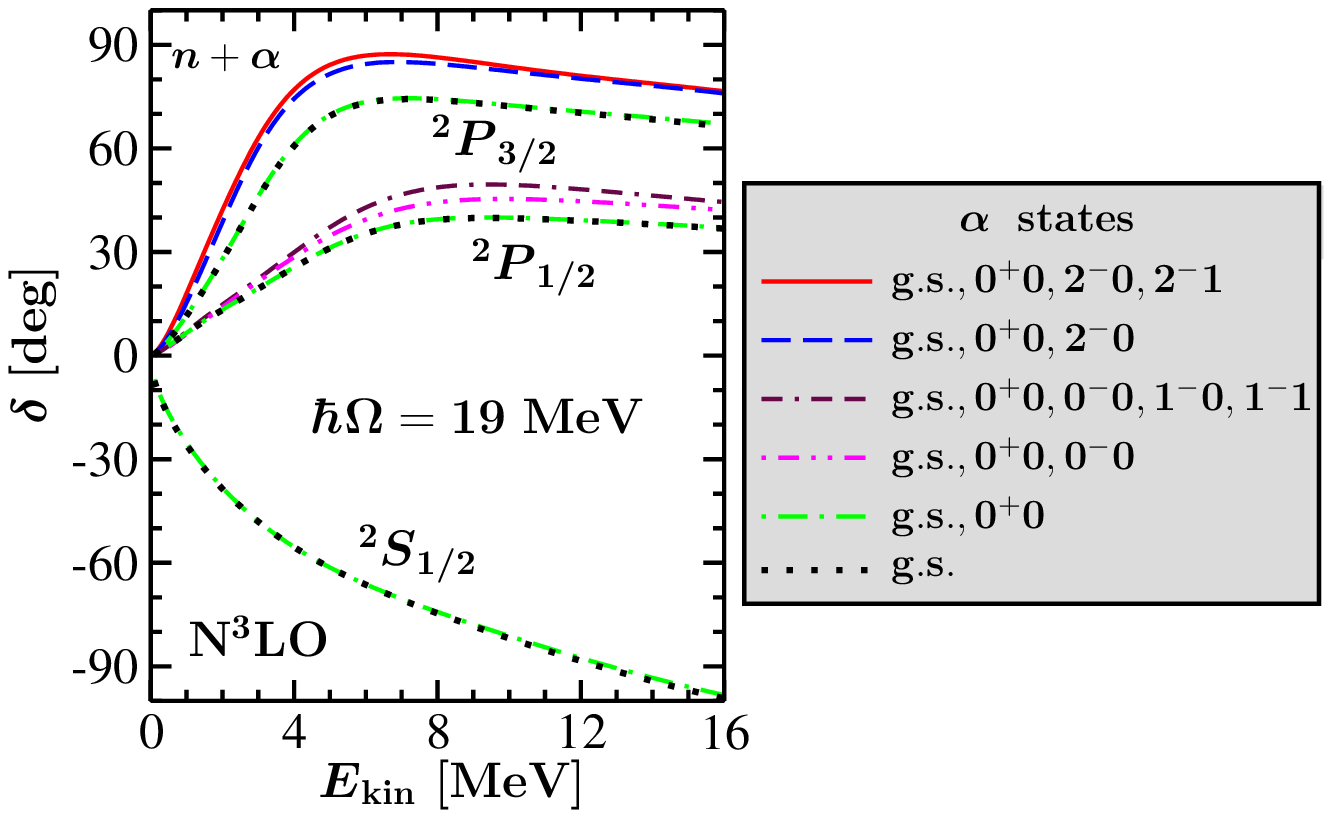}}
\end{minipage}
\caption{Calculated $n$-$\alpha$ phase shifts using the $V_{lowk}$~\cite{Vlowk,Hagenpriv} and N$^3$LO~\cite{N3LO} (right panel) NN potentials at $\hbar\Omega=18$ and $19$ MeV, respectively: (left panel)dependence on $N_{\rm max}$ of the $n$-$\alpha$(g.s.) results; (right panel) influence of  the lowest six excited states ($0^+ 0, 0^- 0,1^-0,1^-1, 2^- 0,2^-1$) of the $\alpha$ particle.}\label{na}
\end{figure}

\begin{figure}
\begin{minipage}{8cm}
\centering{ \includegraphics[height=.23\textheight]{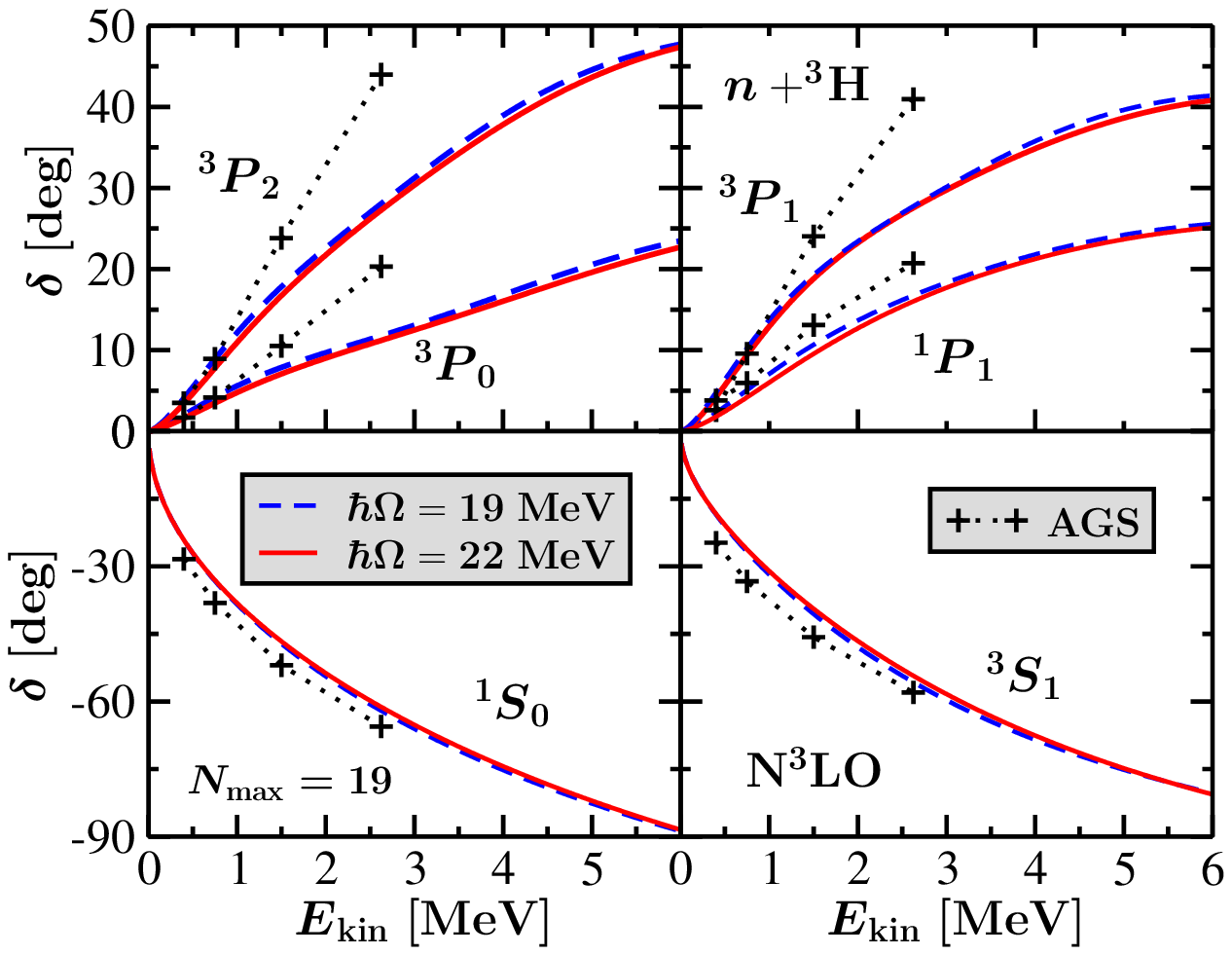}}
\end{minipage}
\begin{minipage}{8cm}
{\includegraphics[height=.23\textheight]{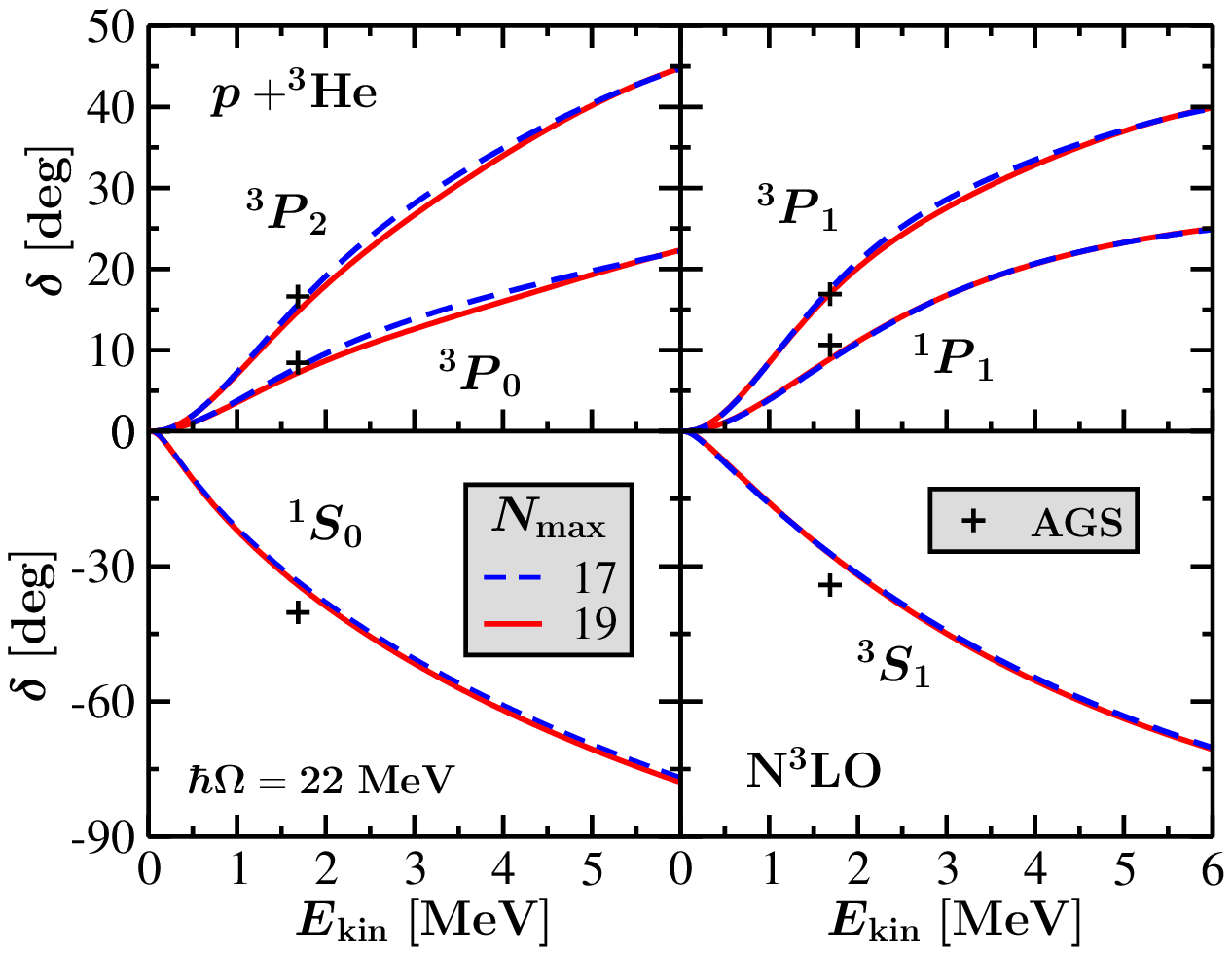}}
\end{minipage}
\caption{Calculated $A=4$ phase shifts using the N$^3$LO NN potential~\cite{N3LO} compared to AGS results of Refs.~\cite{Deltuva, DeltuvaPriv}. In particular: (left panel) $n\,$-${}^3$H
results for $N_{\rm max} = 19$ and  $\hbar\Omega=19$, and $22$ MeV; (right panel) $p\,$-${}^3$He
results for $\hbar\Omega=22$ MeV and $N_{\rm max} = 17$, and $19$. All NCSM/RGM results were obtained in a coupled-channel calculation including only the g.s.\ of the ${}^3$H ($^3$He) nucleus  (i.e., the channels $\nu=\{3\,{\rm g.s.}\,\frac12^+\frac12;\,1\frac12^+\frac12;\,s\,\ell\}$). 
}\label{eigph-n3H-p3He-n3lo} \end{figure}

\begin{table}
 \caption{Calculated $^3$H  and $^3$He g.s.\ energies (in MeV), $n\,$-${}^3$H and $p$-$^3$He phase shifts (in degrees), and $n$-$^3$H total cross section (in barns)  for increasing $N_{\rm max}$ at $\hbar\Omega$ = $18$ MeV, obtained using the $V_{{\rm low}k}$ NN potential~\cite{Vlowk,Hagenpriv}. The $n$-$^3$H ($p$-$^3$He) scattering results were obtained in a coupled-channel calculation including only the g.s.\ of the ${}^3$H ($^3$He) nucleus  (i.e., the channels $\nu=\{3\,{\rm g.s.}\,\frac12^+\frac12;\,1\frac12^+\frac12;\,s\,\ell\}$), see text for explanation of notation.}\label{tab-a}   
\begin{tabular}{c c c c c c c c c c}
&$^3$H&\multicolumn{8}{c}{$n\,$-${}^3$H ($E_{\rm kin}=0.40$ MeV)}\\[1mm]\cline{2-2}\cline{3-10}
\\[-3.5mm]
$N_{\rm max}$&$E_{\rm g.s.}$&$0^+$ ($^1S_0$)&$0^-$ ($^3P_0$)&$1^+$ ($^3S_1$)&$1^-$ ($^1P_1$)& $1^-$ ($^3P_1$)&$1^-$ $(\epsilon)$&$2^-$ ($^3P_2$)&$\sigma_t$\\[1mm]
\hline\\[-3mm]
$9$&$-7.80$  & $-20.2$&$0.93$&$-18.9$&$0.85$&$1.96$&$-18.0$&$3.01$&$0.99$\\
$11$&$-7.96$&$-22.9$&$0.97$&$-20.4$&$1.04$&$2.36$&$-13.0 $&$2.58$&$1.15$\\
$13$&$-8.02$&$-23.7$&$0.87$&$-21.0$&$1.24$&$2.47$&$\,\,-9.0$&$2.30$&$1.22$\\
$15$&$-8.11$&$-24.4$&$1.00$&$-21.8$&$1.40$&$2.44$&$\,\,-9.1$&$2.41$&$1.31$\\
$17$&$-8.12$&$-25.1$&$1.06$&$-22.6$&$1.52$&$2.52$&$-10.4 $&$2.45$&$1.39$\\
$19$&$-8.16$&$-25.6$&$1.01$&$-22.9$&$1.64$&$2.60$&$\,\,-9.7$&$2.37$&$1.43$\\[2mm]
&$^3$He&\multicolumn{8}{c}{$p\,$-${}^3$He ($E_{\rm kin}=0.75$ MeV)}\\[1mm]\cline{2-2}\cline{3-10}
\\[-3.5mm]
$N_{\rm max}$&$E_{\rm g.s.}$&$0^+$ ($^1S_0$)&$0^-$ ($^3P_0$)&$1^+$ ($^3S_1$)&$1^-$ ($^1P_1$)& $1^-$ ($^3P_1$)&$1^-$ $(\epsilon)$&$2^-$ ($^3P_2$)\\[1mm]
\hline\\[-3mm]
$9$&$-7.05$&$-12.6$&$1.14$&$-12.5$&$1.04$&$2.29$&$-17.2$&$3.38$\\
$11$&$-7.22$&$-15.9$&$1.30$&$-13.6$&$1.35$&$2.83$&$-12.5$&$3.05$\\
$13$&$-7.29$&$-16.0$&$1.34$&$-13.9$&$1.73$&$3.15$&$\,\,-8.6$&$2.93$\\
$15$&$-7.37$&$-16.8$&$1.63$&$-14.4$&$2.07$&$3.28$&$\,\,-8.4$&$3.20$\\
$17$&$-7.39$&$-17.0$&$1.87$&$-14.9$&$2.41$&$3.56$&$-10.0$&$3.46$\\
$19$&$-7.42$&$-17.4$&$1.95$&$-14.9$&$2.71$&$3.83$&$-9.16$&$3.51$
\end{tabular}
\end{table}
To study the behavior of our approach with respect to the HO model space, we performed NCSM/RGM scattering calculations for the $A=4$ and $5$ systems, using the $V_{{\rm low}k}$ NN potential~\cite{Vlowk}, which is ``soft" and can be treated as ``bare", and the N$^3$LO NN interaction~\cite{N3LO}, which generates strong short-range correlations, thus requiring the use of effective interactions. In particular, for these convergence tests, we restricted our binary-cluster basis to target-nucleon channel states with the target in its g.s.\ (corresponding to channel indexes of the type $\nu=\{A-1\;{\rm g.s.}\,I_1^{\pi_1}T_1;\,1\frac12^+\frac12;\,s\,\ell\}$). A sample of the results obtained for $V_{{\rm low}k}$ is presented in Table~\ref{tab-a} and in the left panel of Fig.~\ref{na}. The overall convergence is quite satisfactory, with a weak dependence on $N_{\rm max}$. The slightly larger differences presented by the phase shifts of small magnitude in Table~\ref{tab-a} is in part a reflection of the sharp cutoff function used to derive the adopted version of $V_{{\rm low}k}$ from AV18 with a cutoff $\Lambda=2.1$ fm$^{-1}$.  
As emphasized by the left panel of Fig.~\ref{na}, presenting the $n$-$\alpha$ scattering phase shifts, the convergence rate for N$^3$LO (achieved by using two-body effective interactions tailored to the HO model-space truncation), is much slower than that obtain with $V_{{\rm low}k}$. However, a gradual suppression of the difference between adjacent $N_{\rm max}$ values with increasing model-space size is visible, although the pattern is somewhat irregular for the $P$ phase shifts. Although not shown, the $p$-$\alpha$ phase shifts present analogous convergence properties. The situation for the $n$-$^3$H and $p$-$^3$He systems is similar, and is shown in Fig.~\ref{eigph-n3H-p3He-n3lo}. Results in the largest model spaces ($N_{\rm max}=17$ and $19$) are very close. An additional sign of convergence is provided by the rather good frequency independence presented by the $N_{\rm max}=19$ $n$-$^3$H phase shifts. 

The two panels of Fig.~\ref{eigph-n3H-p3He-n3lo} show also the results ($+$ symbols) obtained for the same N$^3$LO NN potential~\cite{N3LO} by Deltuva and Fonseca~\cite{Deltuva, DeltuvaPriv} from the solution of the Alt, Grassberger and Sandhas (AGS) equations. The discrepancy, increasing with the energy, between the two calculations highlights the influence played by closed channels not included in our basis states, that is, target-nucleon channel states with the target above the $I_1^{\pi_1}=\frac12^+$ g.s., and 2+2 configurations, both of which are taken into account in the AGS results.    
Because these states correspond to the breakup of the $A=3$ system, it is not feasible to include them in the current version of the NCSM/RGM approach, which so far has been derived only in the single-nucleon projectile basis. However, we are planning on extending our approach to be able to account for the target breakup, and these development will be discussed in future publications. 
For the time being we can explore the effects of the virtual excitations of the target on the nucleon-$\alpha$ scattering, where much higher energies are needed in order to break up the $\alpha$ particle. This is done in the right panel of Fig.~\ref{na}, which presents the effect of the inclusion of the first six excited states of $^4$He on the $^2S_{1/2}$, $^2P_{1/2}$ and $^2P_{3/2}$ scattering phase shifts. The use of the five different combinations of ground and excited states shown in the legend of Fig.~\ref{na} indicates that the $^2S_{1/2}$ is well described already by coupled-channel calculations with g.s. and first $0^+0$ (the $^2S_{1/2}$ phase shifts obtained in the four larger Hilbert spaces are omitted for clarity of the figure). On the other hand, the negative-parity excited states have relatively large effects on the $P$ phase shifts,  and, in particular, the $0^-0,1^-0$ and $1^-1$ states mostly on the $^2P_{1/2}$ phase shifts, whereas the $2^-0$ and $2^-1$ states on the $^2P_{3/2}$.
\begin{figure}
\centering{\includegraphics[height=.25\textheight]{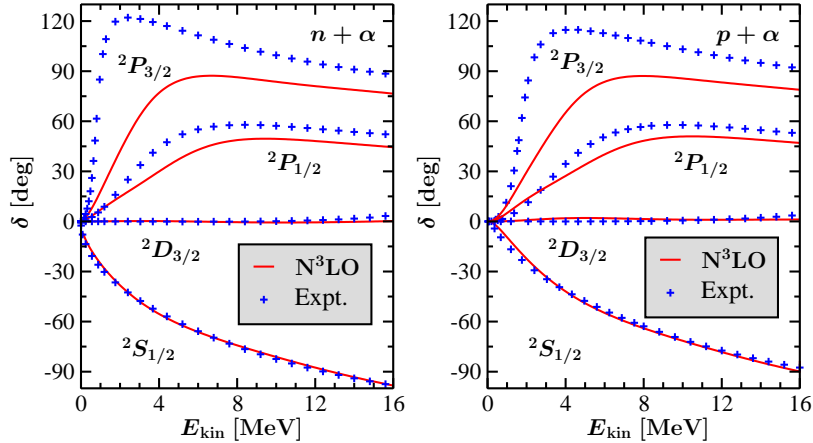}}
\caption{Calculated phase shifts for (left panel) $n$-$\alpha$  and (right panel)  $p\,$-$\alpha$ scattering, using the N$^3$LO NN potential~\cite{N3LO}, compared to an $R$-matrix analysis of data ($+$)~\cite{HalePriv}. Theoretical results include the $^4$He g.s., $0^+ 0$, $0^- 0$, $1^-0$, $1^-1$,  $2^- 0$, and $2^-1$ states.}\label{npaexp}
\end{figure}

The comparison with an accurate $R$-matrix analysis of the nucleon-$\alpha$ scattering~\cite{HalePriv}, presented in Fig.~\ref{npaexp}, reveals that for both neutron (left panel) and proton (right panel) projectiles we can describe quite well the $^2S_{1/2}$ and, qualitatively, also the $^2D_{3/2}$ phase shifts, using the N$^3$LO NN potential. On the other hand, the same interaction is not able to reproduce well the two $P$ phase shifts, which are both too small and too close to each other. This lack of spin-orbit splitting between the $^2P_{1/2}$ and $^2P_{1/2}$ results can be explained by the omission in our treatment of the NNN terms of the chiral interaction, which would provide an additional spin-orbit force. This sensitivity of the $P$ phases to the strength of the spin-orbit force corroborated by the differences among the $V_{{\rm low}k}$ and N$^3$LO results in the left panel of Fig.~\ref{na}:  $^2P_{1/2}$  and $^2P_{3/2}$ are both larger and more separated for $V_{{\rm low} k}$.  The good agreement of the N$^3$LO $^2S_{1/2}$ phase shifts with their $V_{{\rm low}k}$ analogues and with the $R$-matrix analysis can be credited to the repulsive action (in this channel) of the Pauli exclusion principle for short nucleon-$\alpha$ distances, which has the effect of masking the short-range details of the nuclear interaction. 
\begin{figure}
\begin{minipage}{8cm}
\centering{ \includegraphics[height=.25\textheight]{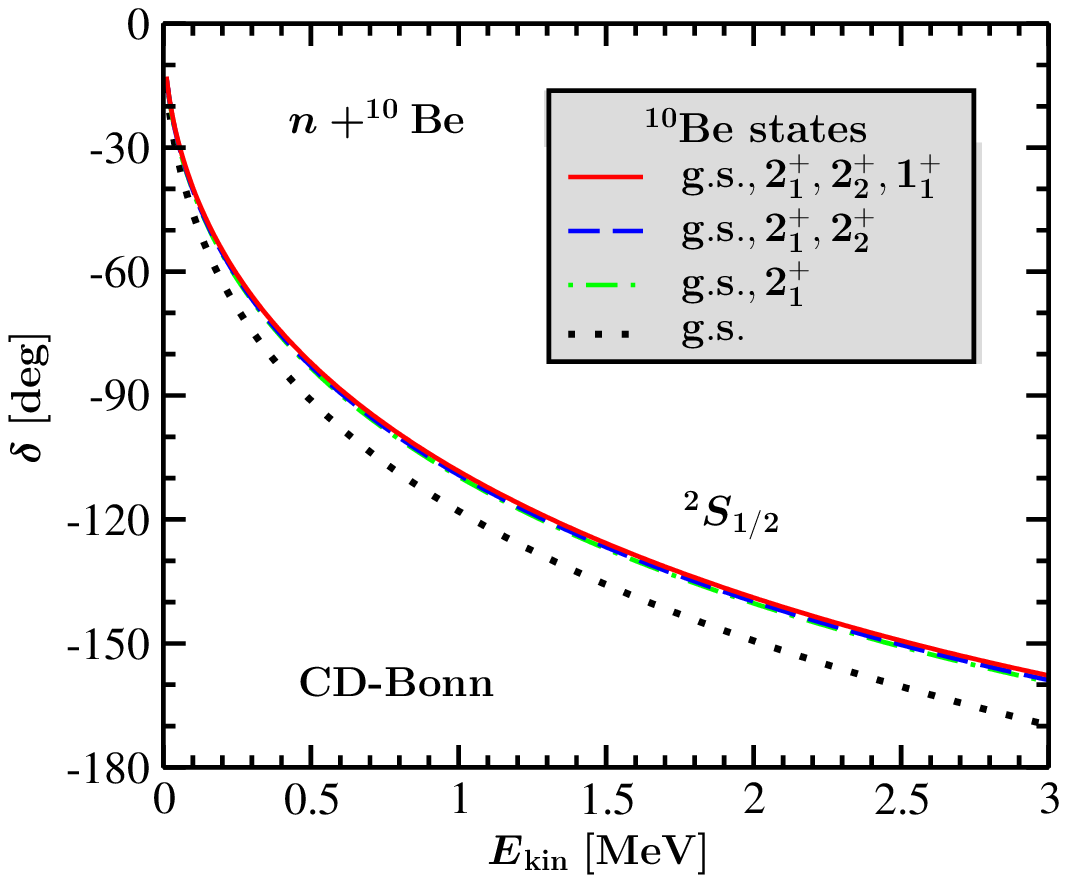}}
\end{minipage}
\begin{minipage}{8cm}
\centering{\includegraphics[height=.25\textheight]{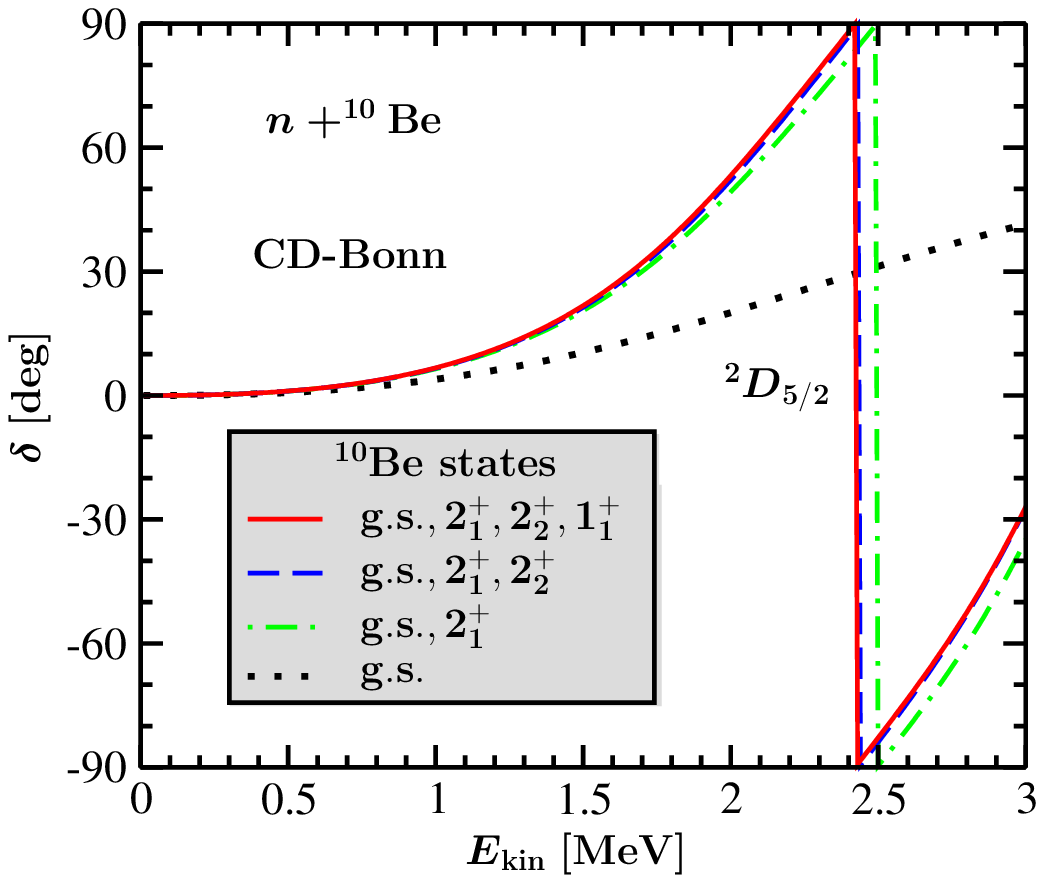}}
\end{minipage}
\caption{Calculated $n\,$-${}^{10}$Be phase shifts as a function of the relative kinetic energy in the c.m.\ frame $E_{\rm kin}$, using the CD-Bonn NN potential~\cite{cdb2k} at $\hbar\Omega=13$ MeV: (left panel) $^2S_{1/2}$ and (right panel) $^2D_{5/2}$ results. The NCSM/RGM results were obtained using $n+^{10}$Be configurations with $N_{\rm max}$ = 6 g.s., $2^+_1$, $2^+_2$, and $1^+_1$ states of $^{10}$Be. The obtained $^2S_{1/2}$ scattering length is $+10.7$ fm.}\label{n10Be}
\end{figure}

Figure~\ref{n10Be} highlights one of the promising aspects of the NCSM/RGM approach, that is the ability (through the use of SD channel states) to perform {\em ab initio} scattering calculations for $p$-shell nuclei. The $^2S_{1/2}$ (left panel) and $^2D_{5/2}$ (right panel) $n$-$^{10}Be$ phase shifts  were obtained in a $N_{\rm max}=6$, $\hbar\Omega=13$ MeV HO model space. The inclusion of the $2_1^+$ excited state of $^{10}$Be has a significant effect on the $S$ and more importantly on the $D$ phase, where it is essential for the appearance of a resonance below 3 MeV.  We note that a resonance has been observed at $\sim 1.8$ MeV with a tentative spin assignment  of $(5/2,3/2)^+$~\cite{ajz90:506rev}. The further addition of the $2_2^+$ and, especially, the $1_1^+$ excited states produces rather weak differences.      We have also extracted the scattering length for the $^2S_{1/2}$ partial wave and found a result of $+10.7$ fm, which is comparable to the value of $+13.6$ fm obtained by Descouvemont in Ref.~\cite{11Be-Volkov}, when fitting the experimental binding energy of $^{11}$Be. 

\subsection{Parity-inverted ground state of $^{11}$Be}
Although we mainly described its scattering applications, the NCSM/RGM is a powerful tool also for structure calculations, particularly for loosely-bound systems. By imposing bound-state boundary conditions to the set of coupled channel Schr\"odinger equations of Eq.~(\ref{r-matrix-eq}), we tested the performance of our single-nucleon projectile NCSM/RGM formalism for the description of one-nucleon halo systems. In particular, because of the well-known  parity-inversion between its two bound states with respect to the predictions of the simple shell model~\cite{tal60:4}, the $^{11}$Be nucleus represents an excellent test ground for our approach.

As discussed in Sec.~\ref{sec:berillium}, large-scale {\em ab initio} NCSM calculations with several accurate NN potentials of the $^{11}$Be low-lying spectrum were not able to explain its g.s. parity inversion~\cite{Fo05}. The explanation for these results can be related to two main causes: (i) the size of the HO basis was not large enough to reproduce the correct asymptotics of the $n$-$^{10}$Be component of the 11-body wave function; and (ii) the NNN force, not included in the calculation, plays an important role in the inversion mechanism.  The second hypothesis was corroborated by the results obtained with the INOY NN potential~\cite{INOY}, which produced the lowest excitation energy of the $1/2^+$ state compared to other NN potentials. By studying the $^{11}$Be bound states in a NCSM/RGM model space spanned by the $n$-$^{10}$Be channel states with inclusion of the $N_{max}=6$ g.s., $2_1^+,2_2^+$, and $1_1^+$ of $^{10}$Be, we are now in the position to address the first hypothesis. Indeed, the correct asymptotic behavior of the $n$-$^{10}$Be wave functions is described naturally in the NCSM/RGM approach.
 
\begin{table*}
\caption{Calculated energies (in MeV) of the $^{10}$Be g.s.\ and  of the lowest negative- and positive-parity states in $^{11}$Be, obtained using the CD-Bonn NN potential~\cite{cdb2k} at $\hbar\Omega=13$ MeV. The NCSM/RGM results were obtained using $n+^{10}$Be configurations with $N_{\rm max}$ = 6 g.s., $2^+_1$, $2^+_2$, and $1^+_1$ states of $^{10}$Be.}\label{11Be}
\begin{tabular}{clccclrclrc}
&&&$^{10}$Be&&\multicolumn{2}{c}{$^{11}$Be($\frac12 ^-$)}&&\multicolumn{2}{c}{$^{11}$Be($\frac12 ^+$)}&\\[0.7mm]\cline{4-4}\cline{6-7}\cline{9-10}\\[-4mm]
&&$N_{\rm max}$&$E_{\rm g.s.}$ &&\multicolumn{1}{c}{$E$}&\multicolumn{1}{c}{$E_{th}$}&&\multicolumn{1}{c}{$E$}&\multicolumn{1}{c}{$E_{th}$}&\\[0.5mm]
\hline
&NCSM~\cite{10Be,Fo05} & $8/9$& $-57.06$&&$-56.95$&$0.11$&&$-54.26$&$2.80$&\\
&NCSM~\cite{10Be,Fo05} & $6/7$& $-57.17$&&$-57.51$&$-0.34$&&$-54.39$&$2.78$&\\
&NCSM/RGM\cite{NCSM_RGM} &&&&$-57.59$&$-0.42$&&$-57.85$&$-0.68$&\\
&Expt. & & $-64.98$&&$-65.16$&$-0.18$&&$-65.48$&$-0.50$& 
\end{tabular}
\end{table*}
The energies of the lowest $1/2^+$ and $1/2^-$ states of $^{11}$Be obtained in the NCSM and in the NCSM/RGM calculations, using the same CD-Bonn NN interaction~\cite{cdb2k} at $\hbar\Omega=13$ MeV adopted in Ref.~\cite{Fo05}, are presented in Table~\ref{11Be}.  The relatively small differences between the $N_{\rm max}=6/7$ and $N_{\rm max}=8/9$ NCSM results, seems to indicate a reasonable degree of convergence for these calculations. The $1/2^-$ state appears to be the g.s., and the $1/2^+$ state is about 2.8 MeV above the $n\,$-$^{10}$Be threshold. A comparison to the NCSM/RGM calculations (obtained in a model space including g.s., $2^+_1$, $2^+_2$, and $1^+_1$ states of $^{10}$Be) shows a rough agreement  for the $1/2^-$ state, whereas for the $1/2^+$ state one observes a dramatic difference ($\sim$3.5 MeV) in the energy. The $1/2^-$ and $1/2^+$ NCSM/RGM states are both bound and the $1/2^+$ state is the g.s.\ of $^{11}$Be. Correspondingly, we obtain a B(E1; $\frac12^-\rightarrow\frac12^+$) value of $0.18$ $e^2$ fm$^2$, which is not far from experiment.

\begin{table}
\caption{Mean values of the relative kinetic and potential energy and of the internal $^{10}$Be energy in the $^{11}$Be $1/2^+$ ground state. All energies in MeV. NCSM/RGM calculation as in Table~\ref{11Be}. See the text for further details.}\label{11Be_gs_analys}
\begin{tabular}{lcccc}
NCSM/RGM & $\langle T_{\rm rel} \rangle$ & $\langle W\rangle$ & $E[^{10}{\rm Be(g.s.,ex.)}]$ & $E_{\rm tot}$\\
\hline
Model Space & $16.65$  & $-15.02$   & $-56.66$ & $-55.03$ \\
Full                  & $\;\,6.56$ & $\;\,-7.39$ & $-57.02$ & $-57.85$
\end{tabular}
\end{table}
To understand the mechanism that makes the $1/2^+$ state bound in the NCSM/RGM, we evaluated mean values of the relative kinetic and potential energies as well as the mean value of the $^{10}$Be energy, and compared them to those obtained by restricting all the integration kernels within the HO model space (i.e., by replacing the delta function of Eq.~(\ref{norm}) with its representation in the HO model space). These results are shown in Table~\ref{11Be_gs_analys}. The model-space-restricted calculation is then similar, although not identical, to the standard NCSM calculation. In particular, as in the NCSM one loses the correct asymptotic behavior of the $n$-$^{10}$Be wave function. We observe that in the full NCSM/RGM calculation both relative kinetic and potential energies are smaller in absolute value. This is an effect of the re-scaling of the relative wave function in the internal region, when the Whittaker tail is recovered. The difference is significantly more substantial for the relative kinetic energy than for the potential energy. As a result one obtains a dramatic decrease of the energy of the $1/2^+$ state, which makes it bound and even leads to a g.s.\ parity inversion. This study shows that a proper treatment of the coupling to the $n\,$-${}^{10}$Be continuum is essential in explaining the g.s.\ parity inversion. However, we cannot exclude that the NNN force plays a role in the inversion mechanism, until accurate calculations with both the NNN force and full treatment of the $n$-$^{10}$Be tail have been performed.   

\subsection{Outlook: {\it Ab initio} no-core shell model with continuum}

It is possible and desirable to extend the binary-cluster $(A{-}a,a)$ NCSM/RGM basis by the standard $A$-nucleon NCSM basis to unify the original {\em ab initio} NCSM and NCSM/RGM approaches. This will lead to a much faster convergence of the many-body calculations compared to the original approaches and, most importantly, to an optimal and balanced unified description of both bound and unbound states. 

In particular, we can generalize the expansion of the many-body wave function given in Eq.~(\ref{trial}) by explicitly including a set of $A$-nucleon NCSM eigenstates:
\begin{equation}
|\Psi^{J^\pi T}\rangle = \sum_{\lambda}c_\lambda^{J^\pi T}|A\lambda J^\pi T\rangle + \sum_{\nu} \int dr \,r^2\frac{g^{J^\pi T}_\nu(r)}{r}\,\hat{\mathcal A}_{\nu}\,|\Phi^{J^\pi T}_{\nu r}\rangle\,, \label{trial_NCSMC}
\end{equation}
where $H_A^{\rm NCSM}|A\lambda J^\pi T\rangle=E_\lambda|A\lambda J^\pi T\rangle$ with $H_A^{\rm NCSM}$ given by, e.g., $H_A$ of Eq.~(\ref{ham}) projected on the $N_{\rm max}\hbar\Omega$ space or by the NCSM effective Hamiltonians (\ref{Ham_A_Omega_2eff}) or (\ref{Ham_A_Omega_eff}) (with the $H_{\rm CM}$ subtracted) also defined on the $N_{\rm max}\hbar\Omega$ space. By projecting the many-body Schr\"odinger equation on the binary-cluster channel states (\ref{basis}) and the NCSM eigenstates, we arrive at a system of coupled equations that can be schematically written as
\begin{equation}
\left(\begin{array}{cc} H^{\rm NCSM} & {\mathfrak{h}} \\ {\mathfrak{h}} & {\mathcal H} \end{array}\right)
\left(\begin{array}{c} c \\ g \end{array} \right) = 
E \left(\begin{array}{cc} 1 & {\mathfrak{g}} \\ {\mathfrak{g}} & {\mathcal N} \end{array} \right)
\left(\begin{array}{c} c \\ g \end{array} \right) \; .\label{eq:NCSMC}
\end{equation}
Here, ${\mathcal H}$ and ${\mathcal N}$ are the Hamiltonian and norm integration kernels defined in Eqs.~(\ref{H-kernel}) and (\ref{N-kernel}), respectively. The ${\mathfrak g}$ are the overlap functions introduced in Eq.~(\ref{cluster_form_factor}) and the ${\mathfrak h}$ are ``vertex'' functions defined by 
matrix elements $\langle A \lambda J^\pi T|H\hat{\mathcal A}_\nu|\Phi^{J^\pi T}_{\nu r}\rangle $ with $H$ the intrinsic Hamiltonian that can be expressed, e.g., as in Eq.~(\ref{Hamiltonian}). It is straightforward to implement this new approach that we name {\it ab initio} NCSM with the continuum (NCSMC).

\section{Conclusions}\label{sec:Concl}

The {\it ab initio} NCSM has evolved into a powerful many-body technique.
In this review, we presented some of recent results obtained within this approach. We discussed, in particular, calculations with chiral EFT NN and NNN interactions for both $s$-shell and $p$-shell nuclei used, on the one hand, as a tool to determine the NNN interaction low-energy constants and, on the other hand, to predict properties of light nuclei. These calculations demonstrate the importance of the NNN interaction for nuclear structure. Recent advances in experimental techniques that allowed precise measurements of radii and moments of exotic isotopes motivated us to perform large basis NCSM calculations for He, Li and Be isotopes. These calculations were overviewed in this paper. We also described efforts to extend the NCSM calculations to larger model spaces and heavier nuclei by means of the importance-truncated calculations and by development of effective interactions for model spaces with a closed core. The most significant new developments, at least in our view, were discussed in the last part of this review. Extension of the NCSM to describe scattering and nuclear reactions via the RGM technique serves as a bridge to a development of a unified {\it ab initio} description of light nuclei with both bound and unbound states described simultaneously and treated on the same footing. Extensions of the NCSM/RGM formalism  to include two-nucleon (deuteron), three-nucleon (triton and $^3$He) and four-nucleon ($^4$He) projectiles are now under way. As a large HO basis expansion is needed in this formalism, not just for the convergence of the target and projectile eigenstates, but also for the convergence of the localized parts of the integration kernels, a combination of this approach with the importance-truncated NCSM is key in a successful application of the NCSM/RGM and the NCSM with  the continuum to heavier nuclei. 

There are other developments in the NCSM calculations that were not covered by
this review that deserves attention.  One of them is the recent analysis of
NCSM wave functions by means of the representations of the symplectic Sp(3,R)
group~\cite{Dy07, Dytrych08Review} and attempts to develop an NCSM code for
calculations within SU(3) $\subset$ Sp(3,R) symmetry-adapted basis. This
approach aims to augment the model space by nuclear collective correlations,
which are required for a microscopic description of monopole and quadrupole
vibrational and rotational dynamics. This would allow, in particular, a
realistic description of giant quadrupole resonances.

Further, the NCSM calculations were also used as a tool for development of the similarity renormalization group evolved NN interactions~\cite{Roth_SRG,Bo07} and the UCOM interactions~\cite{UCOM}. Also, the NCSM method was recently adapted for systems of strongly interacting bosons in a trap~\cite{Christensson09}.

Finally, there are remarkable advances in the development of the code MFD~\cite{MFD}. Calculations on tens of thousands of processors are now possible~\cite{Ma09} and results in model spaces up to $N_{\rm max}=10$ for $^{12}$C and $^{14}$N with NN interactions are within reach~\cite{JPV_private}. Furthermore, calculations in the $N_{\rm max}=8$ model space with the NN and NNN interactions from Section~\ref{sec:chiral_NN_NNN}  have already been performed for $A=7$ and $A=8$ nuclei.

\ack
We would like to thank all the collaborators that contributed to the cited papers and, in particular, Alexander Lisetskiy for input for Section~\ref{sec:core} and Robert Roth for input for Section~\ref{sec:IT}. We also thank D. Furnstahl for useful comments.
Prepared by LLNL under Contract DE-AC52-07NA27344.
This work was supported by the LDRD contract No.~PLS-09-ERD-020, by the
U.S.\ DOE/SC/NP (Work Proposal Number SCW0498) and by the UNEDF SciDAC
Collaboration under DOE grant DE-FC02-07ER41457. 
B.R.B. acknowledges partial support from NSF grants PHY0244389 and PHY0555396 and thanks the GSI Helmholzzentrum f\"ur Schwerionenforschung, Darmstadt, Germany, for its hospitality during the preparation of this manuscript and to the Alexander von Humboldt Stiftung for its support. B.R.B. and S.Q. thank the Institute for Nuclear Theory at the University of Washington for its hospitality and the Department of Energy for partial support during the completion of this work.

\end{document}